# Light communicative materials

Hongshuang Guo,[1] Kai Li,[2]* Jianfeng Yang,[1] Dengfeng Li,[1] Fan Liu,[1] Hao Zeng[1]*

**Affiliation:**

[1] Faculty of Engineering and Natural Sciences, Tampere University, P.O. Box 541, FI-33101 Tampere, Finland.

[2] Department of Civil Engineering, Anhui Jianzhu University, Hefei 230601, China.

*Correspondence to: kli@ahjzu.edu.cn; hao.zeng@tuni.fi.

**Abstract**

The natural interactive materials under far-from-equilibrium conditions have significantly inspired advances in synthetic biomimetic materials. In artificial systems, gradient diffusion serves as the primary means of interaction between individuals, lacking directionality, sufficient interaction ranges and transmission rates. Here, we present a method for constructing highly directed, communicative structures via optical feedback in light responsive materials. We showcase a photomechanical operator system comprising a baffle and a soft actuator. Positive and negative operators are configured to induce light-triggered deformations, alternately interrupting the passage of two light beams in a closed feedback loop. The fundamental functionalities of this optically interconnected material loop include homeostasis-like self-oscillation and signal transmission from one material to another via light. Refinements in alignment facilitate remote sensing, fiber-optic/long-distance communication, and adaptation. These proof-of-concept demonstrations outline a versatile design framework for light-mediated communication among responsive materials, with broad applicability across diverse materials.

## Introduction

The exploration of bioinspired material systems has evolved from static functionalities,[1-3] such as structural colors, roughness, and high strength, towards embracing heightened levels of stimuli-responsive dynamism.[4-6] This paradigm shift encompasses mechanisms such as self-regulation,[7] feedback,[8] entrainment,[9] and synergy,[10] reflecting the adaptive characteristics observed in living organisms. To fully leverage the intricate dynamism akin to life, a roadmap is necessary to delineate the internal feedback mechanisms between the stimulus source and material responsiveness,[11,12,13] as well as their external interactions and communications across different entities.[14]

In a biological context, communication denotes the interactive behavior wherein one organism influences the present or future actions of another. Beyond communication methods reliant on sophisticated biological components (*e.g.* vision-based),[15] there exist three basic forms of communication of particular interest to reductionists. The first involves physical contact, exemplified by species of army ants forming mechanical structures for group transport,[16] and hydrodynamically cooperative cilia generating metachronal waves in liquids.[17] The second occurs through the exchange of biochemical substances, as beetles and flies utilize pheromones for mating.[18] The third form operates via contactless means, illustrated by phenomena like bird echoes and the synchronized flashing of fireflies.[19] In the realm of bioinspired material research, the utilization of out-of-equilibrium[20] soft matters has enabled versatile life-like functions[21,22] and the opportunities for mutual interaction mainly through contact-based methods.[14,23-26] These have spurred the emergence of novel developments across broader disciplinary perspectives, including mechanical intelligence,[27,28] microrobot swarm,[29,30] collective matters,[31-33] and systems chemistry.[34-36] These research efforts underscore the autonomy and adaptive nature of interactive constructs, paving the way for new trends in research grounded in dissipative mechanisms.[37] Current dissipative interactions hinge on the localized interplay between physical and chemical variables, such as the field interaction,[38] mechanical nonlinearity[39] and concentration diffusion inherent in chemical feedback reactions.[40,41] Frequently, these interactions entail diminished spatial transfer distances and temporal delays. A contactless approach facilitating low temporal delay, spatial coverage across long distances, and precise directional control offers promising potential for emulating biological systems in their dynamic multi-body communication.[42]

Here, we propose that photomechanical self-oscillators[43-45] can serve as elementary units for light-mediated communication between individuals, free of spatial restriction. A self-oscillator[46] is a structure capable of self-exciting and sustaining its mechanical motion by absorbing energy from a constant field, operating far from thermodynamic equilibrium. Typically, self-oscillators involve negative feedback,[47] establishing an internal homeostasis-like steady state that exhibits resistance to external disturbances and is self-regulated by the materials' stimuli-responsiveness.[13] In this study, we demonstrate the coupling of self-oscillators through external light beams, enabling light-mediated communication via a closed feedback loop involving negative and positive operators (see Figure 1a). Additionally, we outline a general research pathway for communication among active materials, extending to common materials encountered in everyday life.

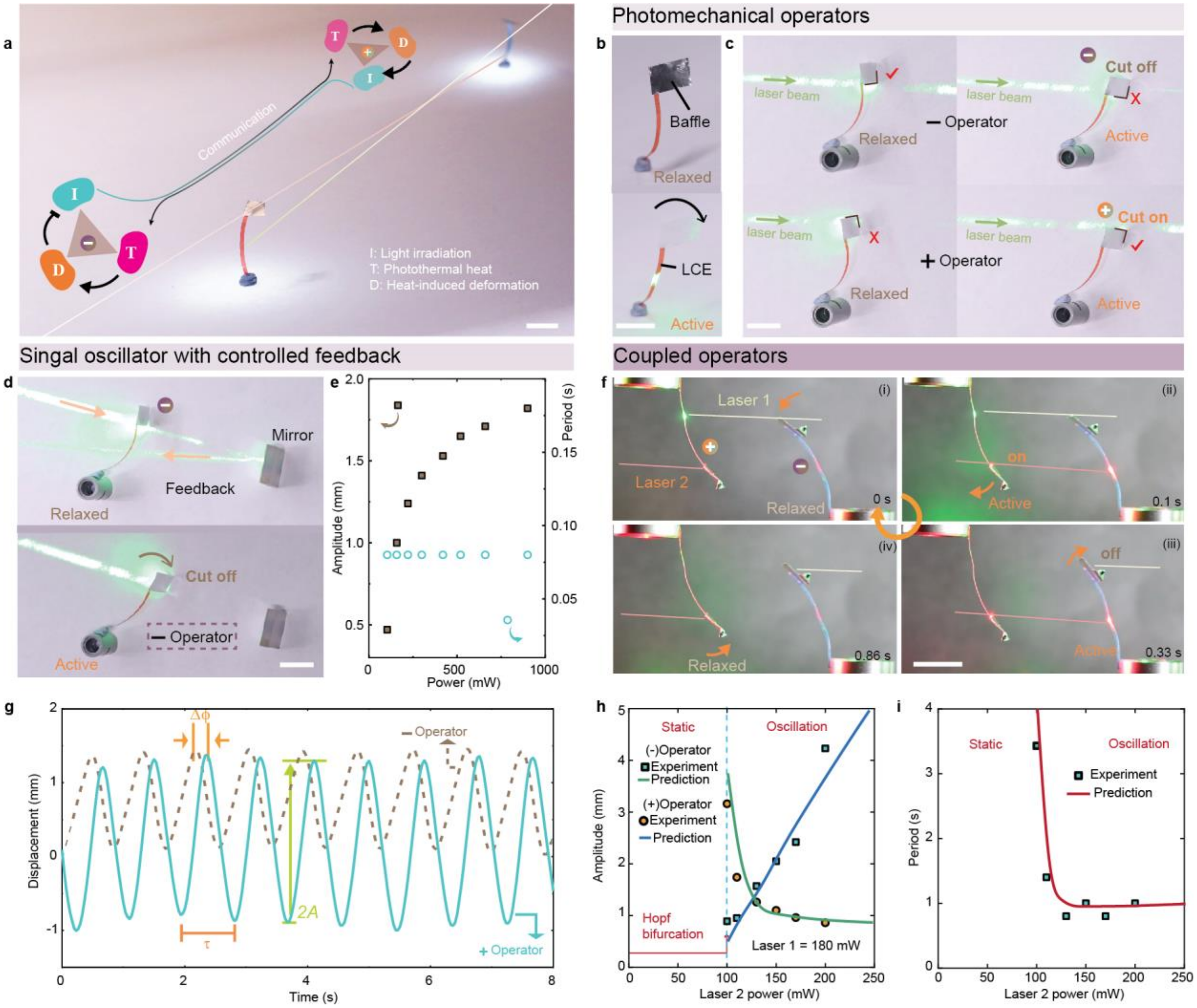

**Figure 1. System concept.** (a) Schematics illustrating light communication between two responsive polymers via a feedback loop. Inset: illustration of feedback, →, induce, –|, prohibit. The green line and red line indicate the propagation of two laser beams. (b) Photographs showing a baffle-actuator system in the relaxed state (top) and deformed state upon light illumination (bottom). The arrow indicates the bending direction of the LCE strip upon actuation. (c) Schematics of baffle-actuator systems depicting the (-)operator (top) and (+)operator (bottom) that cut off or cut on the light beam when actuated. Relaxed/active (-)operator passes/blocks the light; Relaxed/active (+)operator blocks/passes the light. (d) Schematics illustrating an optical feedback system through mirror reflection. The actuator is activated by the reflected laser beam that is controlled by the operator itself. Top: relaxed state (light on). Bottom: active state (light is cut off). (e) Variation of amplitude and period of single oscillator upon increase of excitation power. Light: 2 mm in diameter, 532 nm continuous laser. (f) Photographs of coupling between a negative and a positive operator through two laser beams. The arrow indicates the instant deformation direction of the actuator. The red line indicates a red laser for excitation on blue-colored LCE and, a green light for a green laser on red-colored LCE. Red laser: 635 nm, 210 mW. Green laser: 532 nm, 138 mW. (g) Oscillation data of the coupled oscillators. Displacement shows the change of tip positions of the two baffles. Insets: $\tau$, oscillation period, $A$, amplitude, $\Delta\phi$, phase difference between two operators. Laser 1 (532 nm): 138 mW. laser 2 (635 nm): 210 mW. (h) Experimental and simulated results show the change of oscillation amplitudes of two operators upon varying the excitation power of the red laser beam. Laser 1: 180 mW. (i) Experimental and simulated results showing the variation of the period with excitation power. Laser 1: 180 mW. LCE

sample dimensions: 24 × 2 × 0.1 $mm^3$, baffle size: 5 × 20 × 0.01 $mm^3$. All scale bars are 1 cm. Spot size: 2 mm (532 nm), 3 mm (635 nm).

## Results

### Optically coupled oscillators

The predominant method for creating a photomechanical self-oscillator involves utilizing a self-shadowing mechanism:[48] A light beam induces material deformation, which subsequently shields the incident light, resulting in the cessation of the deformation itself (negative feedback). By autonomously controlling the activation and deactivation of light on the responsive material, oscillation is achieved. In our approach, we employ a baffle-actuator system, as depicted in Figure 1b, to separate the light-active and light-shielding components. A liquid crystalline elastomer (LCE) exhibiting reversible light-induced bending serves as the photomechanical element.[49] The sample was prepared with a splayed alignment, resulting in bending deformation upon thermal stimulation. When exposed to light, the dyes within the LCE convert photon energy into heat, causing the strip to bend toward the contracting surface (planar alignment), regardless of the direction of incidence. Detailed information regarding the chemical structures, preparation process, and photomechanical properties can be found in Supplementary Figs. 1-3. To shield the light, a lightweight aluminum foil is affixed to the end of the LCE actuator. Assembly instructions are provided in Supplementary Fig. 4. By adjusting the position of the light beam relative to the baffle, operators with contrasting functionalities can be programmed. As shown in Figure 1c, when the initial beam is positioned below the baffle edge, the activated LCE moves towards the beam spot, tends to obstruct the light, thus acting as a negative operator. Conversely, when the baffle edge initially blocks the beam, the actuation moves the baffle away from the beam spot, allowing light to propagate, and the system functions as a positive operator.

To implement negative feedback in a single operator, a light beam is directed towards a mirror, reflecting onto the negative operator. The light induces deflection, prompting the baffle to interrupt the input beam. Upon cessation of light excitation, the material relaxes, permitting light beam propagation. Subsequently, the system reverts to its initial state, initiating a new cycle (Figure 1d). Beyond a certain threshold, the operator undergoes self-oscillation fueled by a continuous light beam, with the amplitude increasing along the input power, as illustrated in Figure 1e. The period ($T$) is determined by the resonance of the cantilever system,[50] $T = \frac{2\pi}{\omega_0\sqrt{1-\frac{\bar{\beta}^2}{4}}}$ (Eq. 1), where $\bar{\beta} = \beta\sqrt{l^3/3m\Pi}$, $\beta$ is the damping coefficient, $\Pi$ is the bending stiffness, $m$ is the mass of the baffle, $l$ is the length of the LCE cantilever, $\omega_0 = \sqrt{\frac{3\Pi}{ml^3}}$ is the natural angular frequency. Further details of the modeling are provided in Supplementary Note 2.1. In this basic self-oscillation mode, the periodicity remains insensitive to the fuel power, aligning with other self-oscillators based on the self-shadowing effect.[48,51] Conversely, positive feedback leads to an all-in or all-out bistable state. Detailed explanation of positive and negative feedback mechanisms in this model systems, see Supplementary Fig. 5.

To couple the operators, two laser beams are employed to establish a link between the negative and positive operators, as conceptually depicted in Figure 1a. In the experimental setup, two operators are positioned facing each other, as shown in Figure 1f. Initially, beam 1 approaches the edge of the (-)operator, while beam 2 is obstructed by the (+)operator (step i). When beam 1 interacts with the (+)operator, it unblocks beam 2 (step ii). Subsequently, beam 2 activates the (-)operator, blocking beam 1 (step iii). Consequently, the (+)operator returns its position, blocking beam 2 (step iv), while the (-)operator returns to its original position, unblocking beam 1. The system reverts to state (i), initiating a new cycle. Additional information regarding the structural assembly is available in Supplementary Fig. 4c, while the oscillation kinetics are depicted in Supplementary Fig. 6 and Supplementary Movie 1. It is worth noting that two colors shown in Figure 1c are used to distinguish between the laser beams and actuators: the red laser activates the green LCE, while the green laser activates the red LCE, providing a clearer illustration of the working principle. The coupling mechanism does not depend on the excitation wavelength. In Figures 2-5, only green laser beams (and red-colored LCEs) will be used for the experiments.

The coupling between two operators presents a distinctive case of self-oscillation. As illustrated in Figure 1g, two operators become entangled, demonstrating identical periods and a stable phase shift. Remarkably, the coupled oscillator exhibits an extended period compared to individual ones. To elucidate this phenomenon, we developed a nonlinear dynamics model for two coupled oscillators and conducted numerical simulations (see Supplementary Method 2.2 for modeling details). Our modeling results reveal that the frequency of the coupled oscillator is no longer dictated by the resonance of the individual oscillator but is influenced by the delay in the material's light-response. The period of the coupled system is derived and estimated to be $T \approx -4\tau_{\text{heat}} \ln\left(1 - \frac{\bar{w}_0}{\bar{A}}\right)$ (Eq.2), where $\tau_{\text{heat}}$ is the characteristic time for heat exchange between photothermally-responsive LCE cantilever and environment, $\bar{w}_0 = \frac{w_0}{l}$ is the dimensionless on/off transition critical deflection of LCE cantilevers, and $\bar{A}$ is the dimensionless limit deflection of the LCE cantilevers under longtime illumination. The temporal profiles of two oscillators relative to the on/off states of the dual beams are illustrated in Supplementary Fig. 7. The delay, denoted as $t_d$ in Supplementary Fig. 7, represents the duration required for the movement of the baffle-pair between the on and off positions.

By increasing the input power above a critical value, the system loses stability and transitions from the static state into self-oscillation, giving rise to Hopf Bifurcation as shown in Figure 1h. Augmenting the power of one beam leads to an increase in the amplitude of the corresponding operator directly stimulated by that beam, while concomitantly diminishing the amplitude of the other operator responsible for modulating the ON/OFF state of the beam (Figure 1h). It is notable that both laser beams induce deformations and are simultaneously regulated by their respective opposing operators. Alterations in the intensity of either beam yield a predictable pattern of amplitude/period changes (Supplementary Fig. 8).

The reduction in oscillation period is attributed to the faster deflection of the actuator under higher laser power, resulting in expedited cutoff action and thus a shorter delay ($t_d$), thereby shortening the oscillation cycle. The experimental data quantitatively corroborate the theoretical predictions, as depicted in Figure 1i and Supplementary Fig. 9. Two primary features of the system's dynamic behavior are observed: firstly, period and phase synchronization between the two oscillators, indicating interdependence where cessation of one oscillator halts the motion of the other (Supplementary Fig. 10). Secondly, the system's overall behavior becomes responsive to local environmental changes – variations in beam power or added mass to either operator influence amplitude and frequency (Supplementary Figs. 11, 12). These characteristics offer a unique opportunity for signal transmission between materials via light beams. The distinguishing characteristics between single self-oscillators, interactive oscillators documented in the literature,[23,24] and self-oscillators with light-mediated communication presented in this study are summarized in Supplementary Figs. 13.

**Signal transmission**

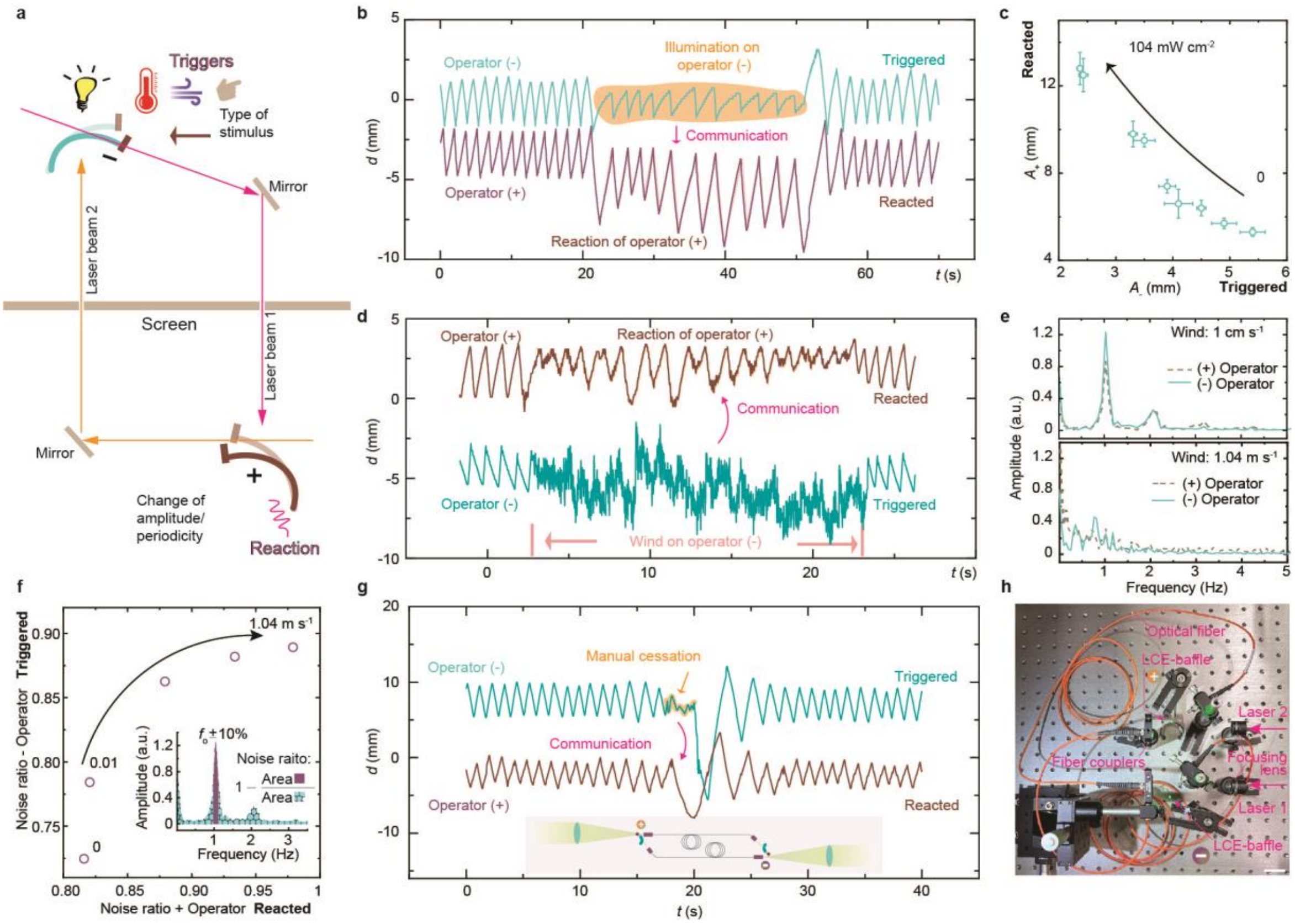

**Figure 2. Signal transmission between coupled oscillators.** (a) Schematic diagram of the communication signal exchange between two oscillations physically isolated by a screen barrier. Types of stimuli onto one of the operators: LED light, heat, wind, and manual stoppage. (b) Oscillation data of the coupled oscillators upon external light disturbance onto (-)operator. External LED light trigger: 635 nm, 69 mW cm$^{-2}$. (c) Amplitude coherence between two oscillators. Externally triggering light: 635 nm LED light source, from 0 to 104 mW cm$^{-2}$. (d) Oscillation data of the coupled oscillators upon wind disturbance onto (-)operator. Wind speed: 1.04 m s$^{-1}$. (e) Fourier transform of oscillation data upon different wind speed disturbances. Light beam for coupled oscillators in (b-e): 532 nm; laser 1, 540 mW, 3 mm spot size; laser 2, 70 mW, 1.2 mm spot size. (f) Noise ratio coherence between two coupled oscillators. Wind disturbance varies from 0 to 1.04 m s$^{-1}$. The inset depicts the definition of noise ratio. (g) Oscillation data of coupled oscillators based on optical waveguiding. The inset depicts the schematic diagram of the assembly of operators and optical fibers. Laser 1 power: 256 mW, Laser 2 power: 200 mW, laser beam are all 532 nm. The size of the light spot for laser 1 is 3 mm, and for laser 2 is 2 mm. (h) Photograph of the fiber-guided coupled oscillator system. LCE sample dimensions: 24 × 2 × 0.1 mm$^3$, baffle size: 5 × 20 × 0.01 mm$^3$. Scale bar: 2 cm.

We established an optical feedback loop by physically isolating the material components using a screen board, as depicted schematically in Figure 2a and Supplementary Movie 2. To facilitate signal transmission, we applied various triggers to one of the operators and monitored the resulting behavioral changes in the other. In the experiment, we applied four types of external triggers:

temperature variations and LED light irradiation to induce additional bending of the soft actuator, mechanical contact to cease oscillations, and wind disturbance to enhance system randomness. Figure 2b illustrates the oscillatory behavior of the system in response to an external disturbance caused by LED light irradiation. Upon application of additional irradiation, the interrupted operator exhibits a decrease in oscillation amplitude, whereas the isolated one displays an increase in amplitude (compared with Figure 1g). Upon cessation of LED irradiation, both operators return to their respective oscillation patterns. Hence, these two materials exhibit communication with a negative correlation (Figure 2c). Further details regarding oscillation under varying intensity triggers are provided in Supplementary Fig. 14. Communication induced by heat disturbance is depicted in Supplementary Fig. 15.

In Figure 2d, airflow induces stochastic oscillations in the directly affected operator, while the feedback loop transmits this randomness as a noise signal to the other. The degree of oscillatory irregularity of the system is quantitatively assessed through Fourier transform spectra, as illustrated in Figure 2e. Under calm or mildly breezy conditions, both operators exhibit spectra with distinct peaks at their fundamental frequencies. However, as wind velocity increases, the amplitude peak around the fundamental frequency diminishes. For detailed oscillation data under varying wind speeds and associated Fourier transform spectra, please refer to Supplementary Fig. 16. In addition, we introduce a noise ratio metric, calculated as 1 minus the ratio between the spectral area at the fundamental frequency within a ±10% bandwidth and the total spectral area encompassing all frequencies, as depicted in the inset of Figure 2f. The noise ratio of the interconnected operators exhibits a positive correlation in their communication (Figure 2f), suggesting that heightened fluctuations in one operator, caused by wind perturbations, result in increased noise in the other.

Laser beam coupling enables long-distance communication. Supplementary Fig. 17

demonstrates light communication between two materials positioned on separate tables, with a six-meter gap between them. Fiber optics offer a compact alternative for transmitting light energy or signals, enhancing durability and facilitating systematization. Figure 2h presents a proof of concept of material communication using fiber optics, where light beams traverse two fibers. These beams are directed through the operators, entering the fibers via focusing lenses. The outputs of the fibers then illuminate the corresponding operators, as illustrated in the schematics inset of Figure 2g and Supplementary Movie 3. Manual interruption or release of one operator leads to cessation or resumption of oscillation in the other (Figure 2g).

## Conceptual generalization

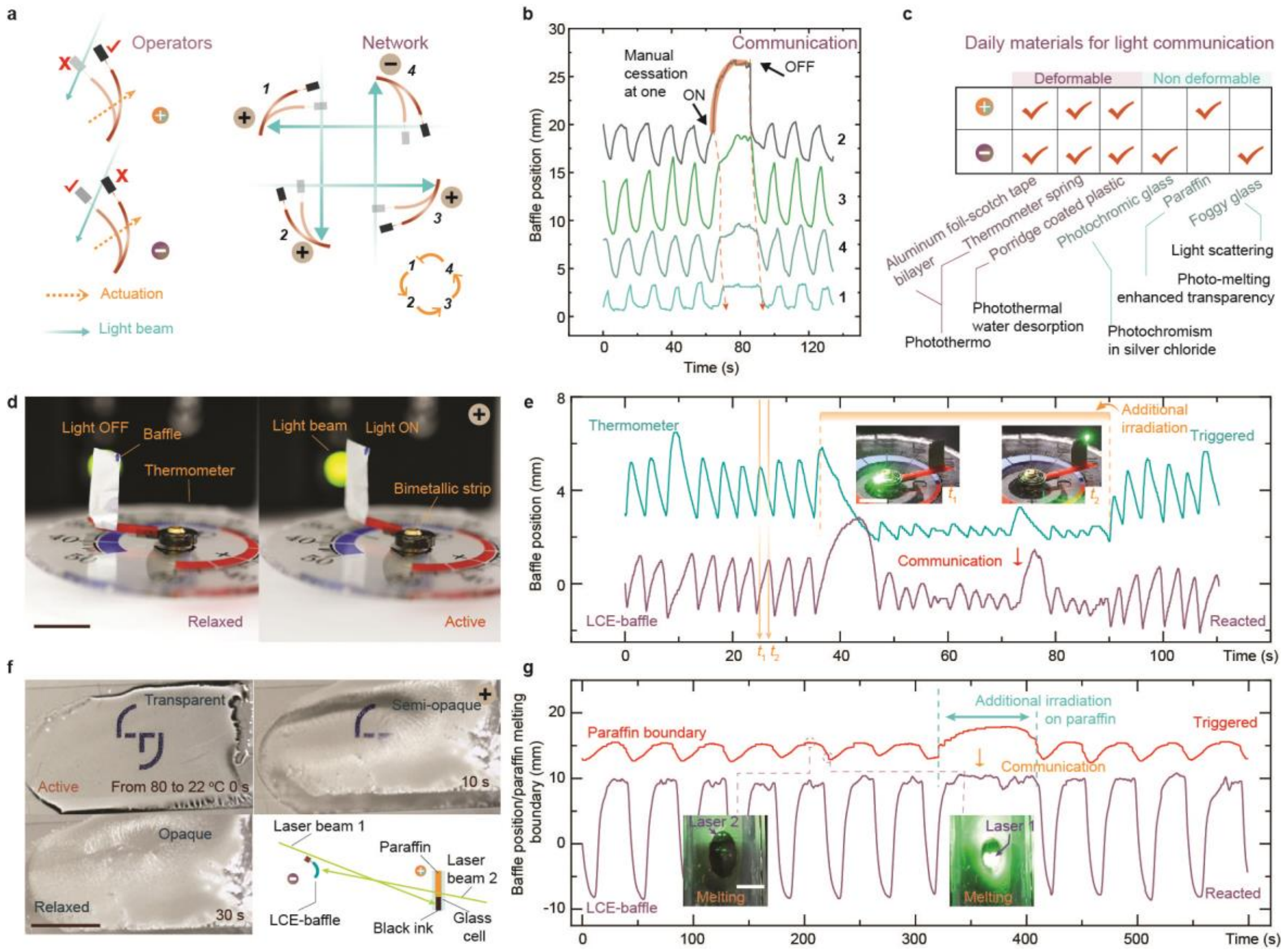

**Figure 3. Light communicative network and conceptual generalization.** (a) Left: Schematic diagrams illustrating the work-principle of individual operators. Dashed arrows indicate the actuation direction of the operator, while solid arrows indicate the propagation of the laser beam. Right: the design of a network composed of four coupled operators. (b) Oscillation data of a four-unit-coupling network. Dashed lines

indicate the sequence of reactions transferred to the neighboring operators upon manual cessation. Four laser beams: 532 nm, 320 mW, 2 mm spot size. (c) List of daily materials built-in with stimuli-responsiveness. (d) Photographs of a (+)operator made of thermometer spring and aluminum baffle. (e) Oscillation data of the coupling between LCE-baffle and a thermometer spring. Insets are the snapshots of the thermometer deformations at different oscillation states. Laser excited on LCE: 532 nm, 43.6 mW, 2 mm spot size. Laser excited on thermometer: 532 nm, 930 mW, 3 mm. (f) Photographs of paraffin on glass substrate upon cooling from 80ºC. Insert: schematic diagram of the coupling between paraffin and LCE-baffle. (g) Oscillation data of the coupling between a paraffin plate and LCE-baffle. Insets show photographs of paraffin during oscillating. Laser excitation on LCE: 532 nm, 80 mW, 1.2 mm spot size. Laser excitation on paraffin: 532 nm, 1440 mW, 2 mm spot size. LCE sample dimensions: 24 × 2 × 0.1 $mm^3$, baffle size: 5 × 20 × 0.01 $mm^3$. All scale bars are 1 cm.

A single (-)operator forms the basis of the individual self-oscillator (Figure 1), while the coupling between (-) and (+) operators establishes a two-element network, facilitating light communication (Figure 2). Figure 3a illustrates the expansion of this system concept into a light-based communication network with an increased number of units. This networking strategy follows a succinct design principle: each communicator constitutes one operator (+ or -), with light beams facilitating connectivity between operators to expand the network. In Figure 3a, a series connection of four units is depicted, accompanied by oscillation data recorded as shown in Figure 3b. Similar to the behavior observed in the previous coupled network, manual cessation or activation at one unit triggers the cessation or reactivation of the remaining operators sequentially. For comprehensive oscillation data in the coupled network with varying unit numbers, please refer to Supplementary Figure 18 and Supplementary Movies 4 and 5. Additionally, Supplementary Figure 19 presents other regulatory network designs inspired by biochemical oscillators.[52]

The light-induced deformation observed in LCEs serves as a specific example of stimuli-response, known as photomechanics, in synthetic materials. This phenomenon of stimuli-response is prevalent in both natural and human-made systems, many of which can be exploited to develop feedback mechanisms, establish networks, and facilitate communication using light beams. Figure 3c illustrates a collection of everyday materials endowed with light responsiveness, categorized into deformable and non-deformable materials. Deformable materials include aluminum foil (commonly used for baking in ovens) and Scotch tape, capable of forming a bilayer structure that undergoes deformation upon temperature increase. This deformation is attributed to the differential (photo-)thermal coefficients between the two layers,[53] akin to the operating principle of a thermometer spring. A porridge thin film demonstrates expansion upon absorbing water in humid environments. Upon photothermal heating, this film releases its water content, resulting in surface shrinkage toward the light source and inducing film bending. These deformable materials offer opportunities for constructing operators based on the principles depicted in Figure 1. Another category involves materials capable of effectively controlling light propagation without the need for mechanical structures. For example, photochromic glass containing particles of silver chloride darkens upon exposure to UV light,[54] serving as a negative operator. Paraffin, initially opaque at room temperature, turns transparent when heated, acting as a positive operator. Furthermore, when a glass piece is positioned over a container of water and the water is photo-heated, fog formation on the glass scatters light, rendering the entire system a negative operator. Supplementary Figs. 20-22 elaborate on various stimuli-responsive concepts in everyday materials and the design principles underlying operator functionality. Subsequently, we briefly highlight an example from each category.

In Figure 3d, the responsiveness of a thermometer spring is depicted, showcasing its function as a positive operator. The metallic spring rotates at a rate of 3° per Kelvin, and upon application of black ink, it responds to light through a photothermal effect. This action adjusts the position of a baffle atop its pointer, thereby modulating the light beam. Consequently, negative and positive feedback can be optically looped between an LCE-baffle (-operator) and a thermometer (+operator), as demonstrated by the previously illustrated principle. The coupled system demonstrates self-sustained oscillation and facilitates communication, as depicted in Figure 3e and

Supplementary Movie 6. Additional data on disturbed oscillation resulting from external stimuli can be found in Supplementary Fig. 23. For further details on optical alignment, sample preparation, and structural orientation, refer to Supplementary Fig. 24. Figure 3f showcases a positive operator based on the transparency modulation of paraffin. Above its melting temperature (approximately 80 °C), the paraffin becomes transparent to light, while upon cooling, it turns opaque, hindering light propagation. To induce photothermal heating, black ink was incorporated into the paraffin, and the two laser beams were aligned to couple between an LCE-baffle and the paraffin. The optical setup is depicted in the inset of Figure 3f. Figure 3g presents data on the coupled oscillations of the system, with the boundary position indicating the melting region of the paraffin: a higher position value corresponds to a larger melting area allowing light transmission, while a lower value indicates a smaller melting region resulting in light blockage. Insets in Figure 3f display photographs depicting the corresponding states of the paraffin. The oscillation and communication between the LCE and paraffin are facilitated by the self-regulation of light propagation between the baffle and paraffin. Further details on optical alignment, paraffin preparation, and structural orientation can be found in Supplementary Fig. 25. Additional information on oscillation can be found in Supplementary Fig. 26 and Supplementary Movie 7.

**Adaptation**

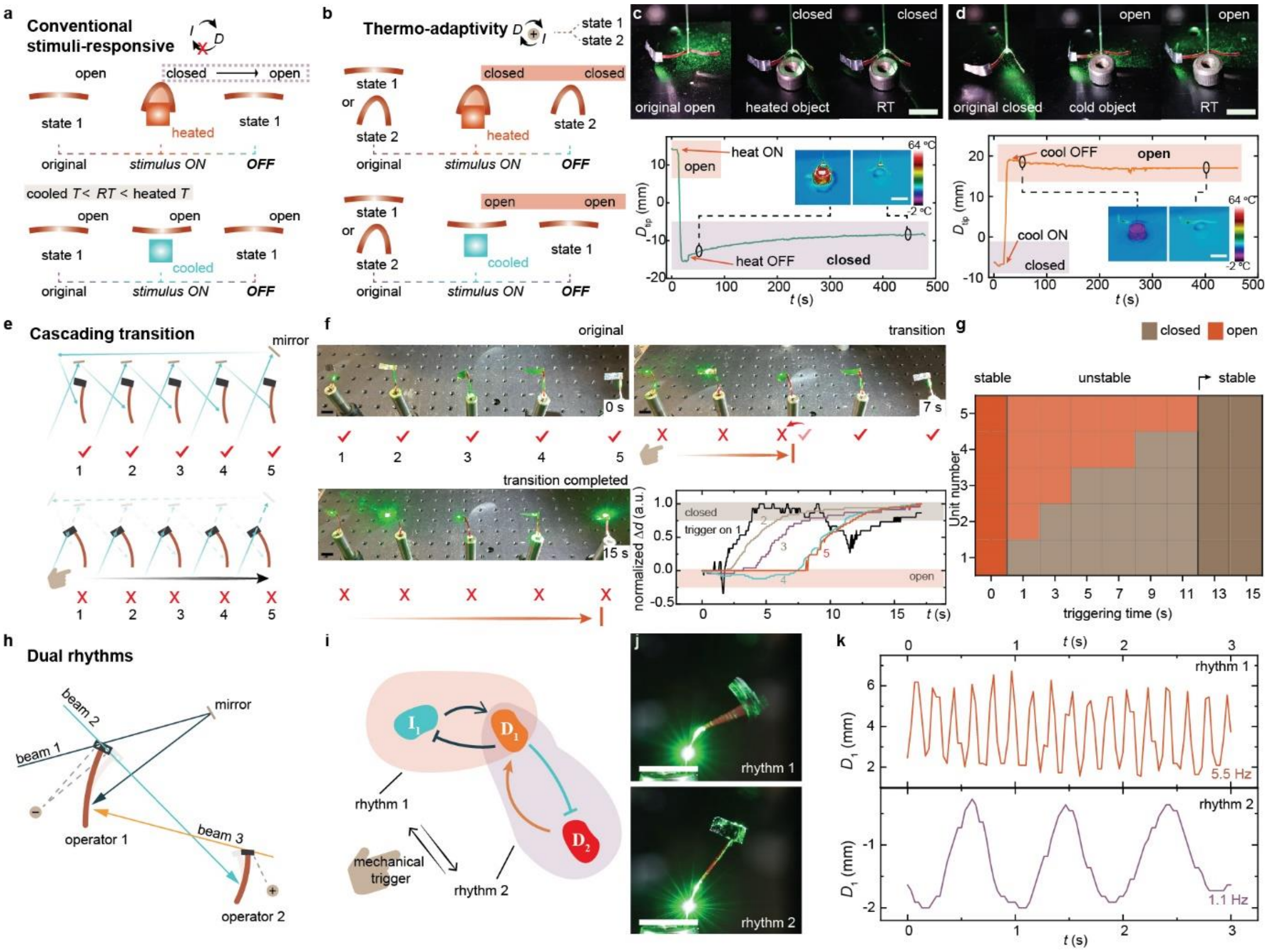

**Figure 4. Adaptation enabled by light-communicative network.** (a) Working principle of conventional stimuli-responsive materials. (b) Concept of material state is induced by a positive feedback loop, where adaptation refers to the change in state in response to external stimuli. (c) Top: Photographs of an operator-based gripper initially in an open state, adapting to a hot object and changing to a closed state. Bottom: Displacement data of the actuator tip, $D_{tip}$, in response to heat from an 80 °C object. Insets: Thermal camera images of the system before and after cooling to room temperature. (d) Top: Photographs of an operator-based gripper initially in a closed state, adapting to a cold object and opening. Bottom: Displacement data of $D_{tip}$ in response to cold stimulus from a 0 °C object. Insets: Thermal camera images of the system before and after warming to room temperature. (e) Schematic diagram of the optical pathway in a network of 5 positive operators, showing two stable states (all-open and all-closed). (f) Snapshots of five operators responding to a mechanical trigger on the first operator. Lower right: Change in baffle position ($\Delta d$) for each operator. (g) System bifurcation, with the state change of each operator recorded as different trigger intervals are applied to the first operator. (h) Design of a dual rhythmic self-oscillator based on two feedback loops. (i) Schematic of the feedback mechanism in the dual rhythmic system. $I_1$, intensity of beam 1. $D_1$, displacement of operator 1. $D_2$, displacement of operator 2. (j) Snapshots of the operator 1 oscillating at different frequencies. (k) Displacement data of the oscillator. Laser used in (c, d): 532 nm, 220 mW, 2 mm. Laser used in (e -g): 532 nm, 100 mW, 4 mm. (h-k) Beam 1 excited on operator 1: 532 nm, 180 mW, 2 mm; beam 2 spot on operator 2: 532 nm, 180 mW, 2 mm; beam 3 spot on operator 1: 532 nm, 160 mW, 2 mm. All scale bars are 1 cm.

Responsive materials can self-adapt to external stimuli, and the concept of adaptation has been demonstrated in various forms. Here, we explore the potential of using optical feedback loops in communicative materials to program three distinct types of adaptivity: bi-stability, cascading transitions, and dual rhythms.

Conventionally, stimuli-responsive materials can change their shapes, as illustrated in Figure 4a. For example, a thermally sensitive polymer strip bends when heated and returns to its original, unbent state after removing the heat. Here, we define "adaptation" as the ability of a material to change its state in response to external stimuli and maintain the altered state even after the stimulus is removed. The material's deformation is not directly powered by the stimulus but rather operates as a self-powered, multi-stable system that can switch between stable states upon external disturbances. In other words, it is the bifurcation of the dissipative system that enables dynamical change of condition by self-organizing into different states. Figure 4b demonstrates an example of thermo-adaptation, which differs from the conventional principle of stimuli-responsiveness by emphasizing the material's ability to maintain its state after the stimulus is turned off. To achieve bi-stable states, we design the system with a positive feedback loop, as positive feedback enables an "all-in" or "all-out" bistable mechanism. Supplementary Fig. 27 illustrates this operator and the surrounding optical excitation pathways, where the LCE strip serves as a proof-of-concept gripping device. In this example, a gripper closes when the LCE strip bends and reopens when the strip unbends. Figure 4c depicts a scenario where the gripper, initially open (with a baffle blocking the light beam so the LCE is inactive), closes when exposed to a heated object (80 °C). The bending of the structure unblocks the light beam, delivering energy to the actuator within the positive feedback loop. As a result, the gripper remains closed even after the object's temperature cools to room temperature. Note that there are two stable states of a positive feedback-driven system, as they can be mutually switched by mechanical triggers. Figure 4d presents the opposite case, where the gripper starts in a closed state (with the baffle unblocking the light beam, bending the LCE strip). When exposed to a cold object (0 °C), the gripper opens, and the baffle blocks the light beam. After the object warms to room temperature, the light beam remains blocked, and the gripper maintains its open state.

A series of positive operators connected in a closed feedback loop can result in a cascading transition of material states. This means that switching the state of one operator can trigger a sequential shape-change in others until all operators complete their transition. Figure 4e illustrates an example of five positive operators coupled by five laser beams in a closed loop. The system exhibits two stable states: (1) all operators are open, allowing all the light beams to pass through, and (2) all operators are closed, blocking all beams. Figure 4f captures snapshots of the transition, starting with all operators in the open state. A mechanical trigger is applied to the first operator to switch it OFF, which sequentially drives the 2nd to 5th operators into the closed state (Supplementary Movie 8). Displacement data shown in the lower right corner of Figure 4f highlights the kinetics of the shape-morphing process, revealing a time delay between the transitions of sequential operators. For more detailed displacement data during the cascading transition, see Supplementary Fig. 28. State (shape) transition can be seen as bifurcation of the system, which depends on the strength of the external stimulus, particularly the triggering period. Figure 4g shows that a triggering period of less than 13 seconds cannot induce a complete transition. In such cases, the recently switched operator reverts to its previous state, reversing the sequence. However, for a triggering period of 13 seconds or more, all operators switch to the closed state, and the system stabilizes in its new configuration. In this case, bifurcations signify a change in the network's morphology, characterized by the shape-switching along operators. The bifurcation diagram of the all-closed to the all-open state transition is also shown in Supplementary Fig. 29.

The examples above illustrate the concept of adaptation in bi-stable states, demonstrated through shape changes in a single operator and sequential shape morphing in spatially separated operators. Oscillation frequency serves as another physical metric for assessing the state's condition. In the following, we demonstrate that bifurcations signify a change in the system's behavior, *i.e.,* transitions between oscillatory rhythms.

Figure 4h presents the optical design of a material system incorporating two negative feedback loops. A negative feedback loop induces homeostasis-like self-oscillation with a well-defined frequency (see Figure 1e). The baffle on operator 1 contains two holes that act as distinct negative operators, controlling the passage of beam 1 & 2. Beam 1 is directly reflected back to operator 1, inducing a self-oscillating system in accordance with the principles illustrated in Figures 1d & 1e. Beam 2 interacts with another negative operator, which modulates beam 3. Beam 3, in turn, excites operator 1, following the coupled oscillator mechanism outlined in Figures 1f–1i. Each feedback loop includes a single negative operator and produces a stable self-oscillation frequency. A mechanical trigger allows switching between two oscillation rhythms, as schematically depicted in Figure 4i and Supplementary Movie 9. Governed by the first feedback loop, the system self-oscillates at a high frequency (5.5 Hz), as shown by the photograph and displacement data in Figure 4j. After switching to the second feedback loop, which couples two operators, the system exhibits a lower frequency due to the increased delay within the loop (see Eq. 2 and Figure 1i). The displacement data showing different frequencies and transitions between rhythms are provided in Figure 4k and Supplementary Fig. 30. In summary, the examples above illustrate the adaptation concept enabled by optical feedback loops in responsive materials. We demonstrated self-adaptation through shape changes and rhythm switching in response to external stimuli, as well as the system's capacity to maintain these altered states (shape or rhythm) after the stimulus is removed.

**Discussion and conclusion**

Research on bioinspired materials has evolved from focusing on static functionalities to embracing increasing levels of dynamic responsiveness and interactive behaviors.[4-6,31] These advancements emphasize the autonomy and adaptability of biological systems, aiming to introduce new interaction behaviors and functions in synthetic responsive materials. For example, the work presented includes methods for mechanical interaction in synchronization,[23] self-oscillation driven by negative feedback from heat diffusion,[12,13] and spatial patterns induced by feedback from chemical reactions.[40] In all these systems, feedback plays a crucial role in maintaining the system far from thermodynamic equilibrium, enabling interaction. However, the predominant methodologies rely on contact-based approaches, such as mechanical interaction[23] and the exchange of reactive substances.[55] These approaches are limited by temporal delays, short transfer distances, and a lack of directionality. In contrast, the method reported in this study couples two or more self-oscillators through an optical feedback loop, providing high-directionality in network connections and enabling material interaction over long distances with minimal delay. This feedback loop can selectively engage different oscillating elements, allowing for adaptation at the system level. The demonstrated principle also has broad applicability across a range of responsive materials.

The light-mediated communication concept provides a method for remote sensing and electrical control of soft actuators over long distances. Details see in Supplementary Note 2.3 Sensing and control.

In this letter, we present a comprehensive approach to achieving light communication among non-equilibrium materials. We introduced an experimental model consisting of a soft actuator and a baffle, where a photomechanical liquid crystalline elastomer actuator and aluminum foil determine the operator's function. Commencing with a single negative operator, we demonstrate the mechanism of optical feedback in self-oscillation. Subsequently, we couple a positive and a negative operator to establish a closed loop of optical feedback. We observe that the frequency of the coupled system depends on the delay of the material's light responsiveness, rather than the resonance of the single oscillator. This coupled feedback loop renders two materials interdependent in oscillation, facilitating signal communication between them. We utilized manual triggers – mechanical contact, heat, light, and wind – to perturb the oscillation of one operator, which then autonomously conveyed the disturbance signal to the coupled operator through the feedback network. While our focus did not center on achieving quantitative control over material deformation via optical excitations, these experiments served as a conceptual illustration of primitive communication between non-equilibrium materials using light. Such light-mediated communication is untethered, with high directionality, and free of spatial restriction. We provided evidence supporting the systematic development of this concept, demonstrating examples of communication over meter-scale distances, signal transmission via optical fibers, series connections with an increased unit number, and the design of complex networks inspired by biological oscillators. We emphasized the widespread occurrence of stimuli-responses in both natural and synthetic materials, exemplified by light-mediated communication utilizing commonplace materials such as a thermometer spring and paraffin. We highlighted three adaptive activities that are enabled by using optical feedback loops, including bi-stable states, cascading transition and dual rhythm. We also introduced the opportunities in robotic application, in which deformation-electric signal transduction, remote sensing and remote control are demonstrated.

This work presents a facile approach to attain temporally and spatially complex dissipative systems using simple stimuli-responsive materials.

## Methods

**LCE Material preparation in brief.** 1,4-Bis-[4-(6-acryloyloxyhexyloxy) benzoyloxy]-2-methylbenzene (99%, RM82) was obtained from SYNTHON Chemicals. 6-Amino-1-hexanol and dodecylamine from TCI, 2,2-Dimethoxy-2-phenylacetophenone were obtained from Sigma-Aldrich, Disperse Red 1, and Disperse Blue 14 were obtained from Merck. ST06512 [4-(6-(Acryloyloxy)hexyloxy)phenyl 4-(6-(acryloyloxy)hexyloxy)benzoate] was obtained from SYNTHON Chemicals. GDA (tetraallyloxyethane) and DPA (dipropylamine) were obtained from TCI. EDDT [2,2′-(ethylenedioxy)diethanethiol], PETMP [pentaerythritol tetrakis(3-mercaptopropionate)], BHT [2,6-di-tert-butyl-4-methylphenol], Irgacure 651 [2,2-dimethoxy-2-phenylacetophenone], were obtained from Sigma-Aldrich. All chemicals were used as received.

**LCE strip actuator.** Liquid crystal cells were made by adhering two glass substrates, one coated with a homeotropic alignment layer (JSR OPTMER, spun at 4000 RPM for 1 min, followed by baking at 100 °C for 10 min and then at 180 °C for 30 min), and the other with polyvinyl alcohol (PVA) rubbed on (5% water solution, spun at 4000 RPM for 1 min, and baked at 100 °C for 10 min) for uniaxial alignment. Microspheres with a diameter of 100 µm (Thermo Scientific) were employed as spacers to determine the film thickness. The liquid crystal mixture consisted of 0.3 mol RM82, 0.05 mol 6-Amino-1-hexanol, 0.05 mol dodecylamine, and 2.5 wt% of 2,2-Dimethoxy-2-phenylacetophenone, mixed at 95 °C. This mixture was introduced into the cell by capillary action at 95 °C and then gradually cooled down to 63 °C (at a rate of 1 °C per minute). The cell was then kept in an oven at 63 °C for 24 hours to allow for an aza-Michael addition reaction for oligomerization. Subsequently, the sample underwent UV irradiation (360 nm wavelength, 180 mW $cm^{-2}$ intensity) for 20 minutes to initiate polymerization. Finally, the cell was opened using a blade, and strips were cut from the film.

**LCE fiber actuator.** The precursor mixture consisting of 0.36 mmol ST06512, 0.32 mmol EDDT, 0.04 mmol PETMP, 0.04 mmol GDA, 1.5 wt% Irgacure 651, 1.0 wt% BHT, and 0.5 wt% DPA was prepared and thoroughly homogenized at 80 °C. The mixture was then injected into silicone tubes and maintained at 80 °C in an oven for 24 hours to allow the thiol-Michael reaction to form oligomers. The resulting oligomers were mechanically stretched and subsequently polymerized under UV light (365 nm, 180 mW $cm^{-2}$, 20 min) to obtain fibers. Finally, the fibers were dyed in a Disperse Red 1-isopropanol solution.

**Daily materials sources.** The materials used in constructing the bilayer experimental setup, namely aluminum foil and Scotch tape, were readily available from local stores. Oats and coffee, selected for their moisture-responsive and temperature-sensitive characteristics, were procured from the same outlet. Additionally, paraffin and a thermometer, for temperature-responsive behavior and thermal actuation, were also sourced locally.

**Sample assembly**. To obtain LCE films of different colors and thus different light sensitivities, the prepared LCE samples were diffused with different dyes. This process entailed placing the samples on a hotplate set to 100 °C and uniformly dispersing powders of Disperse Red 1 and Disperse Blue 14 dyes onto the surface of the LCE. Following a 5-minute diffusion period, any residual powder on the surface was wiped away. The LCE films were then cut to dimensions of $2 \times 24 \times 0.1$ $mm^3$, and aluminum foil was also cut to dimensions of $5 \times 20$ $mm^2$. The cut aluminum foil functioned as a baffle was bonded to one end of the LCE material using glue.

**Stimuli sources**: Two continuous-wave lasers (at 635 nm and 532 nm wavelengths) served as excitation sources, while a CoolLED pE-4000 unit was employed for photopolymerization and served as an external light source for perturbing oscillations. Airflow was generated by a small fan, with wind speed regulated by adjusting the applied voltage and quantified using a Digital Hand-held Wind Speed Gauge Meter (GM816) before each experiment. The heat source consisted of an iron block heated to varying temperatures and positioned at a distance of 2 cm below the sample.

**Light excitation:** We employed a continuous laser (532 nm, 2 W, ROITHNER) for light excitation of red-colored liquid crystal elastomer (LCE), and a continuous laser (635 nm, 1 W, ROITHNER) for excitation of green-colored LCE. Both laser beams were focused using a plano-convex lens (100 mm, THORLABS) to achieve a focal spot approximately 100 microns in diameter near the baffle edge of the operator. Additionally, an LED source (460 nm module, CoolLED pE-4000) was utilized for supplementary photothermal deformation.

**Tracking method.** Canon 5D Mark III camera with a 100 mm lens and an iPhone 15 smartphone were used to capture optical images and videos. Kinovea program was used to track the position of the baffle.

**Nonlinear dynamics modelling.** See details in Supplementary Notes.

**Data availability.** The main data generated in this study are provided in the article and Supplementary Information. The raw data is available in the Fairdata IDA online storage space through below link, https://ida.fairdata.fi/s/NOT_FOR_PUBLICATION_kxFjx5ZAg2dM (the link will be updated with a permanent download address).

## Acknowledgments

We gratefully acknowledge funding from Academy of Finland (postdoctoral researcher No. 347201, research fellow No. 340263 and PREIN, No. 320165). H.Z is thankful to the financial support of the European Research Council (Starting Grant project ONLINE, No. 101076207). K.L. acknowledges the supports from University Natural Science Research Project of Anhui Province (Grant No. 2022AH020029), National Natural Science Foundation of China (Grant No. 12172001), and Anhui Provincial Natural Science Foundation (Grant No. 2208085Y01). J. Y. acknowledges the support from China Scholarship Council (CSC). H. Z. thanks Hang Zhang (Aalto) for the discussion. A heartfelt thank you to Haotian Pi from Aalto University for the inspiring discussion on the topic of Dissipative Structures.

## Author contributions:

H. Z. conceived the idea and supervised the project; H. G. performed experiments with the help of J. Y., F. L. and H.Z.; D. L. and H. G. performed the electric control experiments. H. G., H. Z. and K. L. analyzed the data; K. L. performed the modeling and developed the concept. H.Z. wrote manuscript with the helps from others; All the authors discussed and contributed to the project.

**Competing interests:** The authors declare no competing interests.

**Materials & Correspondence:** Correspondence and material requests should be addressed to K.L. (kli@ahjzu.edu.cn) and H. Zeng (hao.zeng@tuni.fi).

Supplementary Materials for

# Light communicative materials

Hongshuang Guo,[1] Kai Li,[2]* Jianfeng Yang[1], Dengfeng Li,[1] Fan Liu,[1] Hao Zeng[1]*

**Affiliation:**

[1] Faculty of Engineering and Natural Sciences, Tampere University, P.O. Box 541, FI-33101 Tampere, Finland.

[2] Department of Civil Engineering, Anhui Jianzhu University, Hefei 230601, China.

*Correspondence to: kli@ahjzu.edu.cn; hao.zeng@tuni.fi.

**This PDF file includes:**

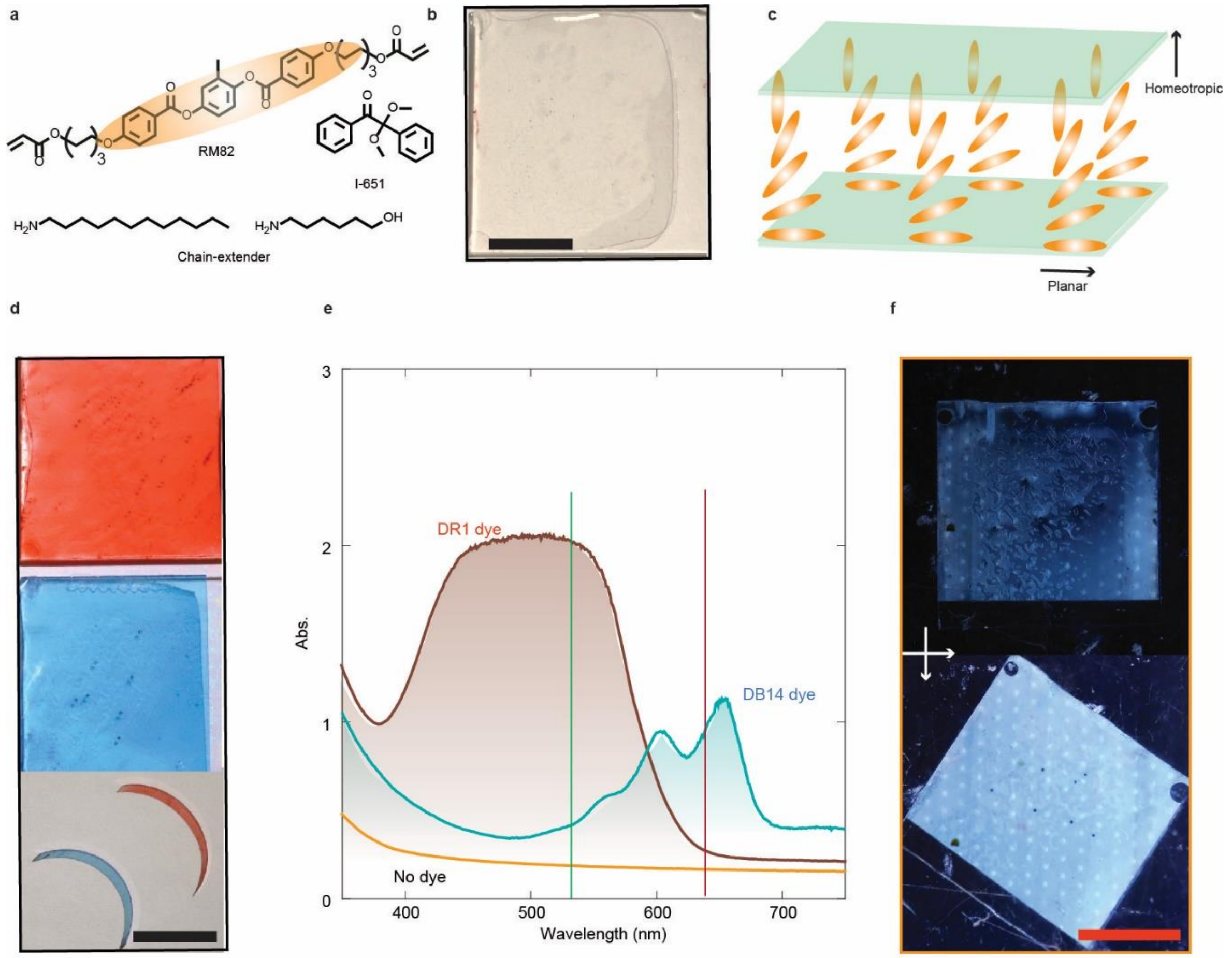

**Supplementary Figure 1. Sample preparation.** (a) Chemical structures of molecules utilized in LCE synthesis. (b) Photographs displaying a dye-free polymerized LCE film on top of a glass substrate. (c) Schematic drawing shows the splayed orientation of molecules inside the LCE. The arrows denote the types of surface treatment and the corresponding liquid crystal alignment at the surfaces. (d) Photographs displaying an LCE film thermally diffused with Dispersed Red 1 dye (top), Dispersed Blue 14 dye (middle), and strip-like actuators cut from corresponding films (bottom). (e) Absorption spectra of the LCE film before and after dye diffusion. The green line represents the wavelength of the 532 nm laser, red line for the 635 nm laser, utilized for sample excitations. (f) Cross-polarized macroscopic images of a dye-free LCE film. Film thickness: 100 µm. All scale bars are 1 cm.

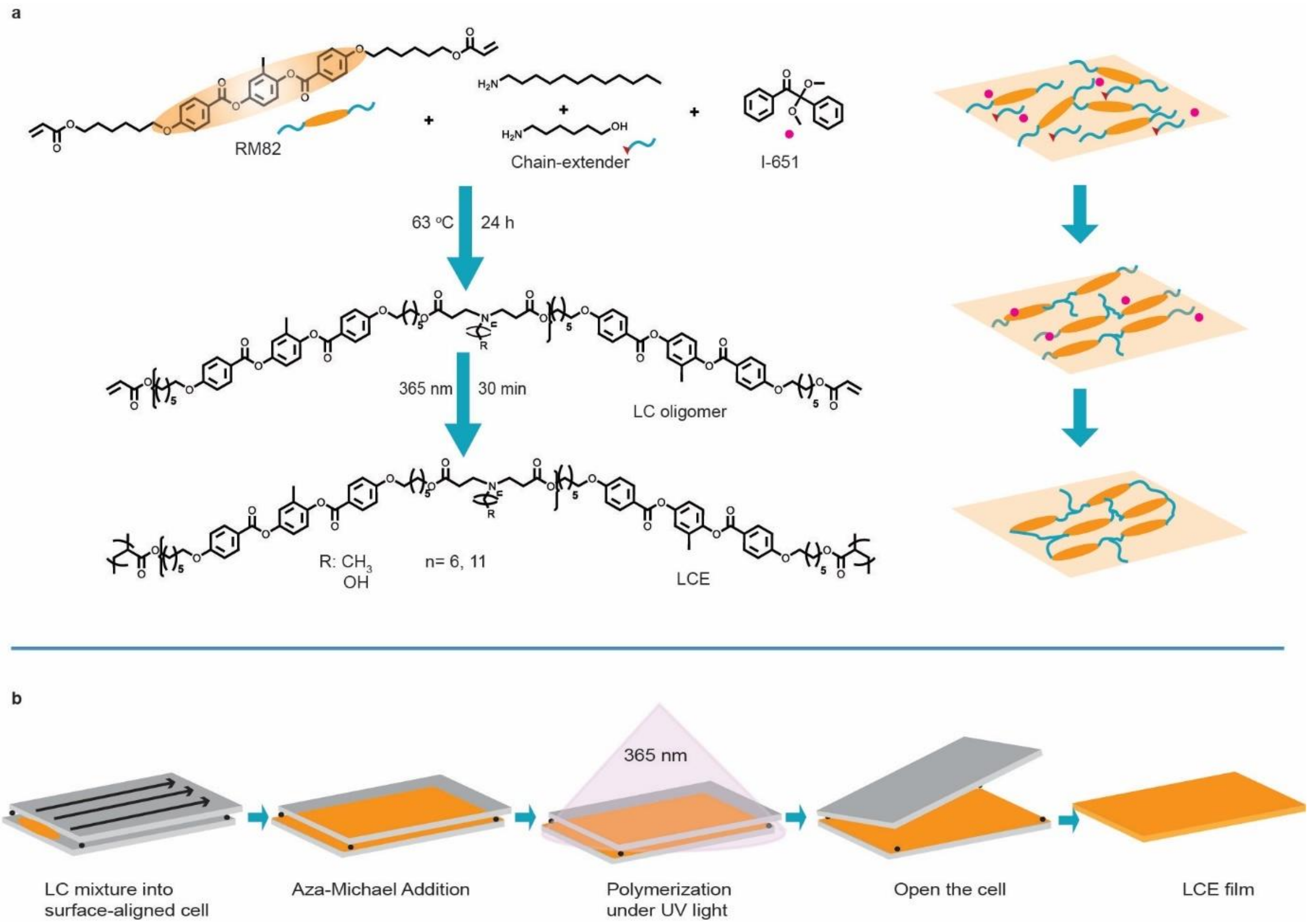

**Supplementary Figure 2. LCE synthesis process.** (a) The steps of the polymerization reaction (left) and a schematic diagram illustrating the polymerization process (right). (b) Liquid crystalline cell fabrication steps. First, RM82, 6-Amino-1-hexanol, dodecylamine, and 2,2-Dimethoxy-2-phenylacetophenone are homogeneously mixed and filled into a cell at 95°C. Secondly, the cell is placed in an oven at 63°C for 24 hours to conduct Aza-Michael addition-based step-growth polymerization, resulting in the formation of the LC oligomer. Thirdly, the remaining diacrylate end-groups are cross-linked upon UV exposure to form the final LCE network. Finally, the cell is opened by using a blade, and the LCE film actuator is removed from the glass substrate.

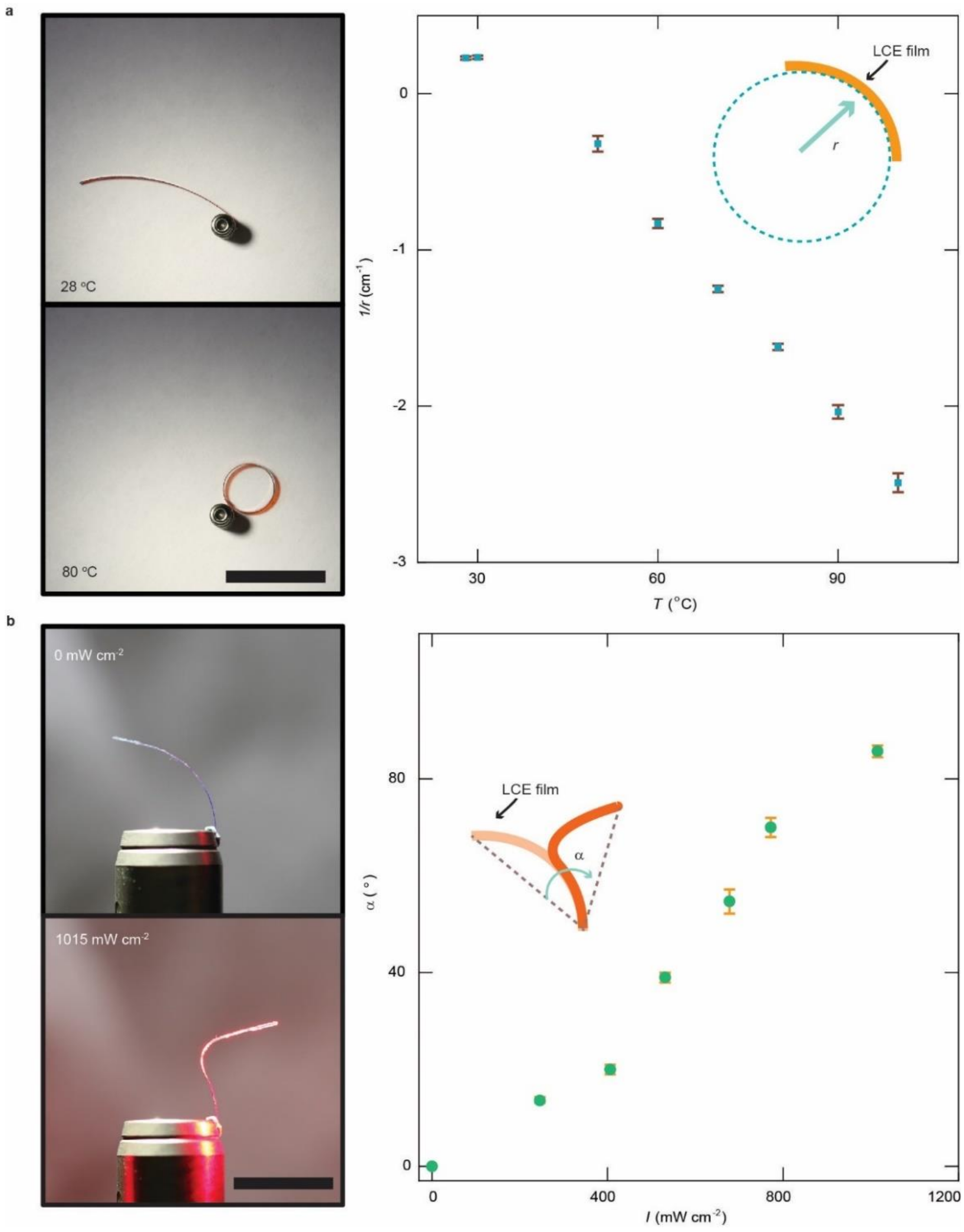

**Supplementary Figure 3. Stimuli-responsiveness of LCE actuator.** (a) Left: Photographs showing the LCE strip geometries at room condition and elevated temperature. Strip size: $15 \times 1 \times 0.1$ mm$^3$. The sample is placed on top of a hot plate and covered with a transparent glass window to attain homogeneous temperature distribution. Right: The curvature variation upon increasing the temperature. Curvature is defined as $1/r$, where $r$ is the radius of the strip, as shown in the inset. (b) Left: Photographs displaying the LCE strip geometries upon different illuminating intensities (635 nm, laser source). Strip size: $15 \times 1 \times 0.1$ mm$^3$. Right: The bending angle of the LCE actuator upon change of illuminating light intensity, $I$. The bending angle ($\alpha$) is indicated in the inset. Error bars represent s.d. for n = 3 measurements. The same sample was measured repeatedly. All scale bars are 1 cm.

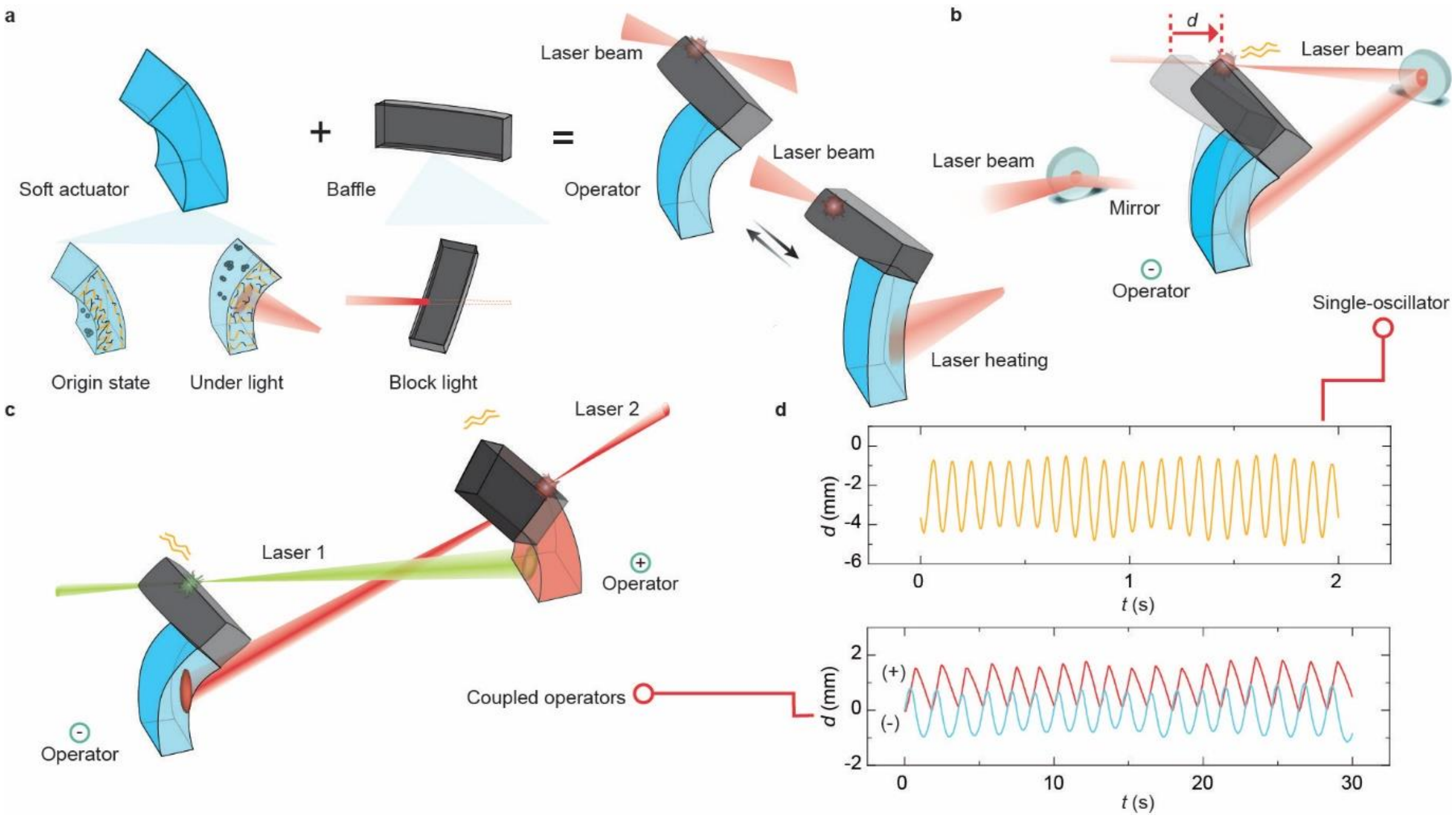

**Supplementary Figure 4. Structure assembly for coupled oscillator systems**. (a) Schematics showing the assembly of the baffle and LCE actuator to form the (-)operator system. (b) Schematics showing the (-)operator controls the propagation of a light beam that has been reflected by a mirror onto the actuator. This forms a negative feedback self-oscillator. *d*, displacement of the baffle tip position. (c) Schematic diagram illustrating the optical setup to achieve the coupling between a negative and a positive operator. (d) Oscillation data of the single self-oscillator and coupled oscillators under constant light beam excitation. Light for single oscillation: 532 nm, continuous laser, 420 mW, spot size is 2 mm. For coupled oscillators, laser 1: 280 mW, 532 nm, spot, 2 mm, laser 2: 320 mW, 635 nm, spot, 3 mm. LCE sample dimensions are 24 × 2 × 0.1 $mm^3$, baffle, 5 × 20 × 0.01 $mm^3$.

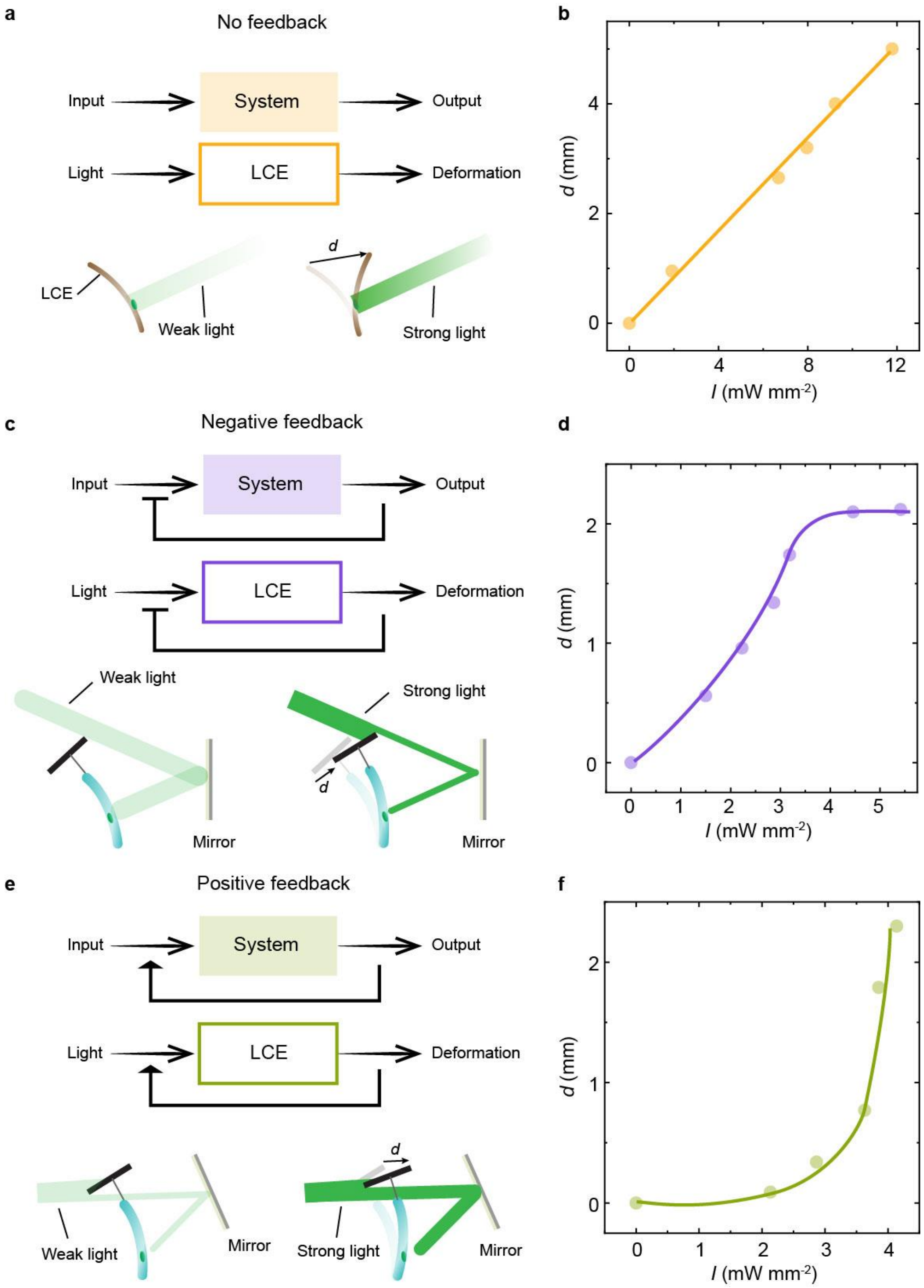

**Supplementary Figure 5. Feedback in optomechanical system.** (a) Schematics of an LCE actuator without feedback. (b) Characteristic curve of tip displacement upon varying the input light intensity with no feedback. (c) Schematics of a baffle-actuator system with negative feedback in photomechanical response. (d) Characteristic curve of tip displacement regulated by negative feedback in response to changes in light input. (e) Schematics of a baffle-actuator system with positive feedback in photomechanical response. (f) Characteristic curve of tip displacement regulated by positive feedback in response to changes in light input. LCE sample dimensions: $24 \times 2 \times 0.1$ $mm^3$, baffle size: $5 \times 20 \times 0.01$ $mm^3$. Incident beam: 532 nm, 2 mm in diameter.

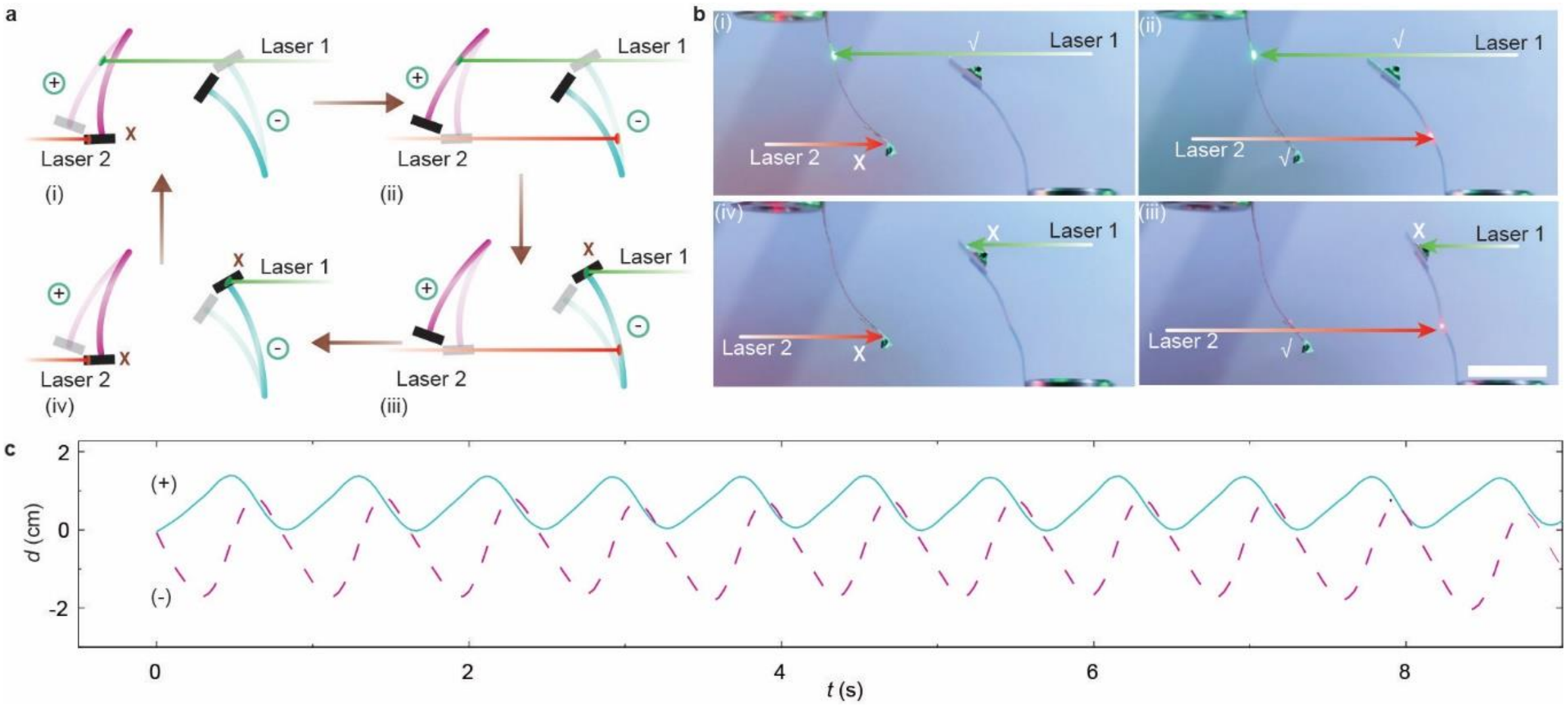

**Supplementary Figure 6. Kinetics of coupled oscillators**. (a) A schematic illustration detailing the deformation steps of two operators. (b) Corresponding images depict the shape-change of actuators within one oscillation cycle. (c) Oscillation data was recorded upon excitation of two constant light beams. Laser 1 (532 nm), 390 mW, laser 2 (635 nm), 460 mW. The spot is 2 mm for laser 1, and 3 mm for laser 2. LCE sample dimensions: 24 × 2 × 0.1 mm$^3$, baffle size: 5 × 20 × 0.01 mm$^3$. Scale bars: 1 cm.

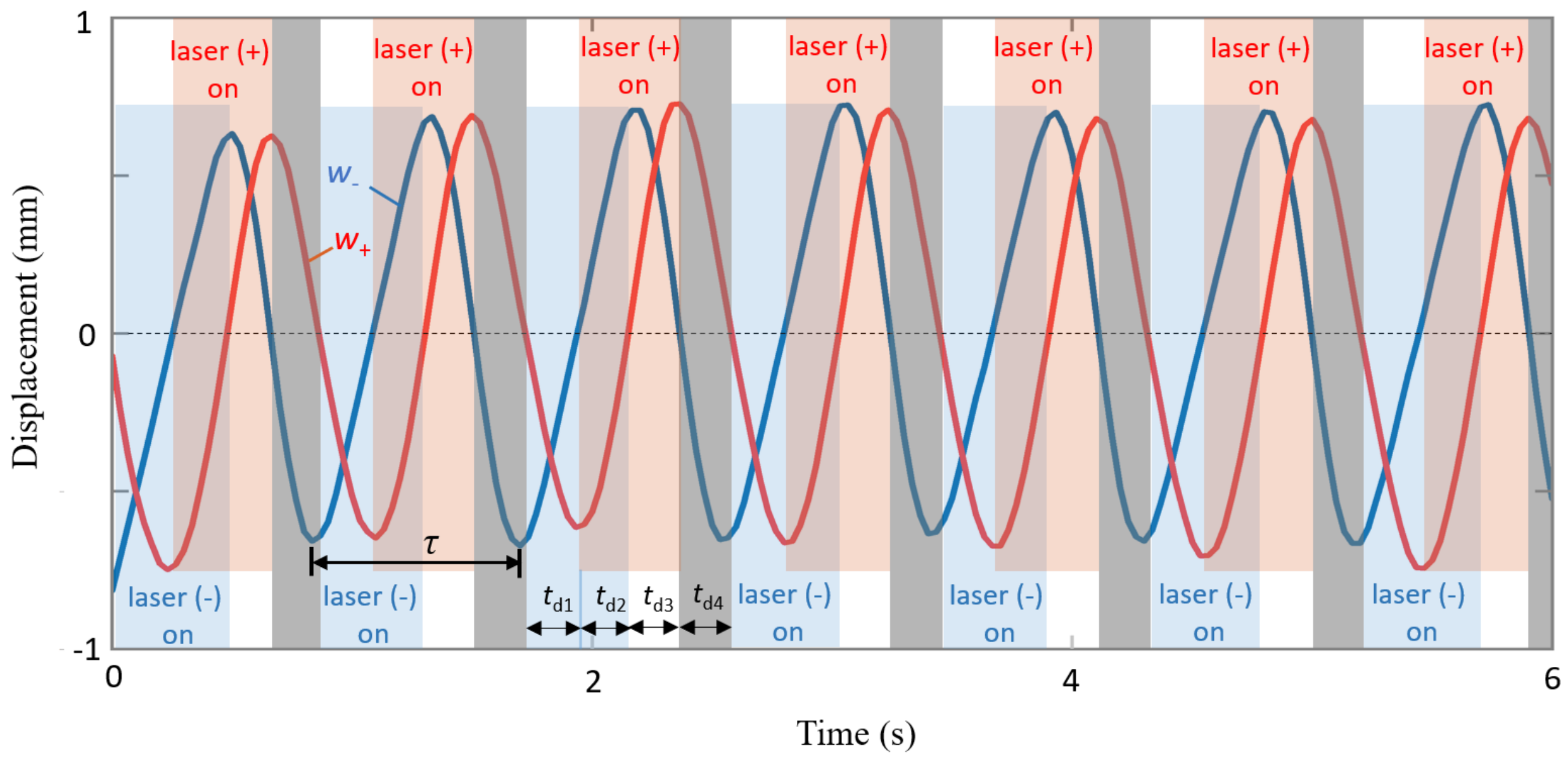

**Supplementary Figure 7. Displacement kinetics explanation.** In the experiment, the parameters are $l_- = 2.5\mathrm{cm}$, $l_+ = 2.5\mathrm{cm}$, $P_- = 210\mathrm{mW}$, $P_+ = 138\mathrm{mW}$. Insets: t, oscillation period, $t_\mathrm{d}$, time delay. $t_{\mathrm{d1}}$: time between laser (-) ON and laser (+) OFF. $t_{\mathrm{d2}}$: time between laser (+) ON and laser (-) ON. $t_{\mathrm{d3}}$: time between laser (-) OFF and laser (+) ON. $t_{\mathrm{d4}}$: time between laser (+) OFF and laser (-) OFF.

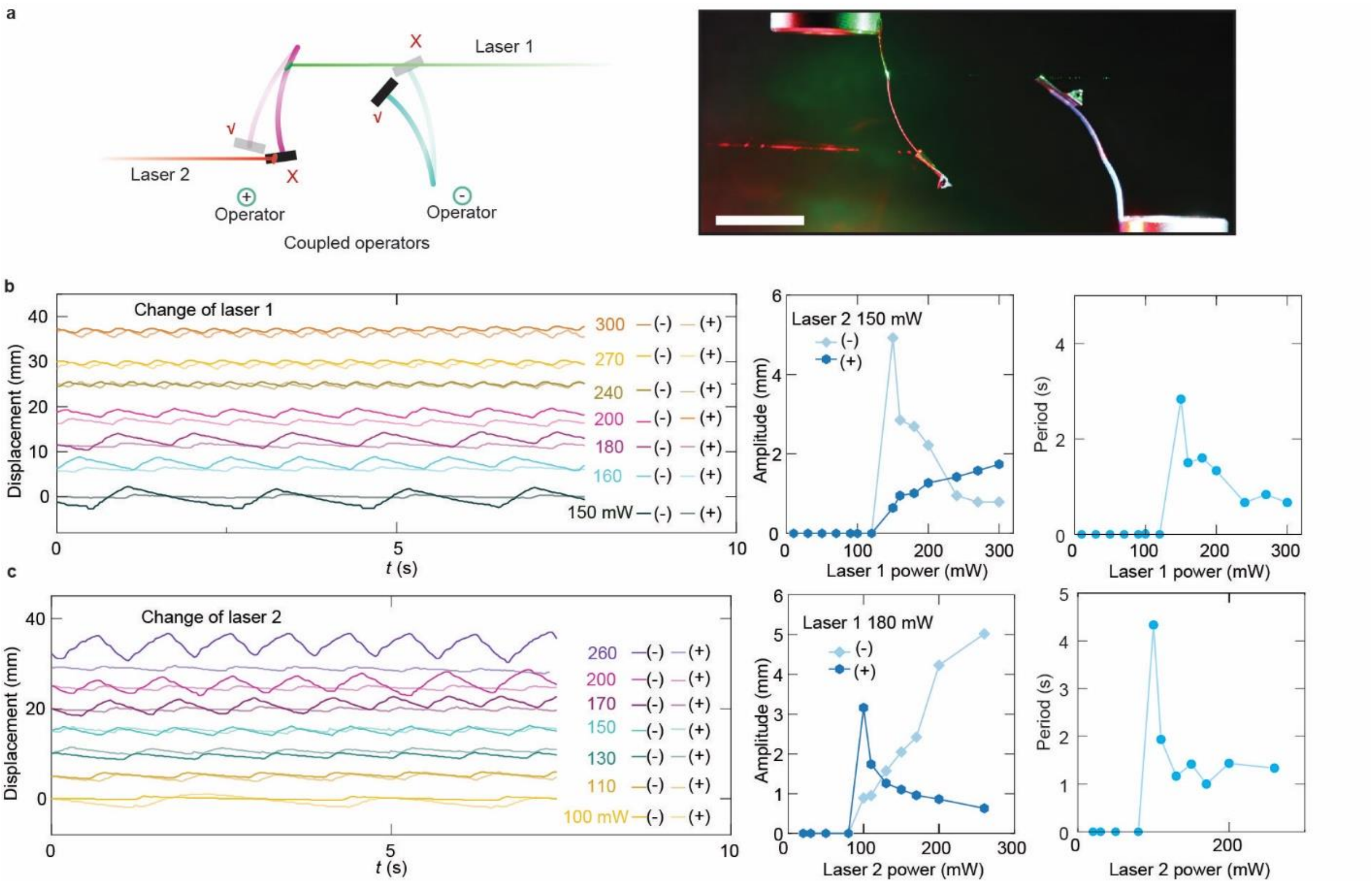

**Supplementary Figure 8. Light power dependent behaviors in the coupled oscillators** (a) Left: A schematic illustration depicts (-) and (+)operators coupled through two laser beams. Right: A photographic depiction showcases the optical setups of the coupled system. (b) Left: oscillation data recorded by varying the power of laser 1 while maintaining the constant power of laser 2. Right: the variation in amplitude and period of oscillation with changes in the power of laser 1. Laser 2: 150 mW, 635 nm, 3 mm spot size. (c) Left: oscillation data recorded by varying the power of laser 2 while maintaining the constant power of laser 1. Right: the variation in amplitude and period of oscillation with changes in laser 2 power. Laser 1: 180 mW, 532 nm, 2 mm spot size. Scale bar: 1 cm.

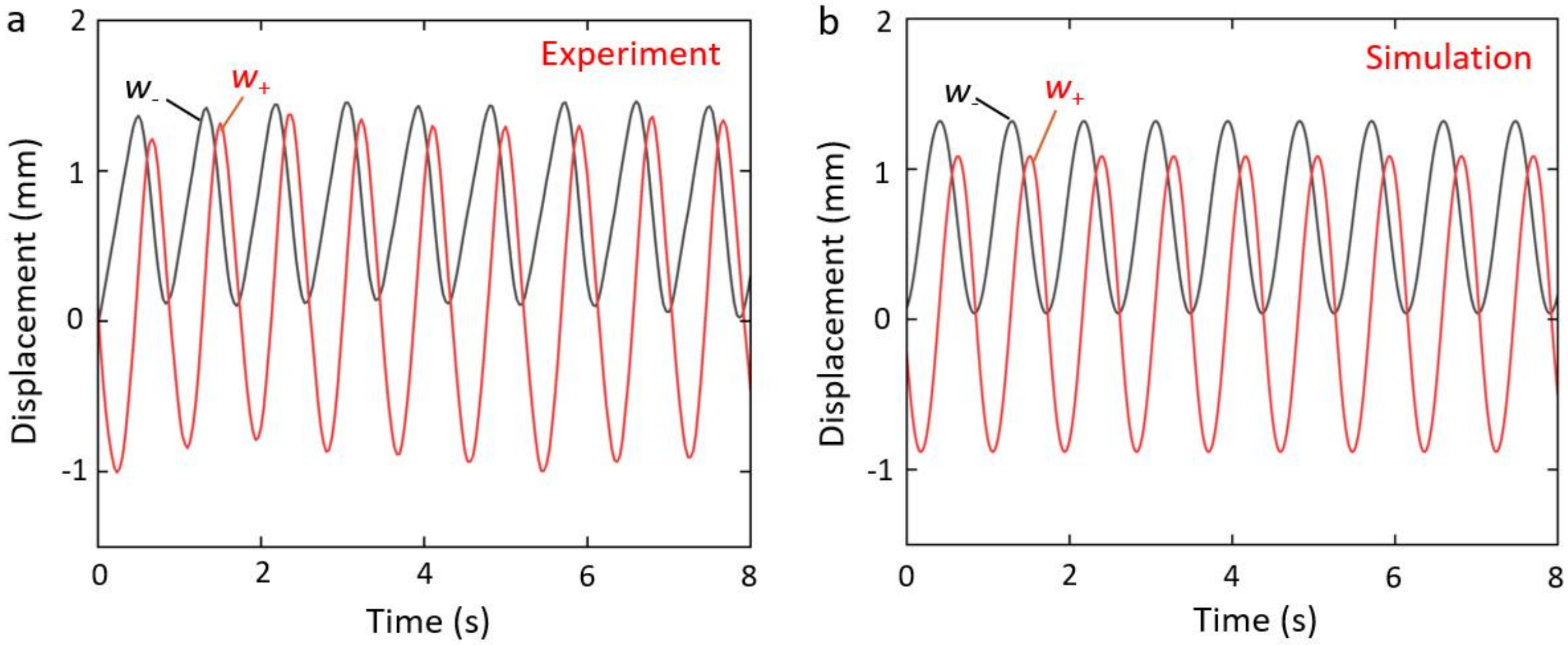

**Supplementary Figure 9. Modelling fitting**. (a) Experimental result oscillation data of the coupled oscillators. Displacement shows the change of tip positions of the two baffles. Laser 1: 138 mW, laser 2: 210 mW. (b) Theoretical prediction. In the simulation, we set $l_-$ = 2.5 cm, $l_+$= 2.5 cm, $w_{0-}$= 0.475 mm, $w_{0+}$= 0.35 mm, $\bar{\beta}$ = 0.2, $\tau_{\text{inertial}}$ = 0.115 s, $\tau_{\text{heat}}$ = 0.287 s, $P_-$ = 210 mW, $P_+$ = 138 mW, $a_- = 0.5$, $a_+ = 1.15$, $P_{0-}$ = 3W, $P_{0+}$ = 3W.

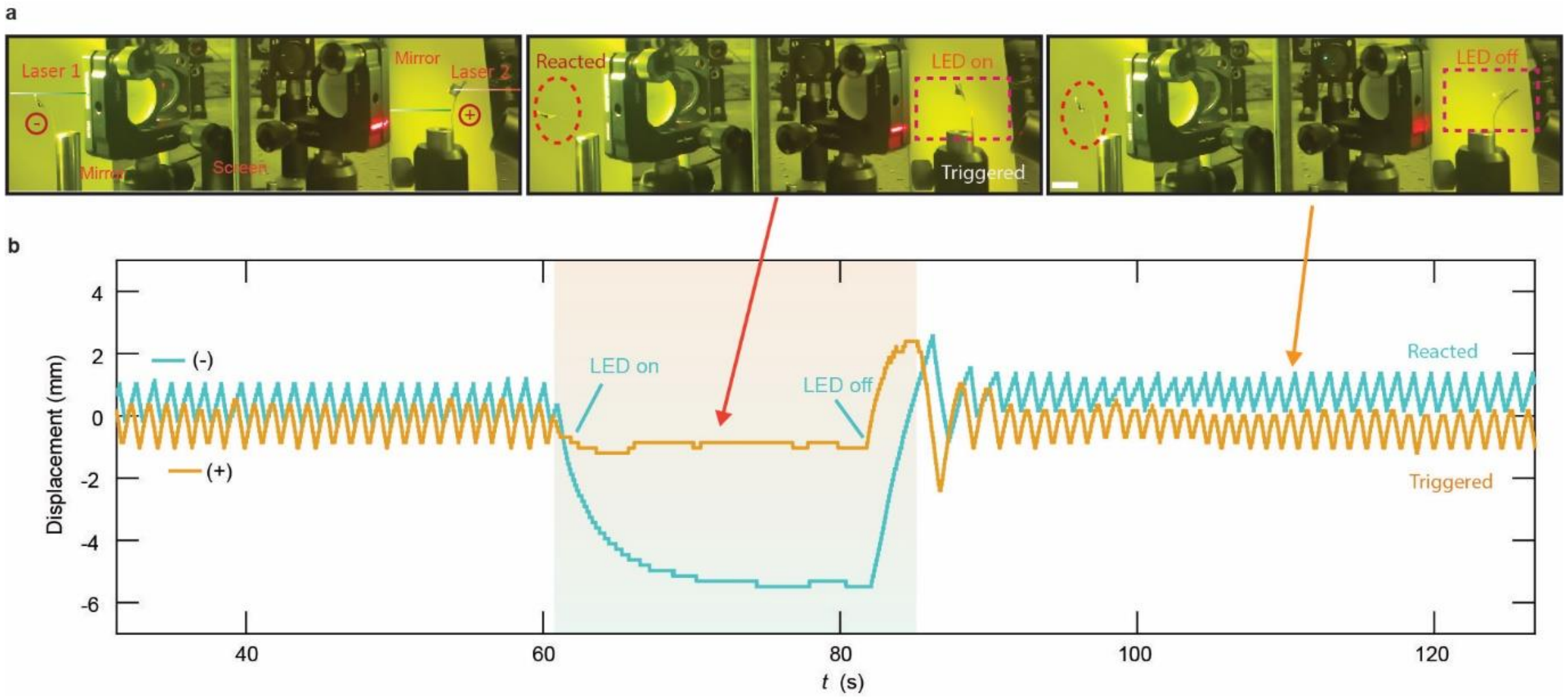

**Supplementary Figure 10. Interdependency of two coupled operators.** (a) Photographic demonstrating physically isolated operators at different oscillation phases. (b) Oscillation data upon manual interruption by illuminating LED light onto the (+)operator. The LED light (635 nm, 15 mW $cm^{-2}$) causes a cessation of vibration. The shadowed area indicates the duration of the light disturbance. Laser spot sizes are 2 mm (laser 1) and 3 mm (laser 2). Laser powers are 70 mW (laser 1) and 540 mW (laser 2). Wavelengths are 532 nm for laser 1 and 635 nm for laser 2. Scale bar: 1 cm.

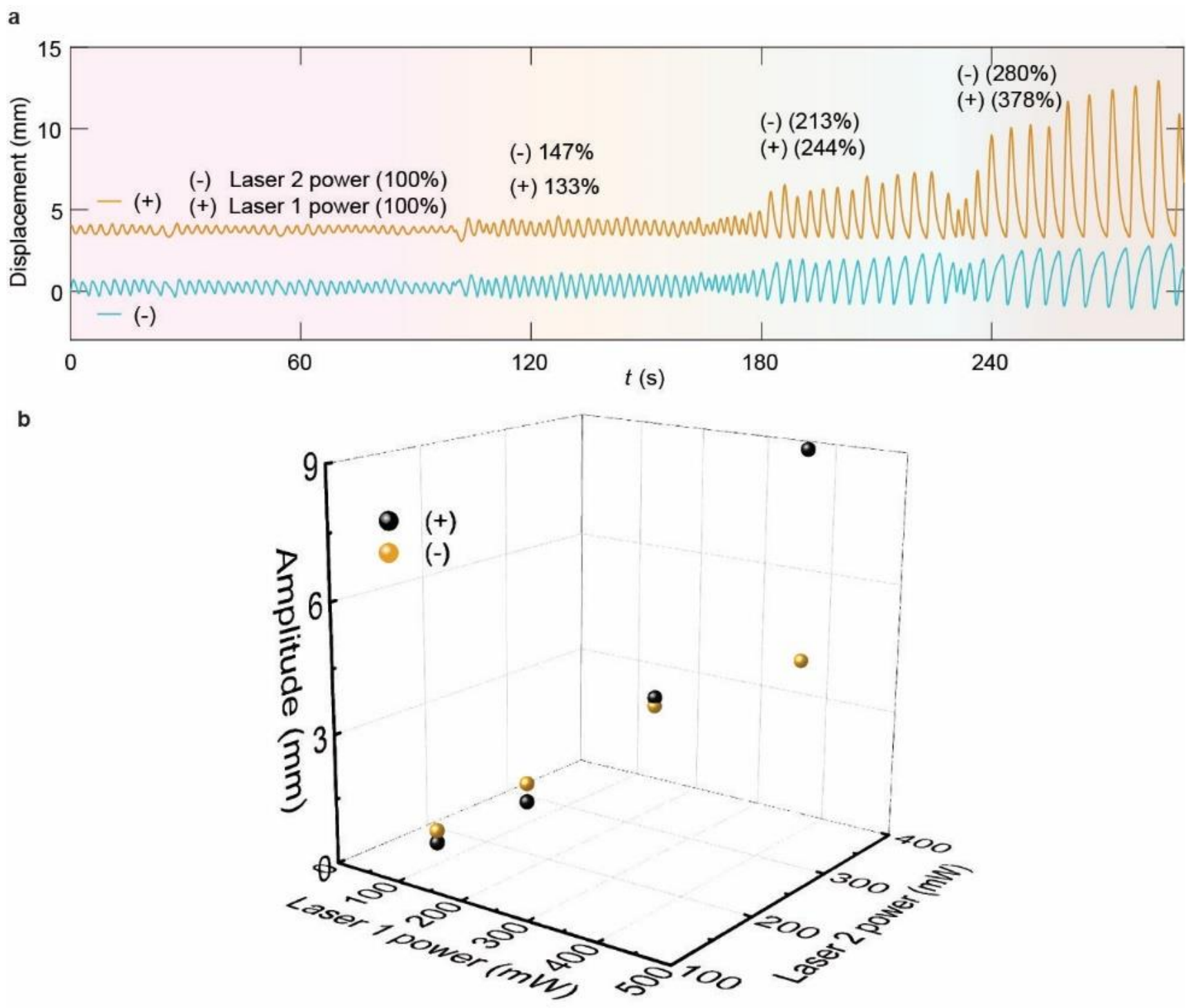

**Supplementary Figure 11. Total power dependency in coupled oscillation.** The powers of two laser beams are increased simultaneously. (a) Oscillation data by changing the power of both laser beams. (b) Corresponding amplitude variations in both operators. Two operators are physically isolated by using a screen board. The initial power of the beams is 150 mW (laser 1) and 90 mW (laser 2). The spot sizes are 2 mm (laser 1) and 3 mm (laser 2). Excitation wavelengths are 532 nm (laser 1) and 635 nm (laser 2).

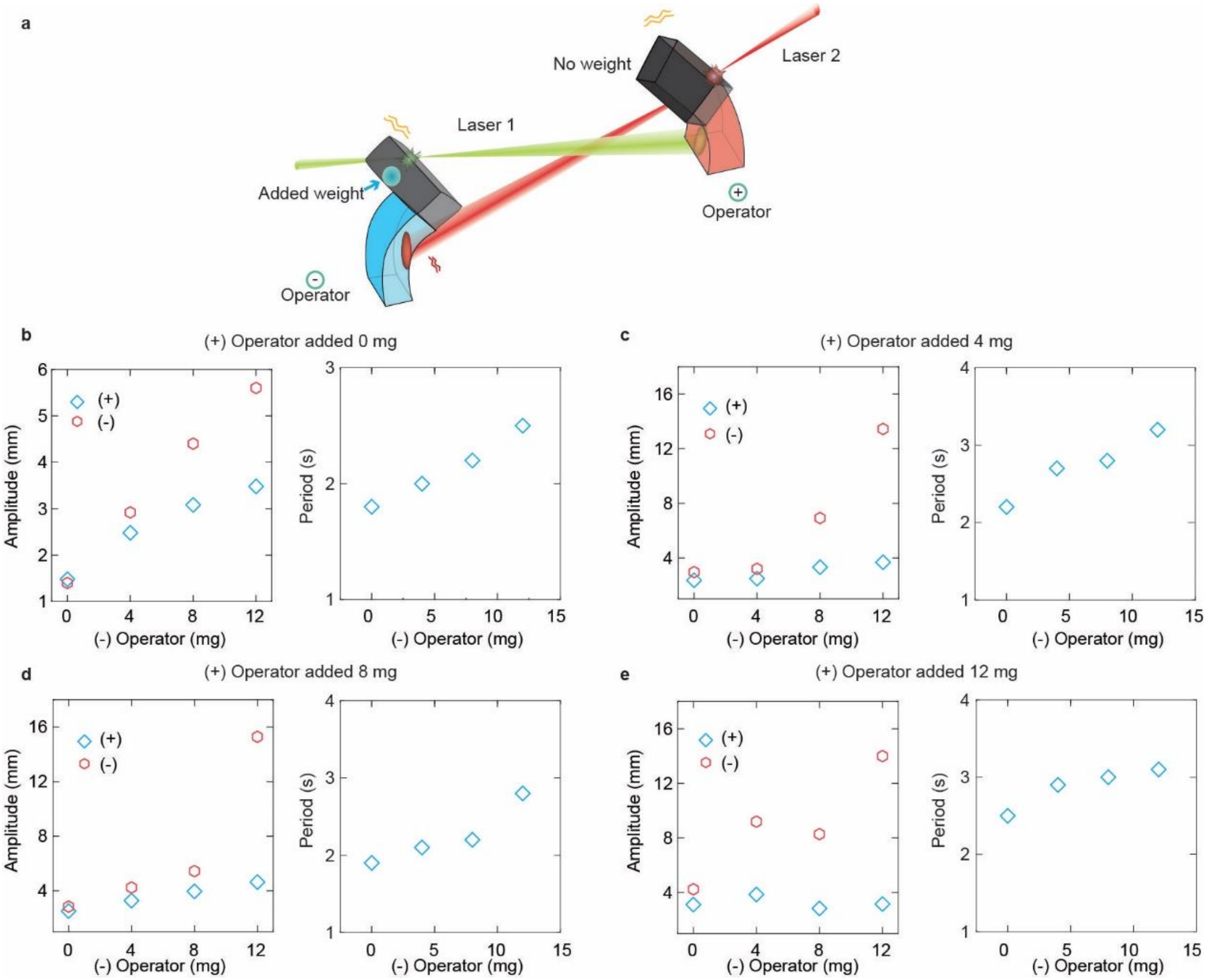

**Supplementary Figure 12. Effect of added mass.** (a) Schematic depiction of two coupled operators. Variation in amplitude and period by adding mass onto (-)operator, while keeping the added mass on (+)operator at 0 mg (b), 4 mg (c), 8 mg (d), 12 mg €. Two operators are physically isolated by using a screen board. Light power is 220 mW for both lasers. Spot sizes are 2 mm for laser 1 and 3 mm for laser 2. Excitation wavelengths are 532 nm (laser 1) and 635 nm (laser 2).

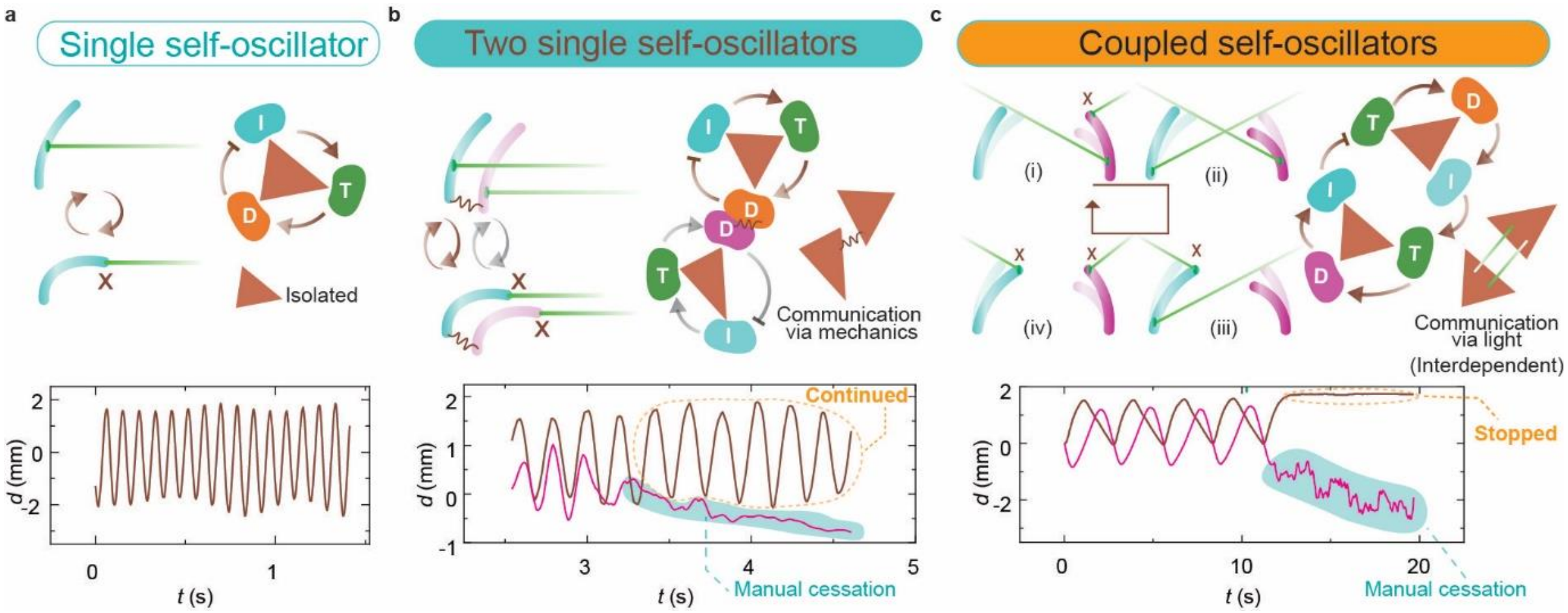

**Supplementary Figure 13. The concepts of interaction in non-equilibrium matters.** (a) The conventional single-piece self-oscillator is built upon a negative feedback loop. Bottom: oscillation data depicts an LCE-based self-oscillator based on the design shown in Fig.1d. (b) In a mechanically connected oscillating system, each oscillator is driven by a feedback loop on its own and interacts with others through physical interaction. Bottom: oscillation data depicts two LCE self-oscillators, in which manual cessation of one-unit unaffecting the oscillation dynamics of the other. (c) The interdependent self-oscillators are coupled through light beams. In this case, two units rely on each other to sustain the motion. Bottom: oscillation data showcases the displacement of two coupled oscillators, in which manual cessation of one-unit stops the motion of the other.

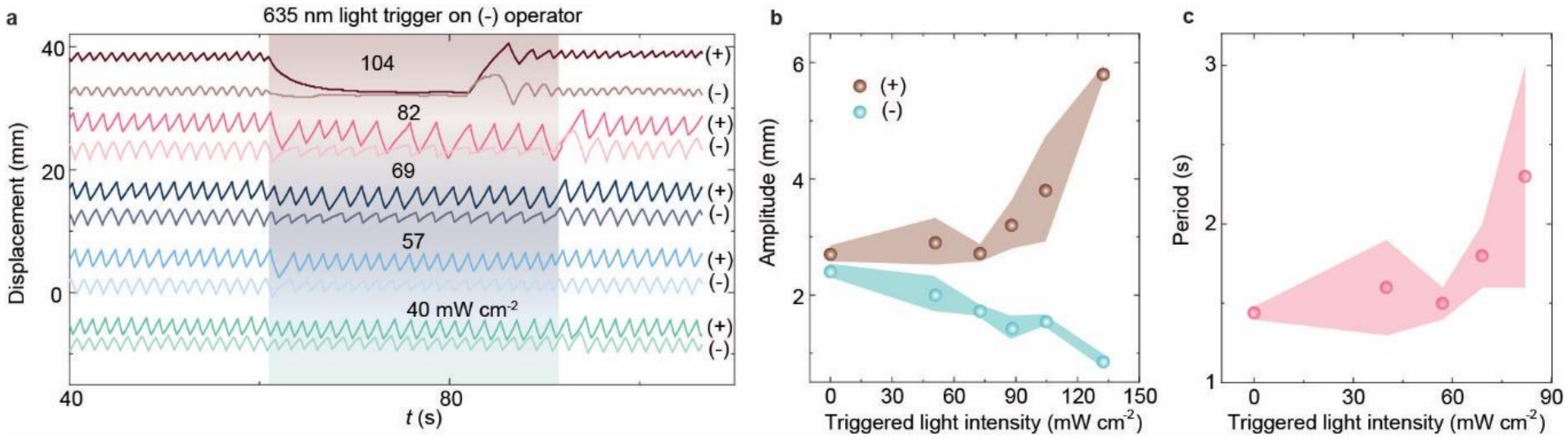

**Supplementary Figure 14. Effect of additional illumination.** (a) Oscillation data upon varying the intensity of external light interruption. External light: 635 nm, LED source, 40 to 104 mW $cm^{-2}$. The light trigger is imposed onto the (-)operator. (b) Amplitude variation of two oscillations with an increase of external light intensity for disturbance. (c) Period variation of the coupled operators by changing external light disturbance. Two operators are physically isolated by using a screen board. The power of the beams is 70 mW (laser 1) and 540 mW (laser 2). The spot sizes are 2 mm (laser 1) and 3 mm (laser 2). Excitation wavelengths are 532 nm (laser 1) and 635 nm (laser 2). Error bars represent s.d. for n = 3 measurements. The same sample was measured repeatedly.

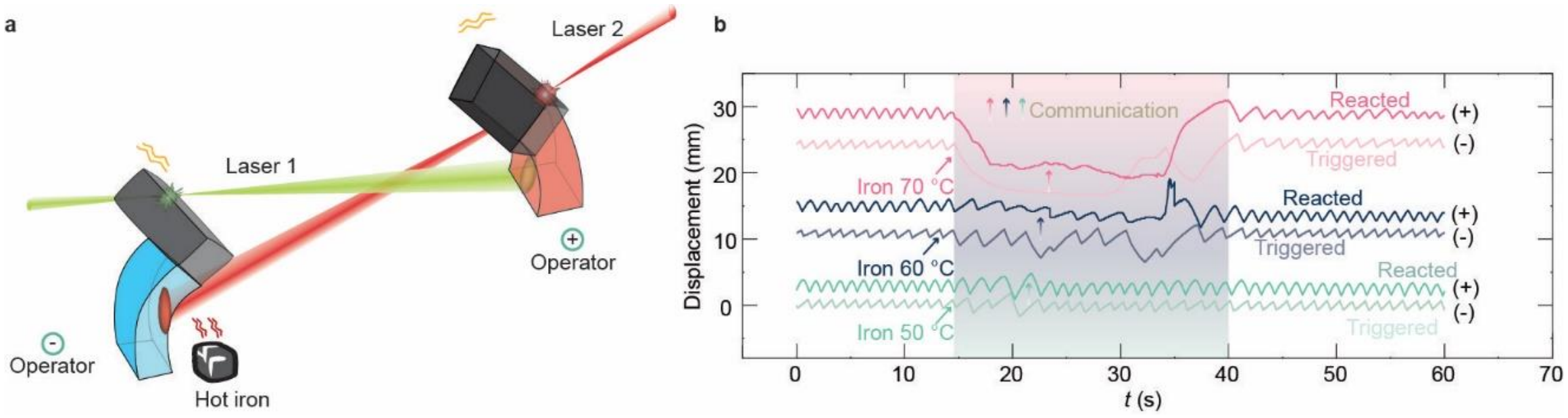

**Supplementary Figure 15. Effect of heat disturbance.** (a) Schematic diagram of the temperature interference in coupled operators. (b) Oscillation data upon different heat interferences. A hot iron (1.5 × 1.5 × 1.5 $cm^3$) was heated (50 to 70 ºC) and placed 2 cm below the (-)operator. Two operators are physically isolated by using a screen board. Laser 1: 532 nm, 70 mW. Laser 2: 635 nm, 540 mW. The spot sizes are 2 mm (laser 1) and 3 mm (laser 2).

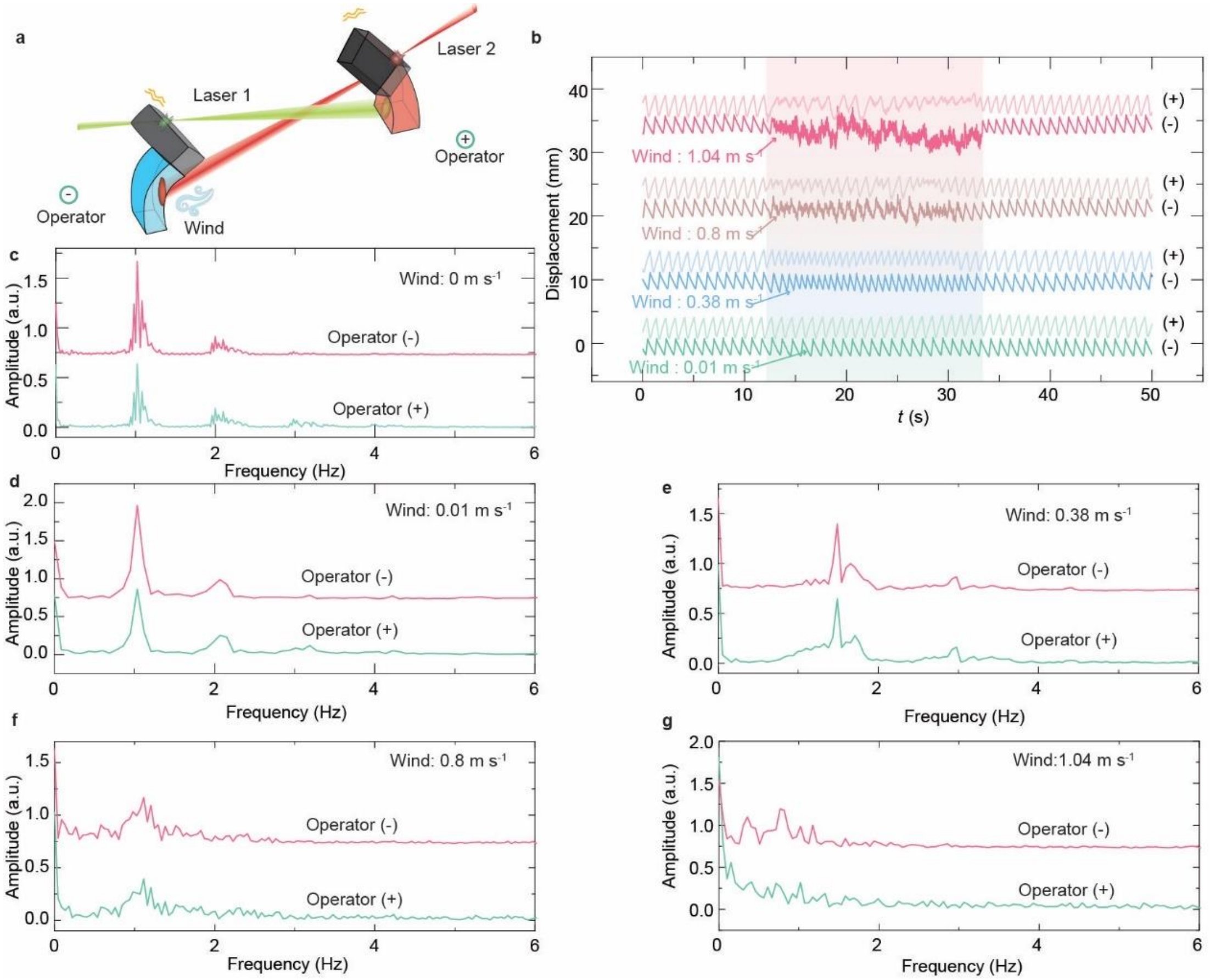

**Supplementary Figure 16. Effect of wind disturbances.** (a) Schematic diagram of the wind interference in the coupled operators. (b) Oscillation data of two operators, while the (-)operator is affected by wind flow with different velocities. Fourier transform spectra of oscillations at static air condition (c), upon 0.01 m $s^{-1}$ (d), 0.38 m $s^{-1}$ (e), 0.8 m $s^{-1}$ (f) and 1.04 m $s^{-1}$ (g) wind interferences. Fourier transform at the frequency domain is calculated from (b). Two operators are physically isolated by using a screen board. Laser 1: 532 nm, 70 mW. Laser 2: 635 nm, 540 mW. The spot sizes are 2 mm (laser 1) and 3 mm (laser 2).

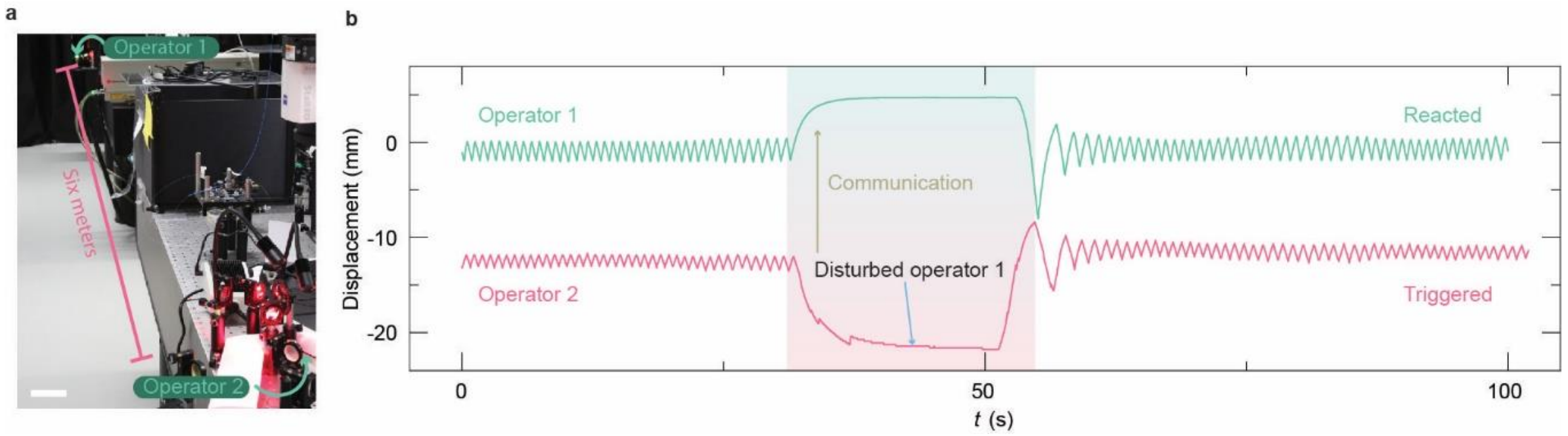

**Supplementary Figure 17. Long distance signal transmission.** (a) Photograph of light communication between two operators set on two optical tables separated six meters away from each other. The arrows point towards the positions of the operator. (b) Oscillation data of long distance coupled operators. One is ceased by manual stoppage. The shaded area represents the duration of the mechanical interference. Laser 1: 532 nm, 300 mW. Laser 2: 635 nm, 460 mW. The spot sizes are 2 mm (laser 1) and 3 mm (laser 2). Scale bar: 10 cm.

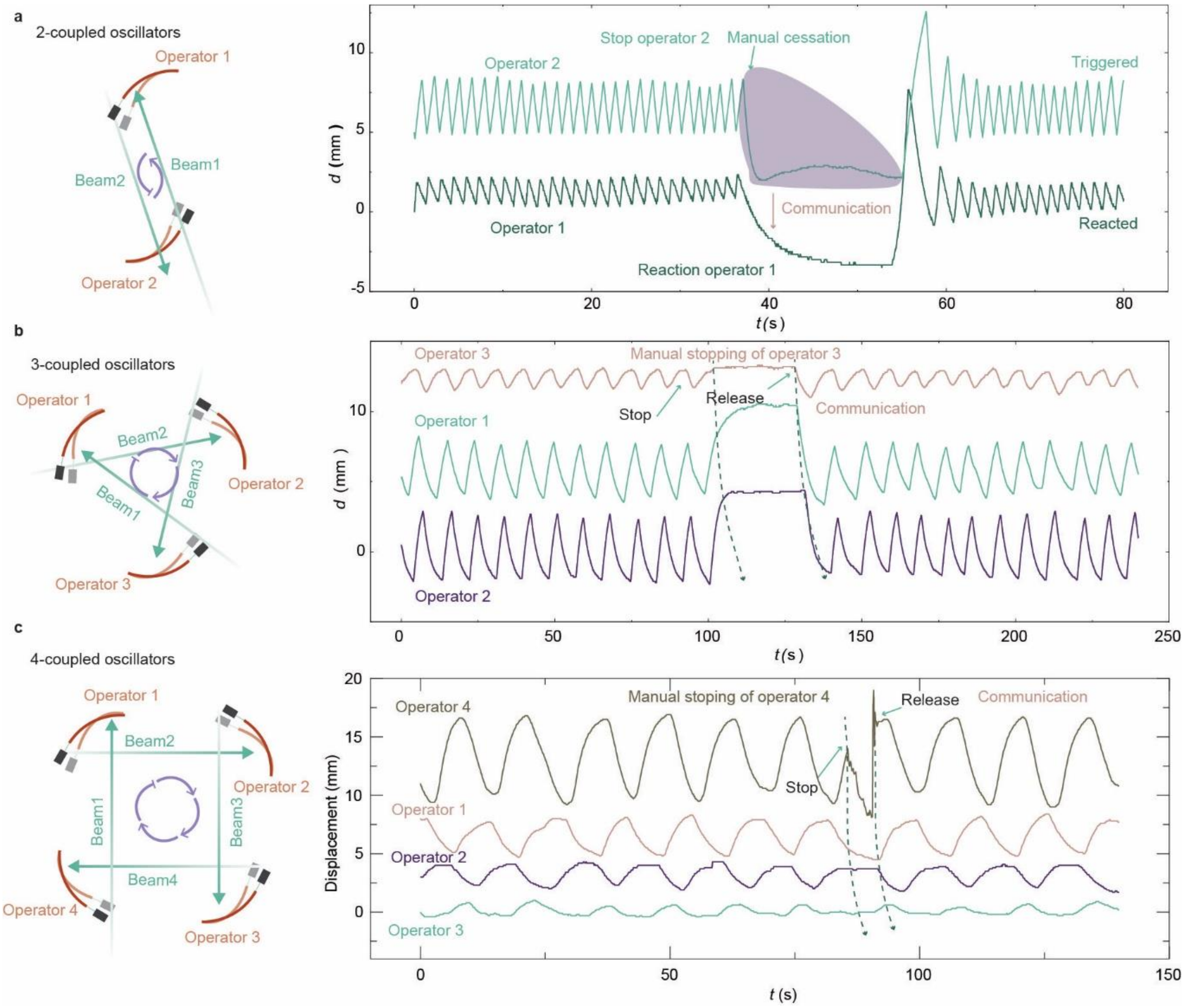

**Supplementary Figure 18. Light communicative network.** (a) Schematic diagram depicting the mechanism of a self-oscillating system consisting of two units, accompanied by oscillation data of the operators. The purple area indicates the duration of the manual stoppage. Laser 1: 532 nm, 70 mW. Laser 2: 635 nm, 540 mW. The spot sizes are 2 mm (laser 1) and 3 mm (laser 2). (b) Schematic diagram of a system composed of three units, along with corresponding oscillation data. All laser: 532 nm, 320 mW. Laser spot sizes: 2 mm. (c) Schematic diagram illustrating a self-oscillating system composed of four units, accompanied by oscillation data. All laser: 532 nm, 320 mW. Laser spot sizes: 2 mm. LCE actuator dimensions in all cases: 24 × 2 × 0.1 $mm^3$. Baffle dimension: 5 × 20 × 0.01 $mm^2$.

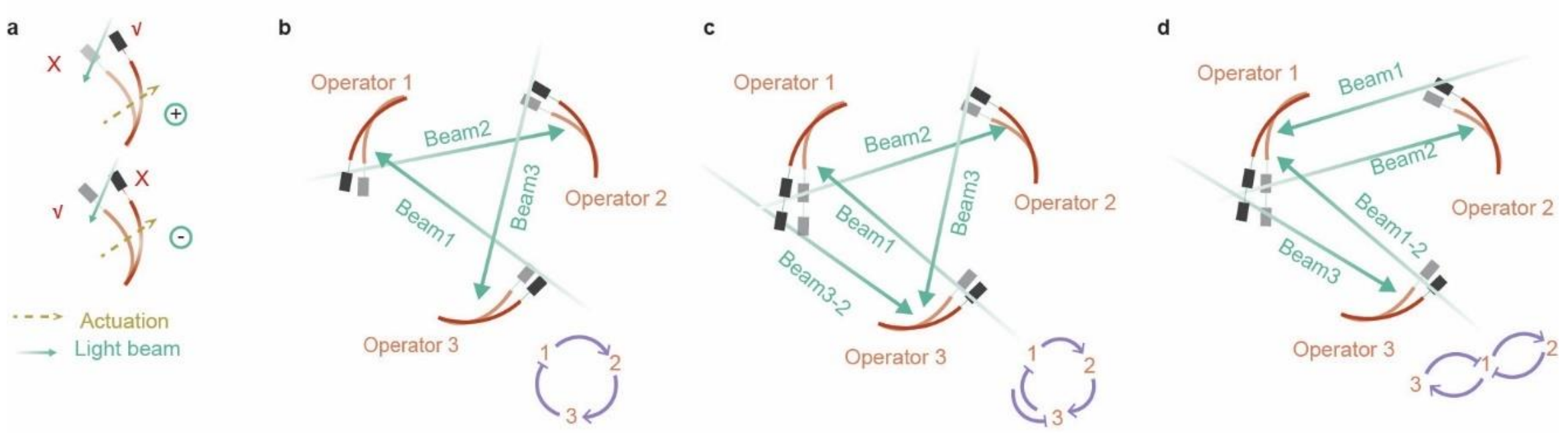

**Supplementary Figure 19. Designs of bio-inspired light communicative network**. (a) The schematic diagram illustrates the mechanism of a single operator. (b) The three-component negative feedback loop represents a basic form of a bio-oscillator, where feedback mechanisms regulate the activity of components within the system. This is the most typical configuration often resulting in rhythmic oscillations based on the interplay of activating and inhibiting signals. (c) The incoherent feedforward loop introduces an additional negative feedback mechanism to enhance the robustness of the oscillatory behavior. It usually makes the system more resilient to external disturbances. (d) In the configuration with two negative feedback loops acting on the same operator, chaos may ensue. This scenario may lead to highly unpredictable and irregular oscillations, characterized by complex dynamics and a lack of stable patterns. The designs are inspired by Ref [1].

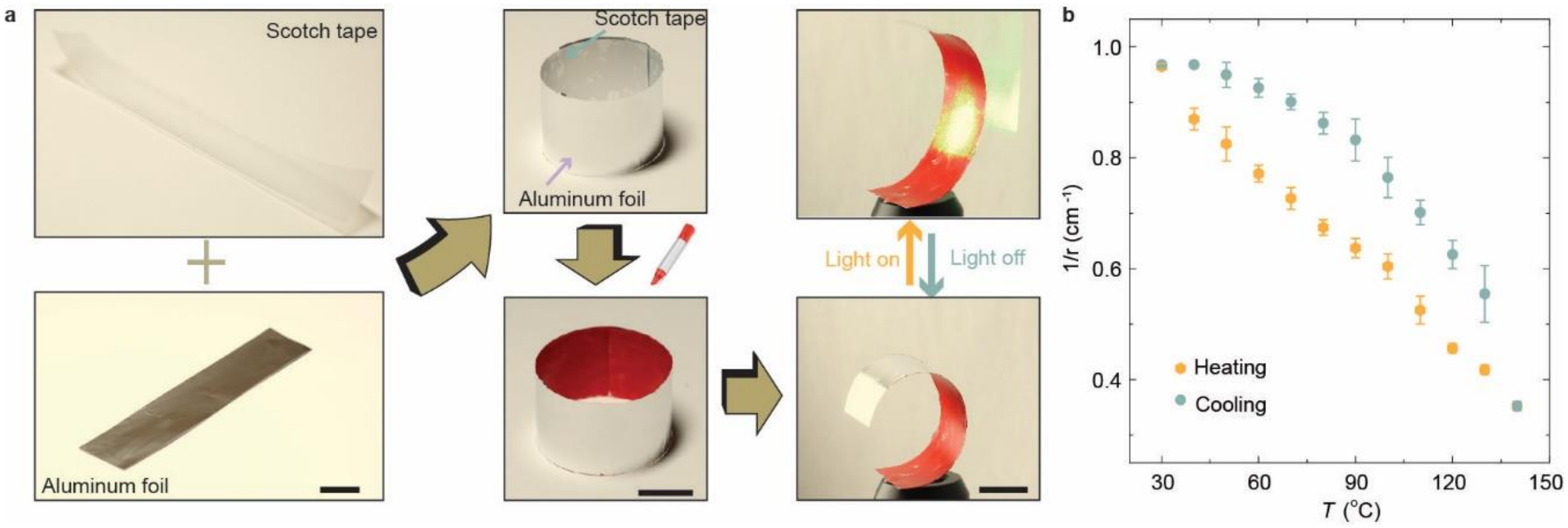

**Supplementary Figure 20. Responsive behaviors in aluminum foil-scotch tap bilayer.** (a) Preparation steps. A thermally responsive bilayer actuator is comprised of two kinds of daily material, *i.e.*, scotch tape (50 μm thick) and aluminum foil (10 μm thick). After sticking two films together and painting with a red colored pen on the scotch tape, the bilayer is annealed on a hotplate at 140 ºC. After annealing, the bilayer exhibits reversible deformation based on the photothermal effect. Light illumination: 460 nm, 50 mW $cm^{-2}$. (b) Curvature (1/r) variation upon temperature change. All scale bars are 1 cm.

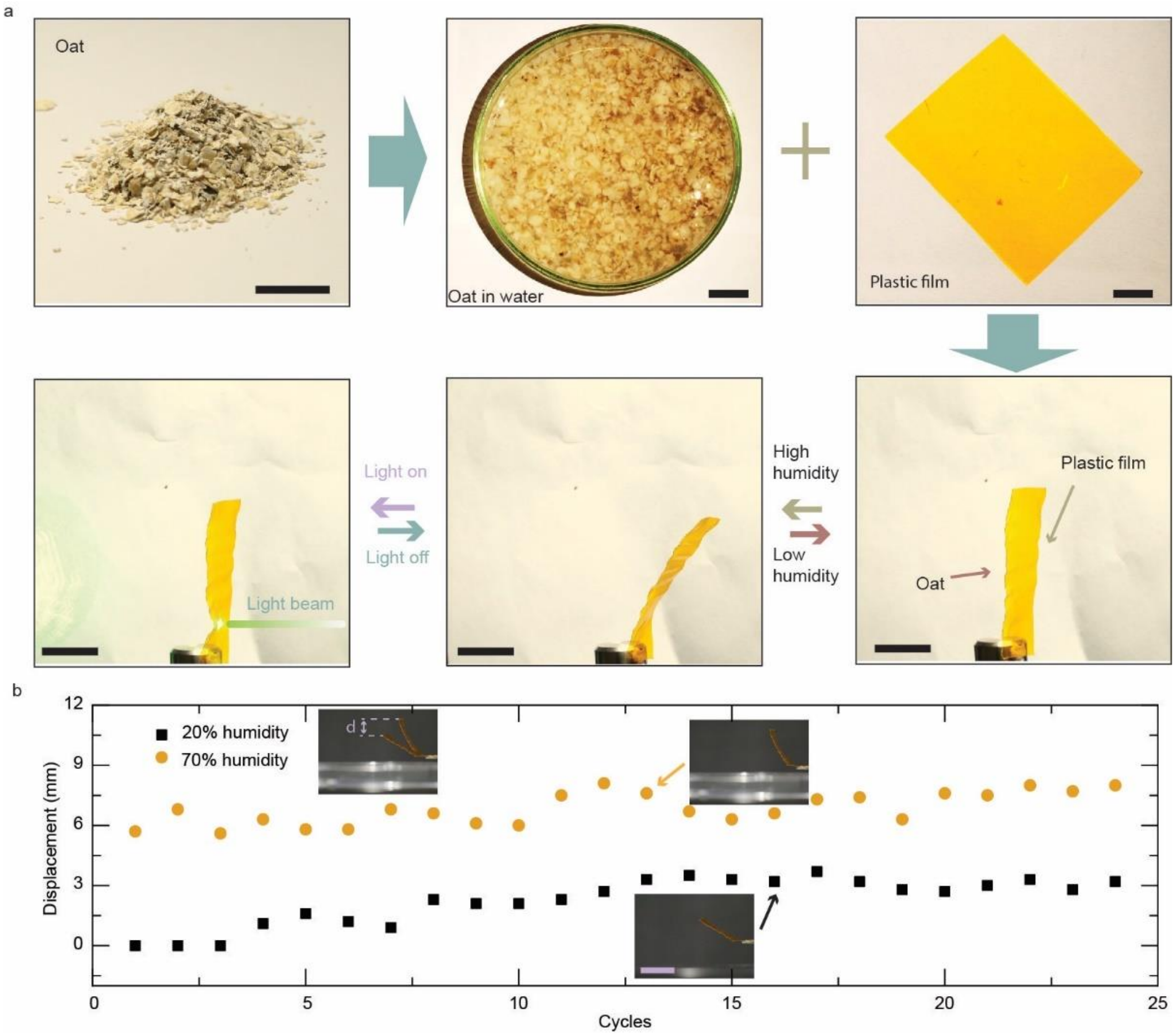

**Supplementary Figure 21. Responsive behaviors in the oat-plastic bilayer.** (a) The depicted process illustrates the preparation of a humidity-responsive film derived from common oats. The procedure involves dissolving oatmeal flakes into water, boiling the water, and drop casting oat solution onto a colored plastic film as a passive layer. After water evaporation, the bilayer exhibits reversible deformation upon a change in ambient humidity. A light excitation can also cause desorption of water content from the oat layer, inducing reversible photo-heat-humidity triggered deformation. Light: 532 nm laser, 90 mW, spot size, 2 mm. (b) The cyclic stability of the humidity-responsive film under multiple cycles of humidity switching between 20% and 70%. The inset defines displacement and shows photos of the film bending at different humidity levels. All scale bars are 1 cm.

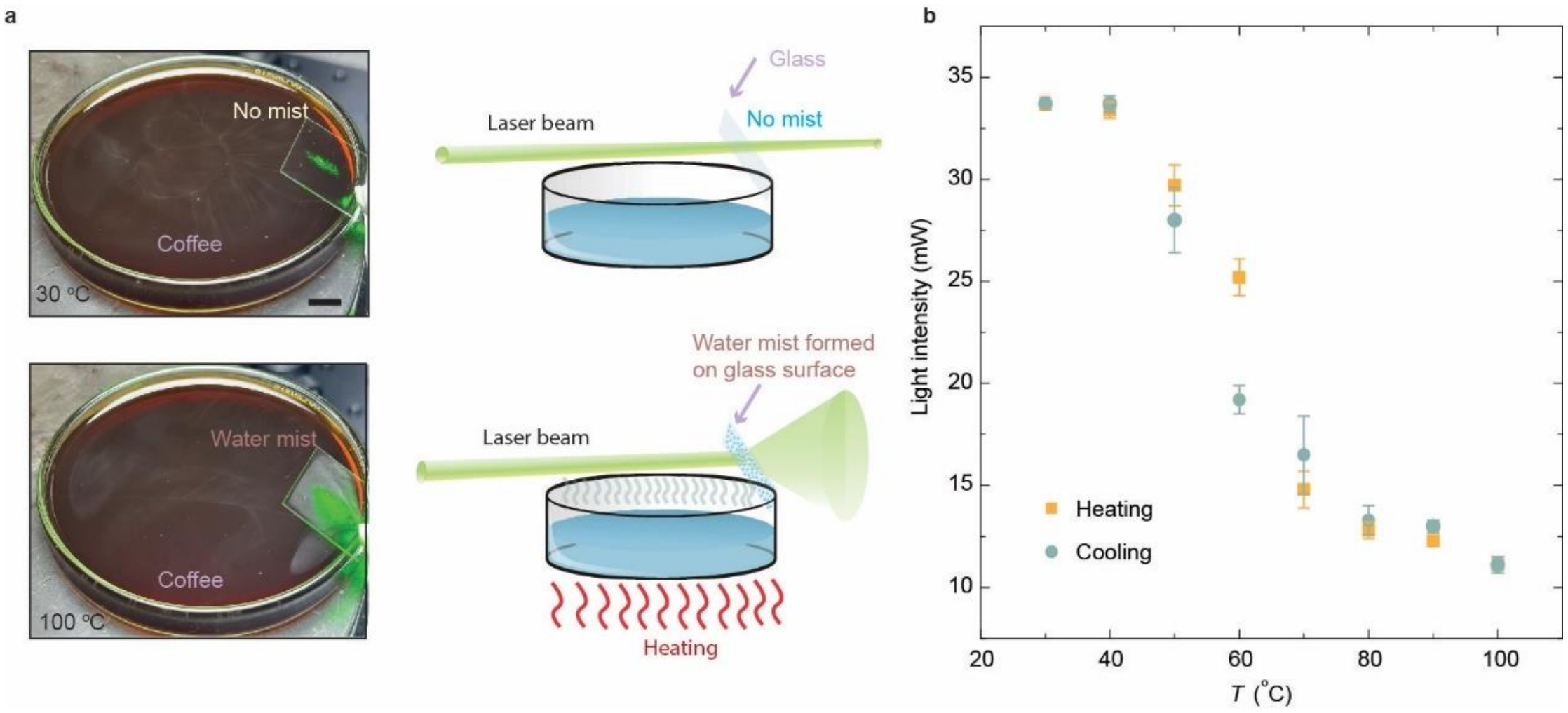

**Supplementary Figure 22. Responsive behaviors in a foggy glass.** (a) Photographs and schematic diagrams depicting the fogging formation on the cover glass slide on top of the coffee container. The moisture from the heated coffee creates a foggy layer that drops the transparency of the cover glass. (b) Light transmission through the cover glass was measured at different coffee temperatures. Light: 532 nm laser beam, input power 33 mW. Error bars represent s.d. for n = 3 measurements. The same sample was measured repeatedly. Scale bar: 1 cm.

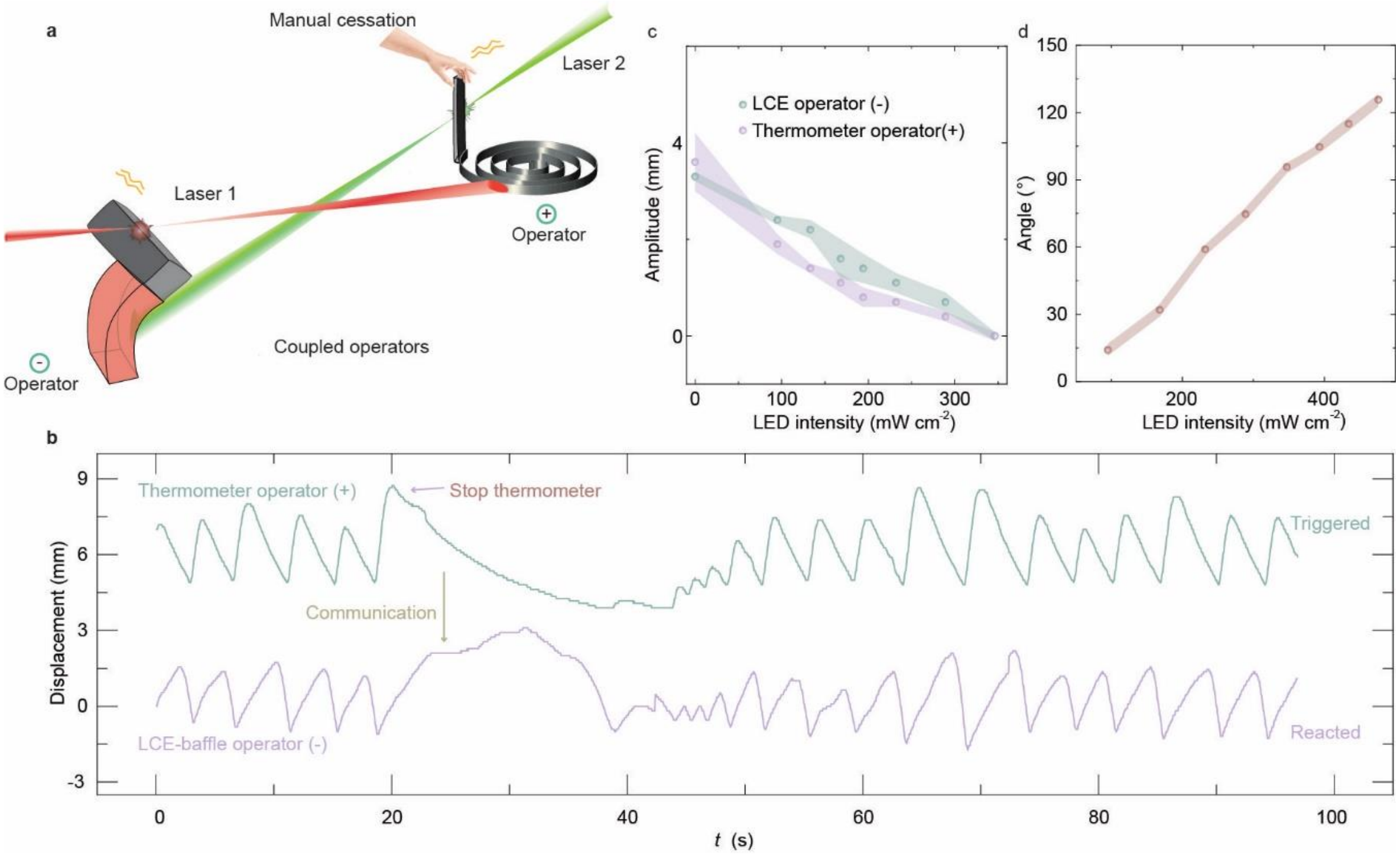

**Supplementary Figure 23. Signal transmission between LCE actuator and thermometer spring.** (a) The schematic diagram illustrates the mechanism of the light communicative system based on the coupling between an LCE-baffle and a thermometer-baffle. The side of the thermometer spring is blackened with a pen to enhance photo-heat absorption. (b) Oscillation data of the coupled system, in which a manual halting at the thermometer and subsequent synchronization between two-oscillator motions are showcased. Laser 1: 532 nm, 930 mW, 3 mm spot size. Laser 2: 532 nm, 44 mW, and 2 mm spot size. (c) Variation of oscillation amplitude upon external light disturbance on the thermometer. Light disturbance: LED source, 460 nm, 0 to 350 mW $cm^{-2}$. (d) The graph presents the photothermal responsive characteristics of the thermometer spring. LCE actuator dimensions: 24 × 2 × 0.1 $mm^3$. Baffle dimension: 5 × 20 × 0.01 $mm^2$. Error bars represent s.d. for n = 3 measurements. The same sample was measured repeatedly.

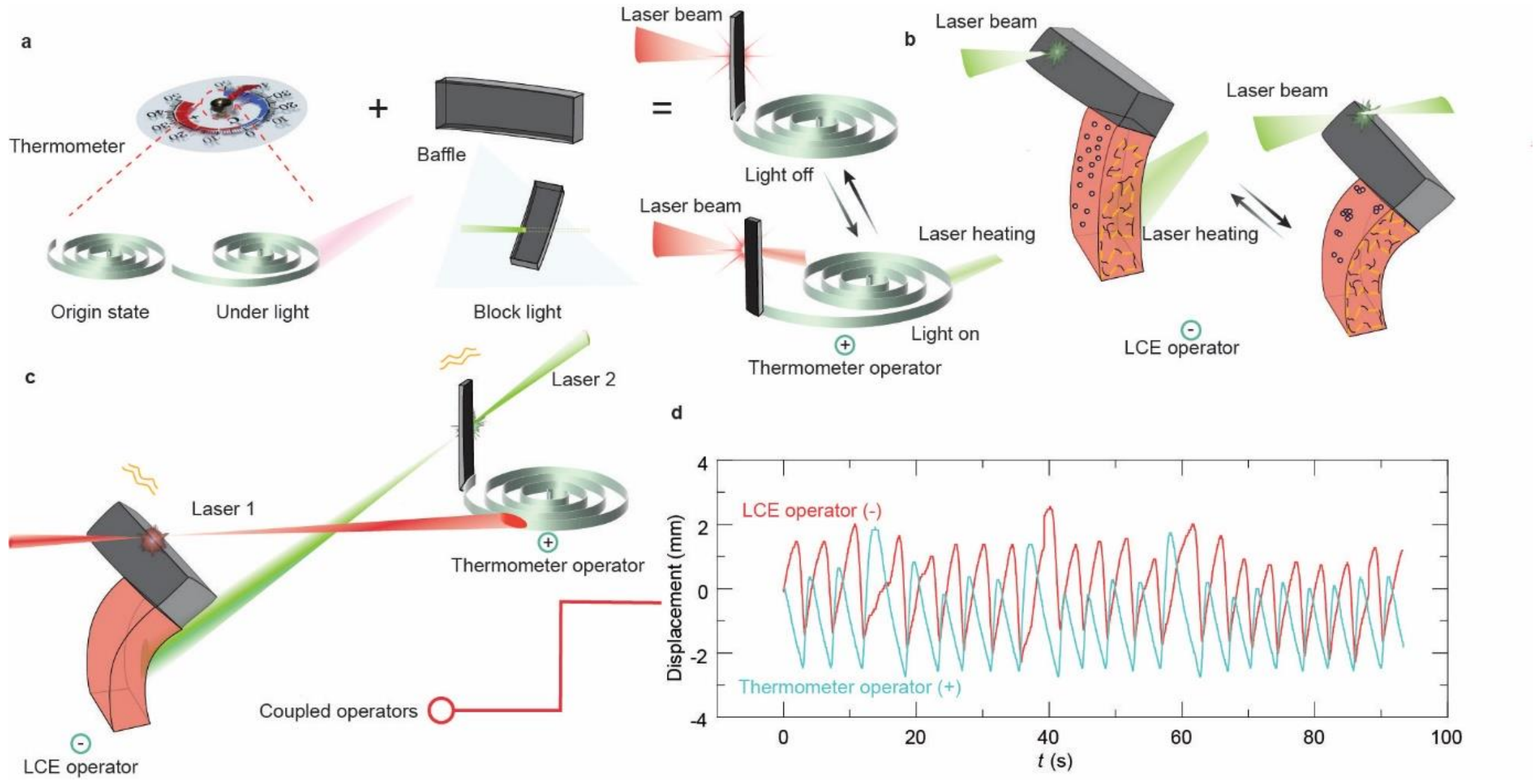

**Supplementary Figure 24. Coupling between LCE and a thermometer spring.** (a) Schematic drawing showing the essential components of the thermometer-baffle operator. The thermometer metal spring undergoes bending upon exposure to light, an aluminum foil sheet acts as a barrier preventing light transmission. (b) A schematic diagram demonstrates the bending behavior of the LCE-baffle (-)operator when subjected to light irradiation. (c) The schematic diagram depicts the coupling between the thermometer and the LCE actuator. (d) Oscillation data of the coupled system. Laser 1 power: 930 mW, Laser 2 power: 44 mW, all beams are 532 nm. The size of the light spot for laser 1 is 3 mm, and for laser 2 is 2 mm. LCE sample dimensions: $24 \times 2 \times 0.1$ mm$^3$, baffle size: $5 \times 20 \times 0.01$ mm$^3$.

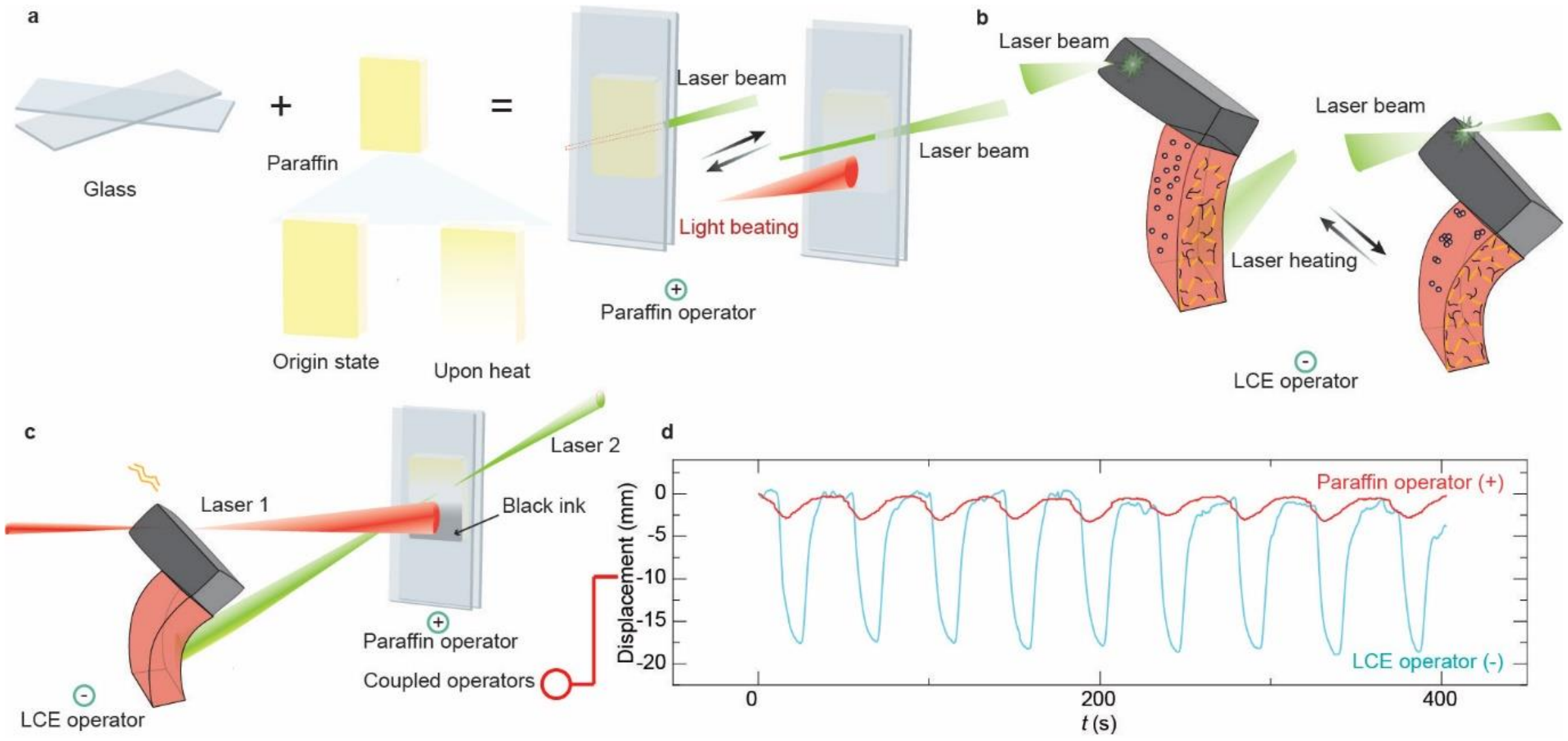

**Supplementary Figure 25. Coupling between LCE and a paraffin plate**. (a) The assembly includes all essential components for operator systems. Glass substrates are used to form a cell to store the paraffin. Paraffin becomes transparent to light when heated. (b) A schematic diagram illustrates the bending behavior of the LCE-baffle (-)operator when subjected to light irradiation. (c) The schematic diagram depicts the coupling between a paraffin plate and an LCE-baffle operator. (d) Oscillation data of the coupled system. Laser 1 power: 1440 mW, Laser 2 power: 80 mW, all beams are 532 nm. The spot size for laser 1 is 2 mm, and 1.2 mm for laser 2. LCE sample dimensions: $24 \times 2 \times 0.1$ mm$^3$, baffle size: $5 \times 20 \times 0.01$ mm$^3$.

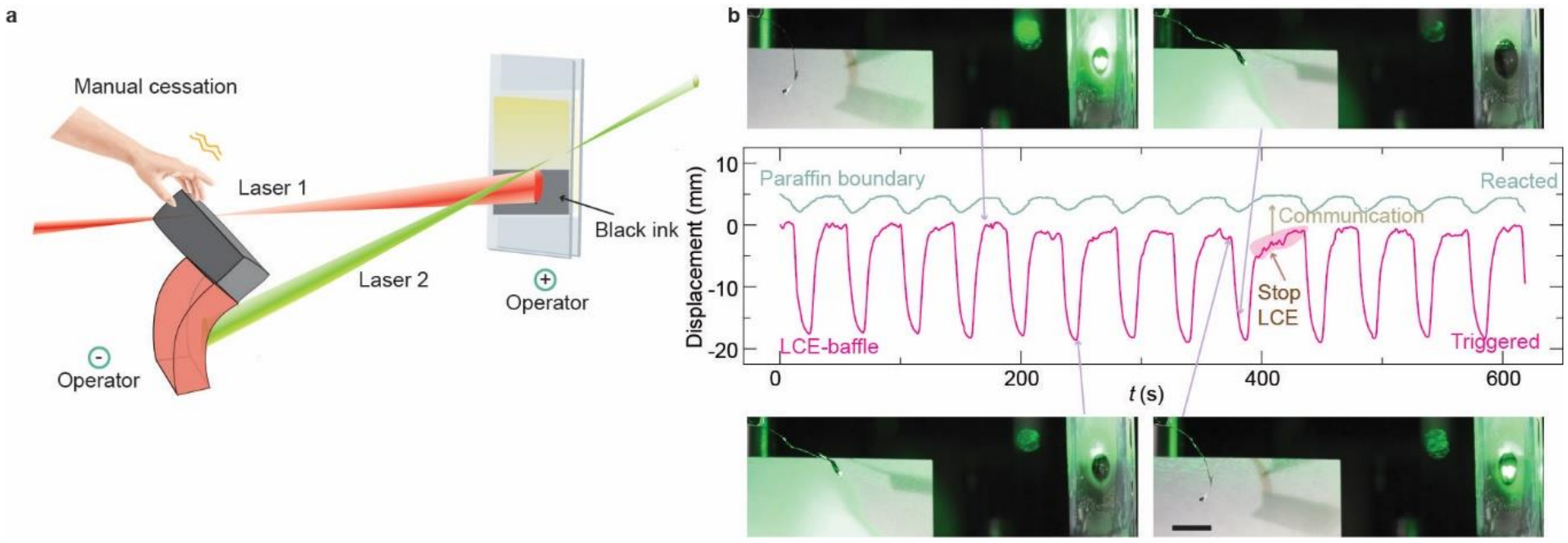

**Supplementary Figure 26. Signal transmission between LCE actuator and paraffin.** (a) Schematic diagram illustrating the mechanism of the light communicative system based on the coupling between an LCE-baffle and a glass cell infiltrated with paraffin. The lower section of the paraffin is blackened with a pen to enhance photo-heat absorption. (b) Oscillation data of the coupled system. Insets are photographs of the bending LCE operator and paraffin at different oscillation phases. Laser 1: 532 nm, 1440 mW, 2 mm spot size. Laser 2: 532 nm, 80 mW, and 1.2 mm spot size. LCE actuator dimensions: $24 \times 2 \times 0.1$ $mm^3$. Baffle dimension: $5 \times 20 \times 0.01$ $mm^2$. Scale bar: 1 cm.

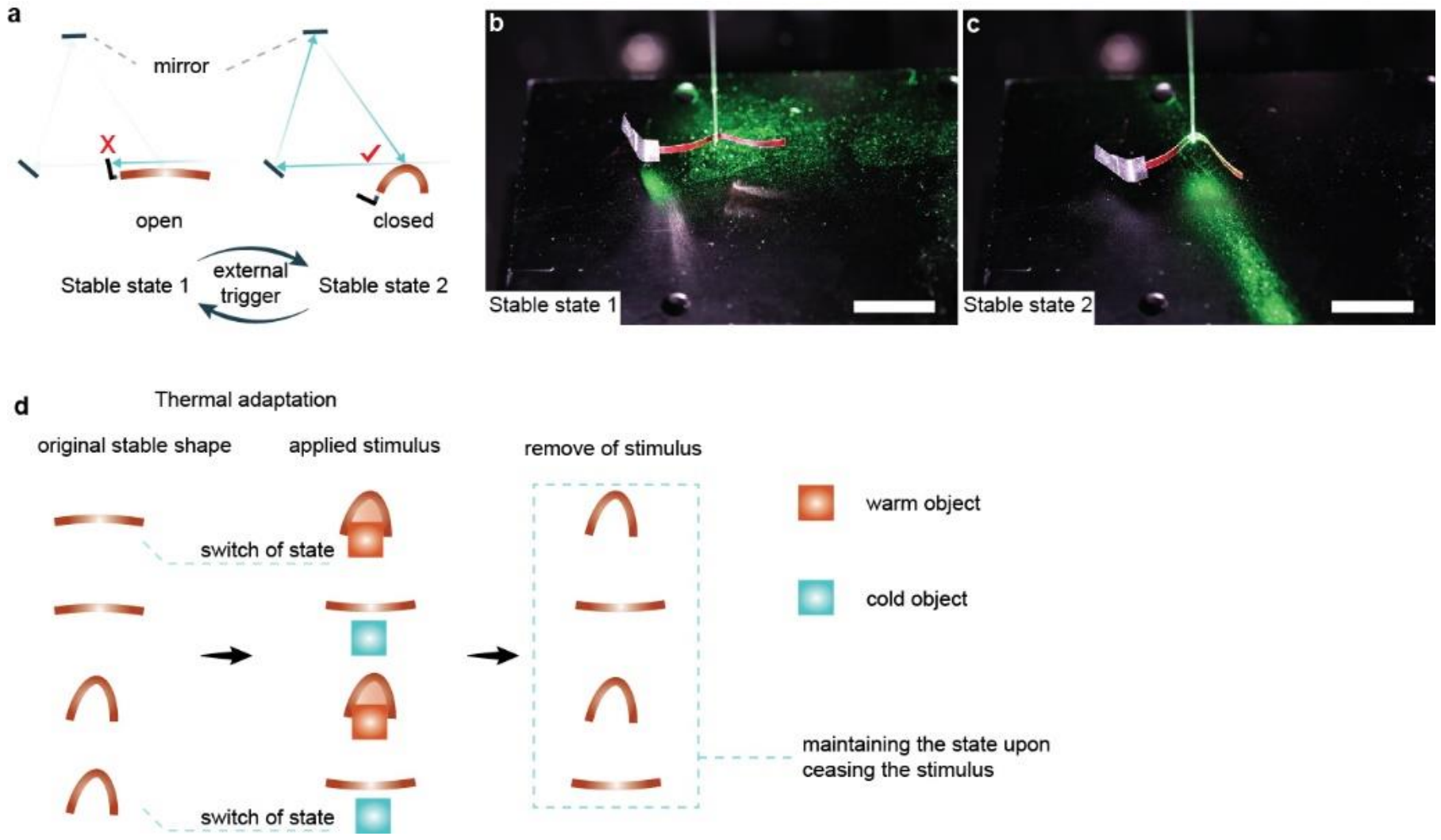

**Supplementary Figure 27. Bi-stability enabled by positive feedback.** (a) Optical design of a positive feedback loop consisting of one positive operator and a single light beam. Photos of the operator in the (b) open state and (c) closed state. (d) Four scenarios illustrating thermal adaptation, showing shape changes in response to heat or cold stimuli, and the capability of maintaining the state after ceasing the stimulus. Scale bar is 1 cm.

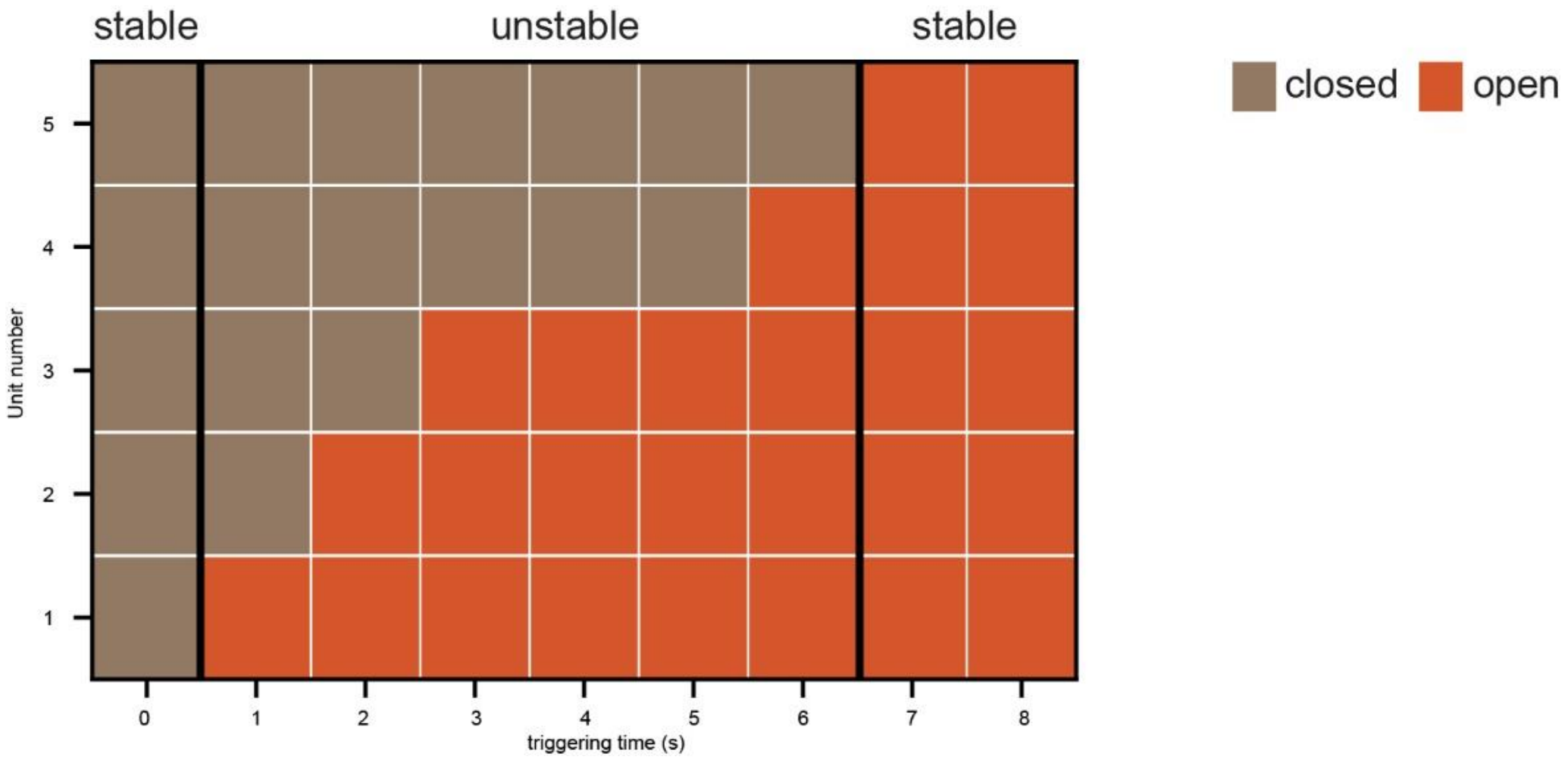

**Supplementary Figure 28. Cascading events in state transition.** The network consists of five positive operators in a closed loop, initially all in the closed state. The state changes of each operator are recorded as varying trigger intervals are applied to the first operator.

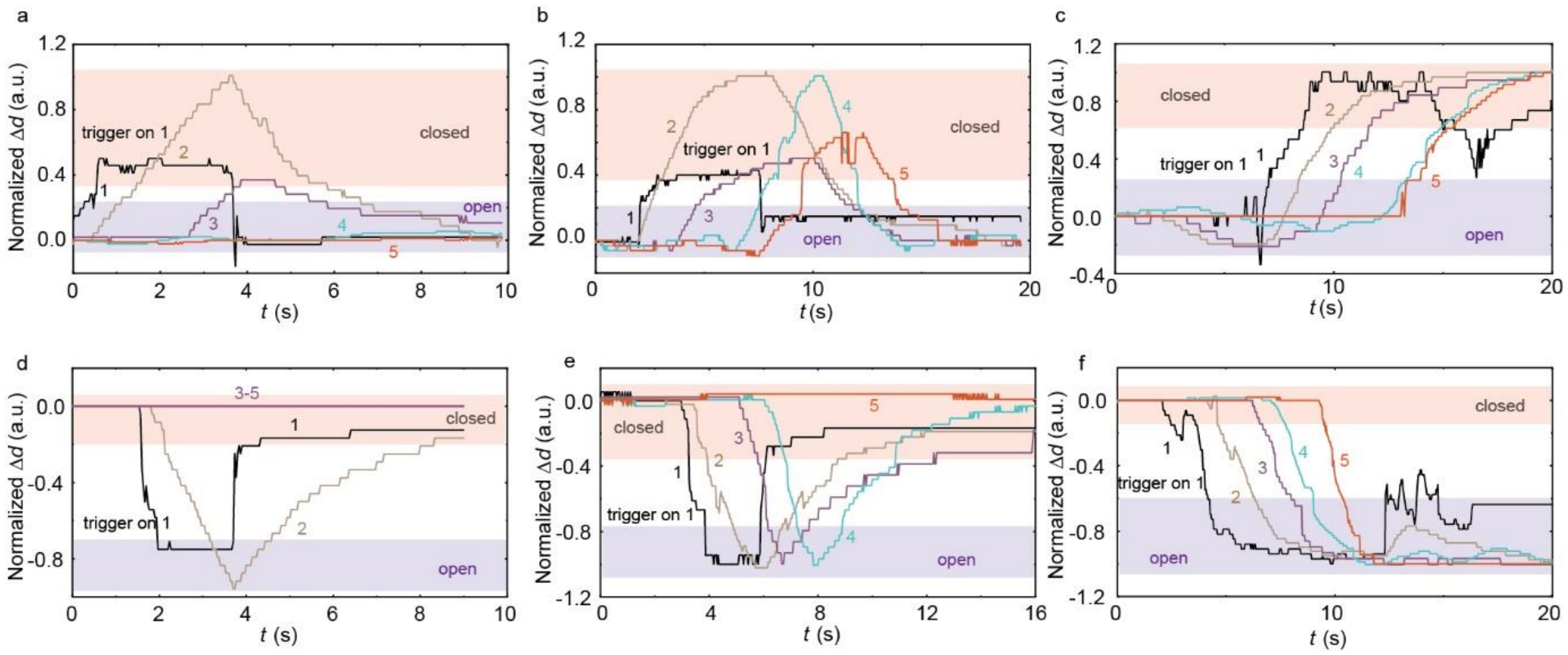

**Supplementary Figure 29. Displacement data during cascading transition.** (a-c) Initially, all operators are in the open state. The change in baffle position ($\Delta d$) for each operator is recorded by applying a mechanical trigger on operator 1 for (a) 2.5 s, (b) 5 s, and (c) 13 s. (d-f) Initially, all operators are in the closed state. The change in baffle position ($\Delta d$) for each operator is recorded by applying a mechanical trigger on operator 1 for (a) 2 s, (b) 4 s, and (c) 7 s.

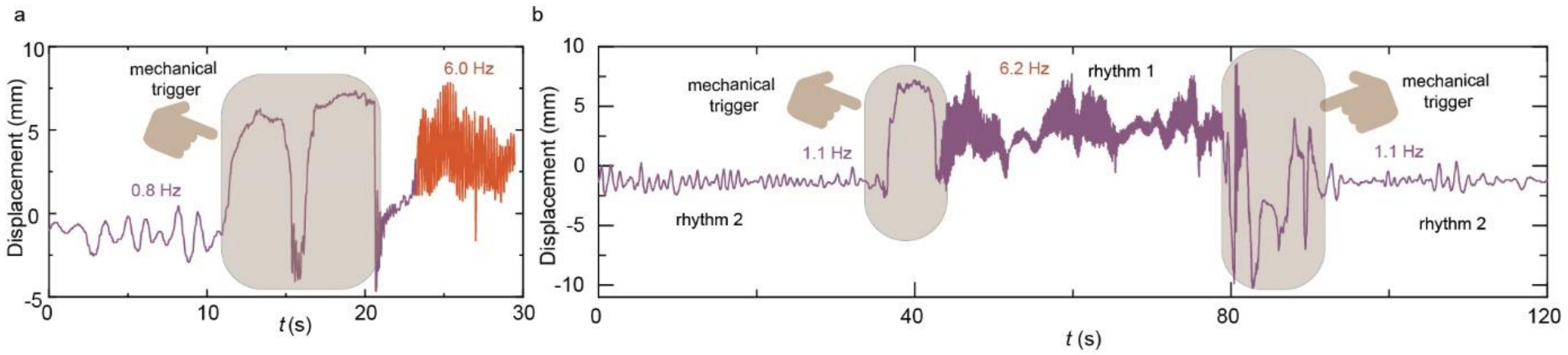

**Supplementary Figure 30. Change in rhythm.** (a) Oscillation data showing the transition from low-frequency to high-frequency oscillation triggered by a mechanical stimulus. (b) Oscillation data illustrating transitions from low to high frequency and back to low frequency, induced by different mechanical triggers.

## 2. Supplementary Note

Theoretical considerations are summarized as follows: In Section 2.1, dynamic equations for the single oscillator coupled with the photothermally responsive LCE model are derived. Numerical solutions of these equations reveal a Hopf bifurcation between static and self-oscillation states, yielding the time history of the single oscillator. Amplitude and period variations for different light powers are determined based on this time history. In Section 2.2, governing equations for two coupled oscillators are derived, emphasizing the time delay mechanism using the time histories of both oscillators. The impact of light powers on amplitude and period is examined. Results align qualitatively with experimental findings and can be understood through the time delay mechanism, as illustrated in Fig. 1 of the main manuscript.

### 2.1 Single oscillator

**Dynamics of the single oscillator. Figure 2.1** illustrates the single oscillator, consisting of an LCE beam, a baffle, a laser beam, and a mirror. Initially, the baffle allows the propagation of light, and the laser beam induces bending in the LCE through photothermal actuation. Consequently, due to LCE deformation, the baffle impedes the light beam. Subsequently, the light-induced bending rebounds, causing the LCE cantilever to unbend, thereby allowing the resumption of light propagation and initiating a new cycle. During the vibration, the baffle is subjected to the spring force $F_s$ of the LCE beam, and the damping force $F_d$, therefore the governing equation for its vibration is written as

$$m\frac{d^2 w(t)}{dt^2} = F_d + F_s, \tag{1}$$

where $m$ is the mass of the baffle, $w$(t) is the end deflection of the LCE cantilever. For simplicity, the damping force is assumed to be proportional to the velocity of the baffle, *i.e.*,

$$F_d = -\beta\frac{dw(t)}{dt}, \tag{2}$$

in which, $\beta$ is damping coefficient. The spring force of the LCE beam is assumed to be elastic bending deformation $w_e$,

$$F_s = \frac{3\Pi}{l^3} w_e(t), \tag{3}$$

in which, $\Pi$ is the bending stiffness, $l$ is the length of the LCE cantilever, and the elastic bending deformation depends on the light-driven bending $w_L(t)$ and current bending deformation and is

calculated as

$$w_e(t) = [w_L(t) - w(t)]. \tag{4}$$

For simplicity, the light-driven bending $w_L(t)$ is assumed to be related to the temperature difference $T_{dif}(t)$ between the temperature $T_{em}(t)$ at the illuminated position and the ambient temperature $T_{am}$, i.e.,

$$w_L = B(P)[T_{em}(t) - T_{am}], \tag{5}$$

in which, $B$ is the light-driven bending coefficient, depending on the detail of the system.

Inserting Eqs. (2)-(5), the governing equation (1) can be rewritten as

$$m\frac{d^2w(t)}{dt^2} = -\beta\frac{dw(t)}{dt} + \frac{3\Pi}{l^3}[BT_{dif}(t) - w(t)]. \tag{6}$$

In Eq. (6), temperature $T_{dif}$(t) depends on the light illumination and is process-related due to the transition between cut-on state and cut-off state of the baffle. The self-oscillation originates from the temperature variation and movement of the baffle.

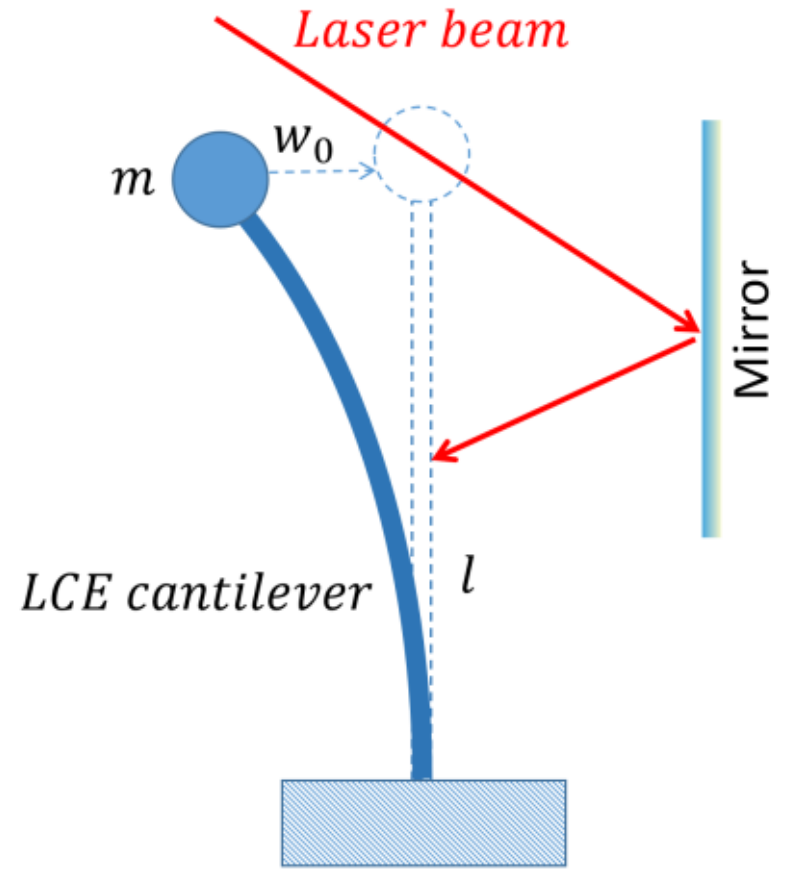

**Figure 2.1. Schematics of a feedback self-oscillator.** The oscillator is composed of a LCE cantilever and a mass block. The mass of the LCE cantilever is ignored.

**Model of a photothermally-responsive LCE beam.** To obtain the temperature $T_{em}(t)$ in Eq. (6), a photothermally-responsively LCE model is adopted. The LCE cantilever converts light energy into heat. Under light illumination, the temperature difference $T_{dif}$ is determined by the heat flux from the light and the heat transfer into the ambient, *i.e.*

$$\rho_c\dot{T}_{em} = q_l - q_{am}, \tag{7}$$

where $\rho_c$ is the specific heat capacity, $q_l$ is the heat absorbed by the LCE cantilever from light per second, which is assumed to be proportional to the light power $P$ of the light beam, i.e.,

$$q_l = \eta P, \tag{8}$$

where $\eta$ is the energy absorption coefficient. The LCE cantilever also exchanges heat with the environment, and the heat flux is assumed to be linear to the temperature difference $T_{\mathrm{dif}}$ between the LCE cantilever and the environment, *i.e.*,

$$q_{\mathrm{am}} = kT_{\mathrm{dif}}, \tag{9}$$

where $k$ is the heat transfer coefficient.

Inserting Eqs. (8) and (9) into (7) the temperature difference $T_{\mathrm{dif}}$ is governed by

$$\dot{T}_{\mathrm{dif}} = \frac{\eta P - kT_{\mathrm{dif}}}{\rho_c}. \tag{10}$$

Solving Eq. (10), in illuminated state, *i.e.* $w < w_0$ or $P \neq 0$, the temperature difference $T_{\mathrm{dif}}$ follows the law:

$$T_{\mathrm{dif}} = T_{\mathrm{Limit}}\left(1 - e^{-t/\tau_{\mathrm{heat}}}\right), \tag{11}$$

while in non-illuminated state, i.e. $w > w_0$ or $P$=0, the temperature difference $T_{\mathrm{dif}}$ follows the law:

$$T_{\mathrm{dif}} = T_{\mathrm{Limit}} e^{-t/\tau_{\mathrm{heat}}}, \tag{12}$$

where, $T_{\mathrm{Limit}} = \eta P/k$ represents the limit temperature difference of photothermally-responsive beam under longtime illumination, and $\tau_{\mathrm{heat}} = \rho_c/k$ reflects the characteristic time for heat exchange between photothermal-responsive beam and environment. Note that the larger $\tau_{\mathrm{heat}}$ indicates the longer time required for attaining the limited temperature difference $T_{\mathrm{Limit}}$ of the photothermally-responsive LCE cantilever.

**Nondimensionalization.** In this system, two time scales are present: inertial characteristic time $\tau_{\mathrm{inertial}} = \sqrt{ml^3/3\Pi}$ and heat time scale $\tau_{\mathrm{heat}} = \rho_c/k$. Actually, $\tau_{\mathrm{inertial}}$ is the reciprocal of the natural angular frequency $\omega_0$. By introducing the dimensionless parameters $\bar{w} = w/l$, $\bar{w}_{\mathrm{L}} = w_{\mathrm{L}}/l$, $\bar{t} = t/\tau_{\mathrm{inertial}}$, $\bar{\beta} = \beta\tau_{\mathrm{inertial}}/m$, $\bar{w}_0 = w_0/l$, and $\bar{\tau}_{\mathrm{heat}} = \tau_{\mathrm{heat}}/\tau_{\mathrm{inertial}}$, Eq. (6) can be rewritten as

$$\frac{\mathrm{d}^2\bar{w}(t)}{\mathrm{d}\bar{t}^2} = -\bar{\beta}\frac{\mathrm{d}\bar{w}(t)}{\mathrm{d}\bar{t}} + \bar{w}_L(t) - \bar{w}(t). \tag{13}$$

Then, Eq. (11) and (12) can also be rewritten as,

in illuminated state, i.e.$\bar{w} \leq \bar{w}_0$:

$$\bar{w}_L(\bar{t}) = \bar{A}\left(1 - e^{-\bar{t}/\bar{\tau}_{\mathrm{heat}}}\right), \tag{14}$$

in non-illuminated state, i.e. $\bar{w} > \bar{w}_0$:

$$\bar{w}_L(\bar{t}) = \bar{A}e^{-\bar{t}/\bar{\tau}_{\mathrm{heat}}}, \tag{15}$$

where $\bar{A} = \frac{\eta P}{kl}B(P)$ denotes the dimensionless limit deflection of photothermally-responsive LCE cantilever under longtime illumination. In general cases, the light-driven limit deflection is nonlinearly related to the laser power. Herein, we assume that $\bar{A}$=a(1−$e^{-P/P0}$) in the following calculations.

Eqs. (13-15) governing the vibration of the single oscillator under light beam.

**Amplitude and period of a single oscillator.** By numerically solving the governing equations (13)-(15), **Figure 2.2** plots the time history of the displacement of the baffle, for different laser powers. In the calculation, we set $l$ =2.5 cm, $w_0$= 0.5 mm, $\bar{\beta}$ = 0.5, $\tau_{\mathrm{inertial}}$ = 0.017 s, $\tau_{\mathrm{heat}}$ = 0.17 s, $a = 0.131$, and $P_0$ = 61 mW. It is shown that the amplitude of self-oscillation increases monotonously with increasing laser power. **Figure 2.3** plots the dependence of self-oscillation amplitude on the laser power. The result shows that there exists a Hopf bifurcation between the static state and the self-oscillation state. Above the Hopf bifurcation point $P_{\mathrm{crit}}$=100 mW, the amplitude of self-oscillation increases with the laser power. In addition, the theoretical prediction aligns with the experimental results, as shown in **Figure 2.3**.

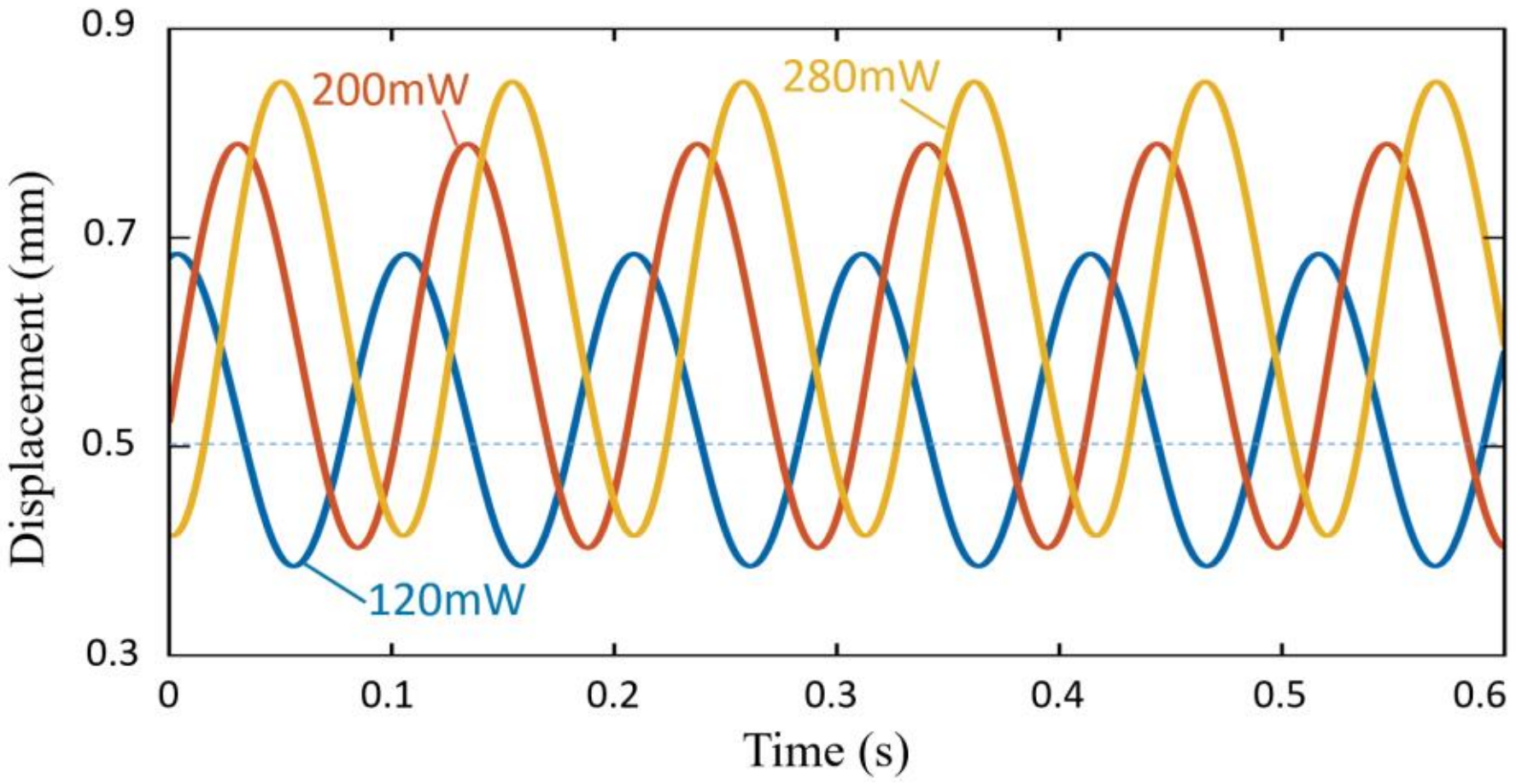

**Figure 2.2. Numerical results.** Time history of the displacement of the mass block, for three different laser powers $P$=120mW, 200mW, and 280mW. In the simulation, we set $l$ =2.5 cm, $w_0$= 0.5 mm, $\bar{\beta}$ = 0.5, $\tau_{\mathrm{inertial}}$ = 0.017 s, $\tau_{\mathrm{heat}}$ = 0.17 s, $a = 0.131$, and $P_0$=61 mW.

Meanwhile, the numerical results show that the period is mainly determined by the natural period $T_0$ = 0.1 s. This theoretical prediction is consistent with the experiment as shown in Figure 3. From Eq. (13), the period of the self-oscillation is derived as,

$$T = \frac{2\pi}{\omega_0\sqrt{1-\frac{\bar{\beta}^2}{4}}}. \tag{16}$$

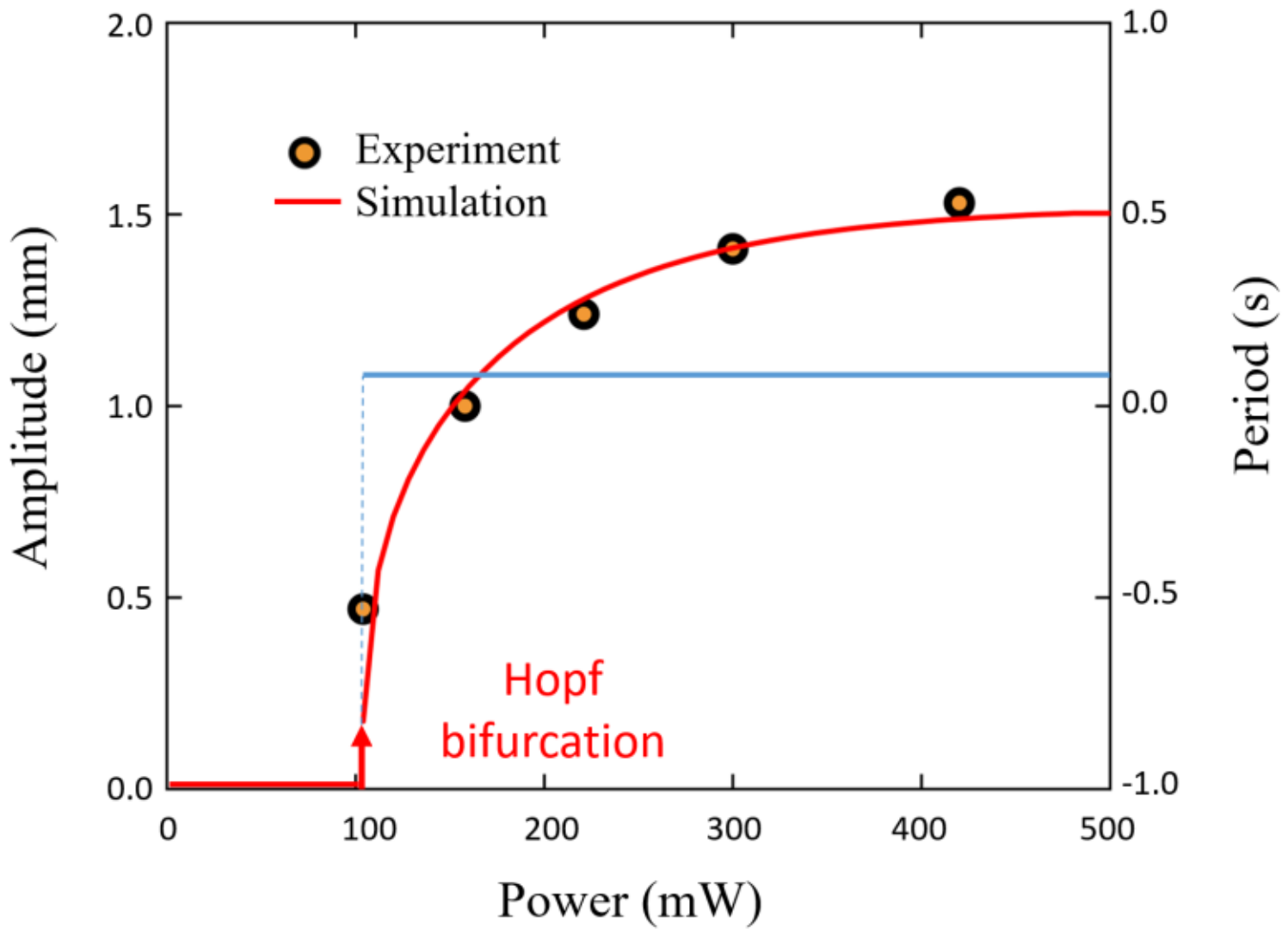

**Figure 2.3. Dependences of amplitude and period of single oscillator on the laser power.** In the simulation, we set $l$ = 2.5 cm, $w_0$= 1.5 mm, $\bar{\beta}$ = 0.2, $\tau_{\text{inertial}}$ = 0.013 s, $\tau_{\text{heat}}$ = 0.13 s, $a = 0.099$, and $P_0$=66 mW. Laser spot size: 2 mm.

## 2.2 Coupled oscillators.

**Governing equations of coupled oscillators.** Two light beams are used to connect the two LCE beams which are coupled, as shown in **Figure 2.4**. Light beam (-) goes near the edge of the baffle on the LCE cantilever (+), light beam (+) is initially blocked by the baffle on the LCE cantilever (-) (**Figure 2.4 a**). When light beam (-) hits the LCE cantilever (-), it unblocks light beam (+) (**Figure 2.4 b**). Then, light beam (+) excites LCE cantilever (+) blocking light beam (-) (**Figure 2.4 c**). As a result, LCE cantilever (-) retains its original position, blocking light beam (+) (**Figure 2.4 d**), and LCE cantilever (+) relaxes to the original position to unblock light beam (-). The system returns to the initial state (**Figure 2.4 a**), and a new cycle starts.

The end deflections of LCE cantilevers (-) and (+) are denoted by $w_-(t)$ and $w_+(t)$. The powers of light beams on LCE cantilevers (-) and (+) are denoted by $P_-$ and $P_+$. For two coupled oscillators, the governing equations are written as

$$\frac{\mathrm{d}^2\bar{w}_-(t)}{\mathrm{d}\bar{t}^2} = -\bar{\beta}\frac{\mathrm{d}\bar{w}_-(t)}{\mathrm{d}\bar{t}} + \bar{w}_{\mathrm{L}\text{-}}(t) - \bar{w}_-(t), \tag{17}$$

$$\frac{\mathrm{d}^2\bar{w}_+(t)}{\mathrm{d}\bar{t}^2} = -\bar{\beta}\frac{\mathrm{d}\bar{w}_+(t)}{\mathrm{d}\bar{t}} + \bar{w}_{\mathrm{L}+}(t) - \bar{w}_+(t), , \tag{18}$$

where

$$\bar{w}_{\mathrm{L}\text{-}}(\bar{t}) = \begin{cases} \bar{A}_-\left(1 - e^{-\bar{t}/\bar{\tau}_{\mathrm{heat}}}\right), \bar{w}_+ < \bar{w}_{0+} \\ \bar{A}_- e^{-\bar{t}/\bar{\tau}_{\mathrm{heat}}}, \bar{w}_+ > \bar{w}_{0+} \end{cases}, \tag{19}$$

$$\bar{w}_{\mathrm{L}+}(\bar{t}) = \begin{cases} \bar{A}_+\left(1 - e^{-\bar{t}/\bar{\tau}_{\mathrm{heat}}}\right), \bar{w}_- > \bar{w}_{0-} \\ \bar{A}_+ e^{-\bar{t}/\bar{\tau}_{\mathrm{heat}}}, \bar{w}_- < \bar{w}_{0-} \end{cases}, \tag{20}$$

in which, $\bar{A}_- = a_-(1 - e^{-P_-/P_{0-}})$, $\bar{A}_+ = a_+(1 - e^{-P_+/P_{0+}})$, $\bar{w}_{0-}$ and $\bar{w}_{0+}$ denotes the on/off transition critical deflections of LCE cantilevers (-) and (+), respectively.

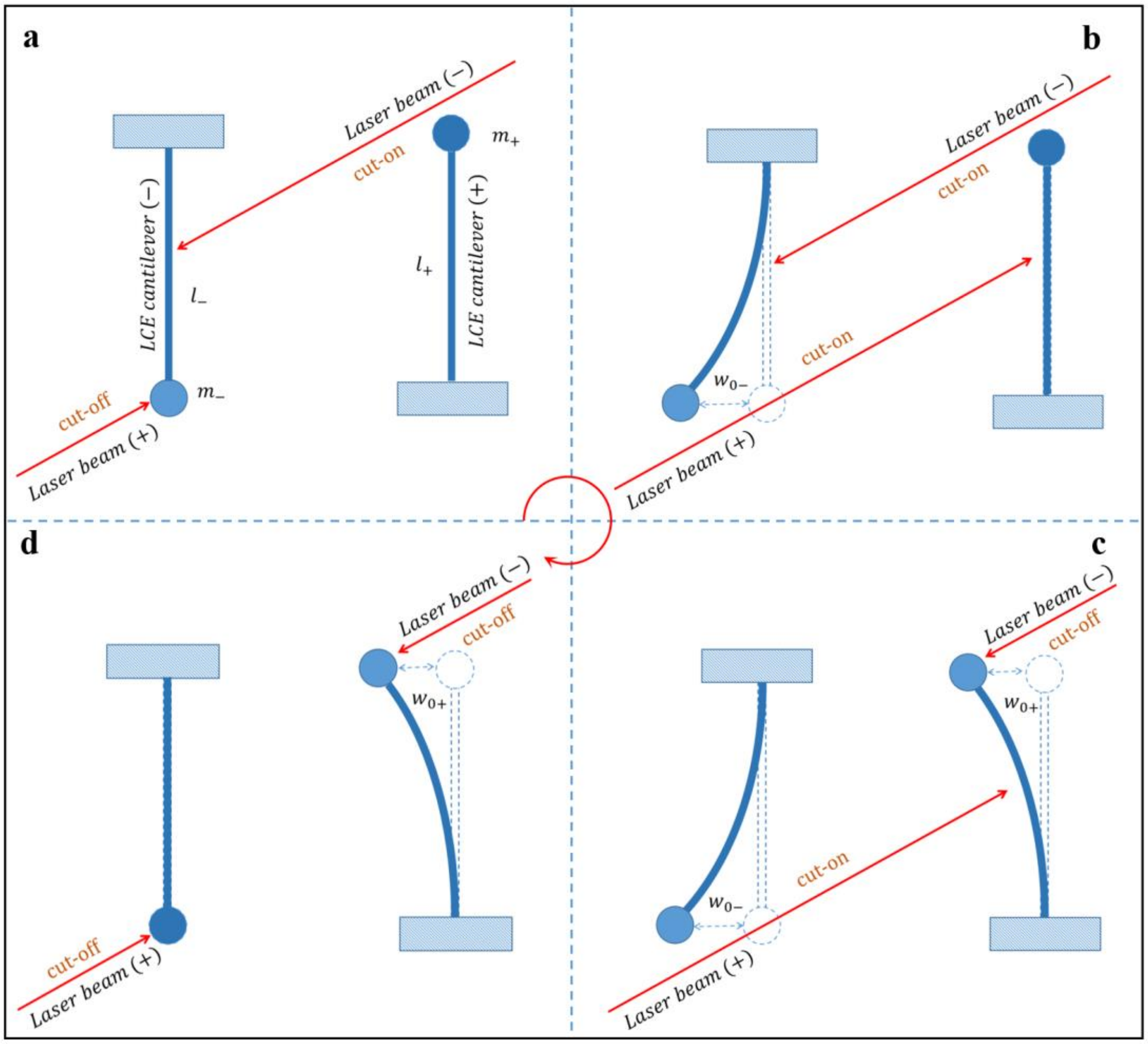

**Figure 2.4. Schematics of two coupled oscillators.** The mass of the LCE cantilever is ignored.

**Delay mechanism of two coupled oscillators.** By solving Eqs. (17)-(20), Figure 5 plots the time histories of the two coupled oscillators. It is shown that there exists a time delay $t_{\mathrm{d}}$ between cut-off position and the equilibrium position, which is much different from the single oscillator discussed above. In the single oscillator, there is no time delay between cut-off position and the equilibrium position.

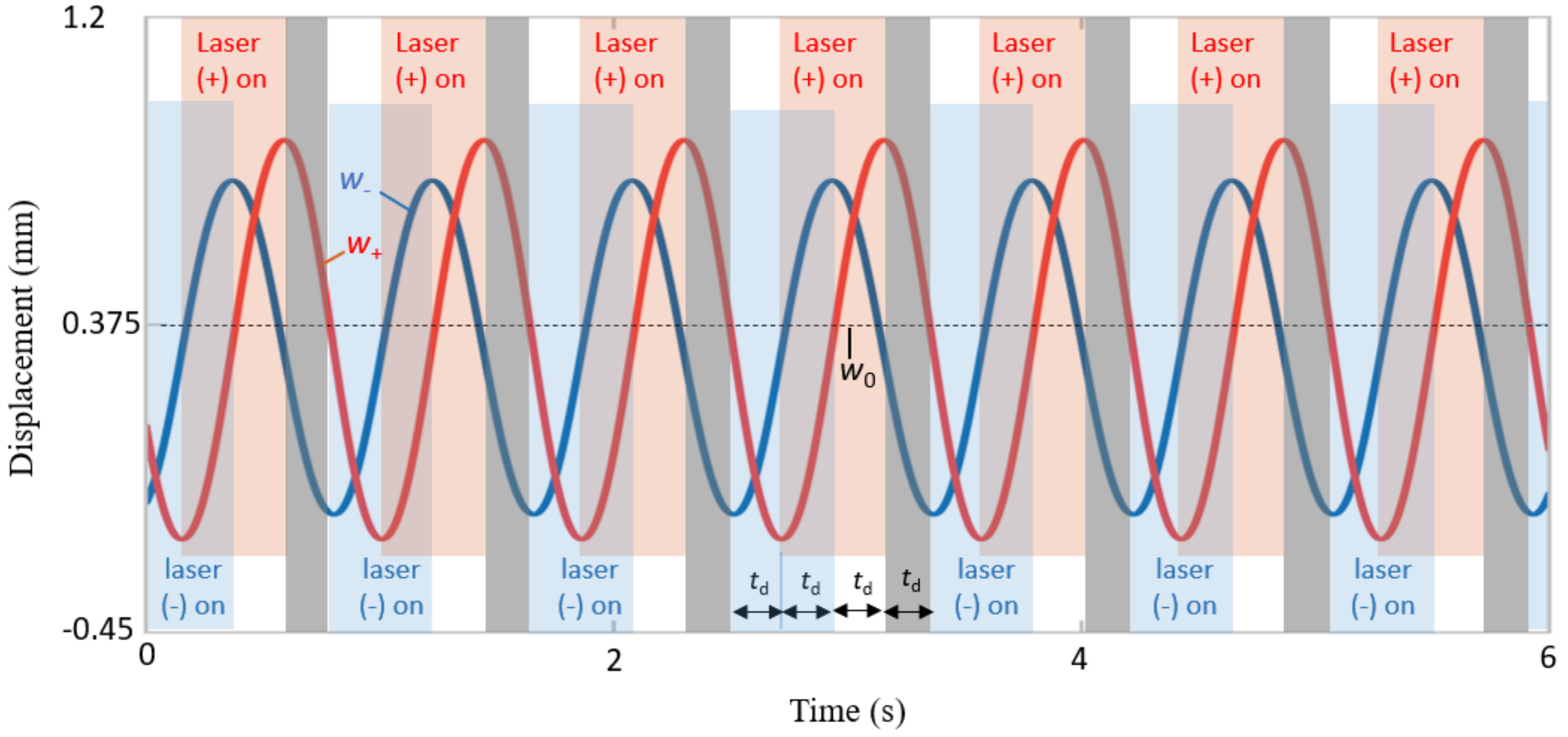

**Figure 2.5. Time histories of two coupled oscillators.** There exists time delay $t_d$ in the oscillations of the two coupled oscillators. In the computation, we set $l_- = 2.5$ cm, $l_+ = 2.5$ cm, $w_{0-} = 0.375$ mm, $w_{0+} = 0.375$ mm, $\bar{\beta} = 0.2$, $\tau_{\text{inertial}} = 0.11$s, $\tau_{\text{heat}} = 0.275$ s, $P_- = 200$ mW, $P_+ = 250$ mW, $a_- = 0.36$, $a_+ = 0.35$, $P_{0-} = 3$W, $P_{0+} = 3$W, spot size for laser (+), 2 mm, size for laser (-), 3 mm.

**Amplitude and period of two coupled oscillators. Figure 2.6** plots the dependence of the amplitude and period of the coupled oscillators on the laser power $P_-$. There exists a Hopf bifurcation between the static state and oscillation state. For laser power $P_-$ smaller than the Hopf bifurcation point $P_{crit}$=100 mW, the LCE cantilever (-) bending is not enough to cut off the light beam (+), and the circle cannot evolve. The prediction is qualitatively consistent with the experimental results.

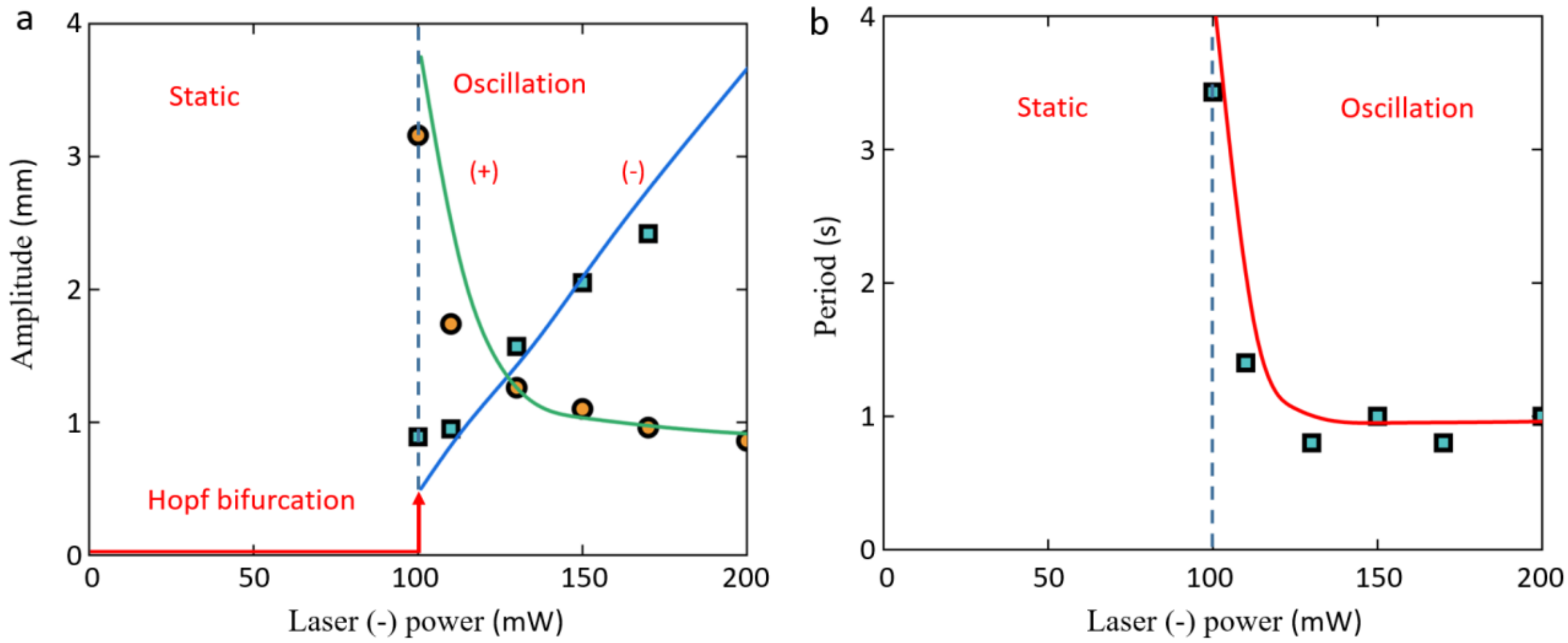

**Figure 2.6. Amplitude and period of two coupled oscillators.** In the computation, we set $l_- = 2.5$ cm, $l_+ = 2.5$ cm, $w_{0-} = 4.25$ mm, $w_{0+} = 4$mm, $\bar{\beta} = 0.8$, $\tau_{\text{inertial}} = 0.017$s, $\tau_{\text{heat}} = 0.34$ s, $P_+ = 180$ mW, $a_- = 3.42$, $a_+ = 1.83$, $P_{0-} = 2$W, $P_{0+} = 2$W. Spot size for laser (+), 2 mm, size for laser (-), 3 mm. Dots: experimental results. Solid lines: theoretical predictions.

Increase in one beam's power leads to the amplitude elevating of the operator such a beam directly excites on, and a decrease in the amplitude of the other operator that controls the beam.

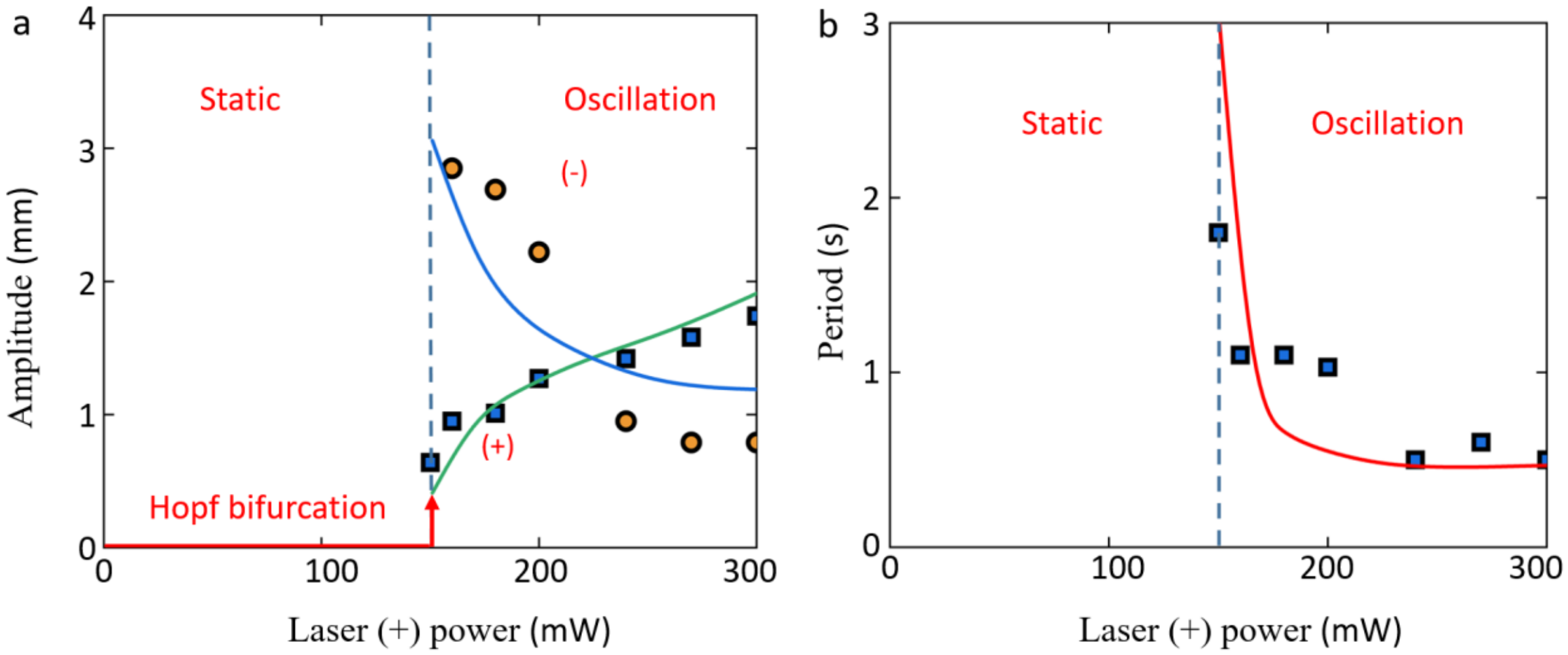

**Figure 2.7. Amplitude and period of two coupled oscillators.** In the computation, we set $l_- = 2.5$ cm, $l_+ = 2.5$ cm, $w_{0-} = 1.625$ mm, $w_{0+} = 3.75$ mm, $\bar{\beta} = 0.8$, $\tau_{\text{inertial}} = 0.02$s, $\tau_{\text{heat}} = 0.4$ s, $P_- = 150$ mW, $a_- = 2.47$, $a_+ = 2.01$, $P_{0-} = 2$ W, $P_{0+} = 2$ W. Spot size for laser (+), 2 mm, size for laser (-), 3 mm. Dots: experimental results. Solid lines: theoretical predictions.

Meanwhile, **Figures 2.6** and **2.7** show that increase in one beam's power leads to the amplitude elevating of the operator such a beam directly excites on, and a decrease in the amplitude of the other operator that controls the beam. In addition, the period decreases with the increase of the laser power $P$. These results can be understood from the effect of laser power on the time delay. The time delay decreases with increasing laser power, due to quick deflection and a faster cut-OFF action, thus shortening the oscillation cycle.

From **Figure 2.5**, it can be seen that the period is approximately four times the time delay. For given critical deflection $\bar{w}_{10} \approx \bar{w}_{20} \approx \bar{w}_0$ and $\bar{A}_- \approx \bar{A}_+ \approx \bar{A}$, the time delay can be approximately estimated from Eqs. (19) and (20) as $\Delta t \approx -\tau_{\mathrm{heat}} \ln\left(1 - \frac{\bar{w}_0}{\bar{A}}\right)$, and the period can be estimated as

$$T \approx -4\tau_{\mathrm{heat}} \ln\left(1 - \frac{\bar{w}_0}{\bar{A}}\right). \tag{21}$$

From Eq. (21), it can also be understood that the period decreases with increasing light power.

## 2.3 Sensing and control

For robotic applications, the light communication concept provides a method for remote sensing and electrical control of soft actuators over long distances. Figure 2.8a illustrates that the deformation of operator 1 affects the transmission intensity of the nearby light beam. By using a photodiode detector, the deformation signal from operator 1 is converted into an electrical signal, which is then received at the location of operator 2, as shown by the displacement and voltage data in Figure 2.8b. This enables remote sensing, such that, for example, a wind disturbance affecting operator 1 can be detected by the variation in the electrical signal at the position of operator 2 (Figure 2.8c). More detailed remote sensing data, including long- and short-period perturbations from wind and mechanical triggers, can be found in Supplementary Fig. 2.9.

Conversely, a tethered voltage signal can induce deformation of operator 1 via an electromagnetic coil, while a light beam can independently control operator 2 over a long distance. The working principle is illustrated in Figure 5d, with setup images provided in Supplementary Fig. 2.10. When a voltage is applied to the electromagnetic coil, the magnetic field generates an attractive force that deforms operator 1, thereby activating the light beam, as shown by the displacement and light power data in Figure 2.8e. The transmitted beam then excites operator 2, causing deformation. Thus, the voltage signal near operator 1 can remotely control the deformation of operator 2 (Figure 2.8f). A negative voltage creates a propelling force that deforms operator 1 in the opposite direction, causing the light transmission to close. Control data based on the negative voltage signal is shown in Supplementary Fig. 2.11. Robotic control can also be achieved by modulating the signal with different bandwidths, as demonstrated in Figure 2.8g and Supplementary Fig. 2.12.

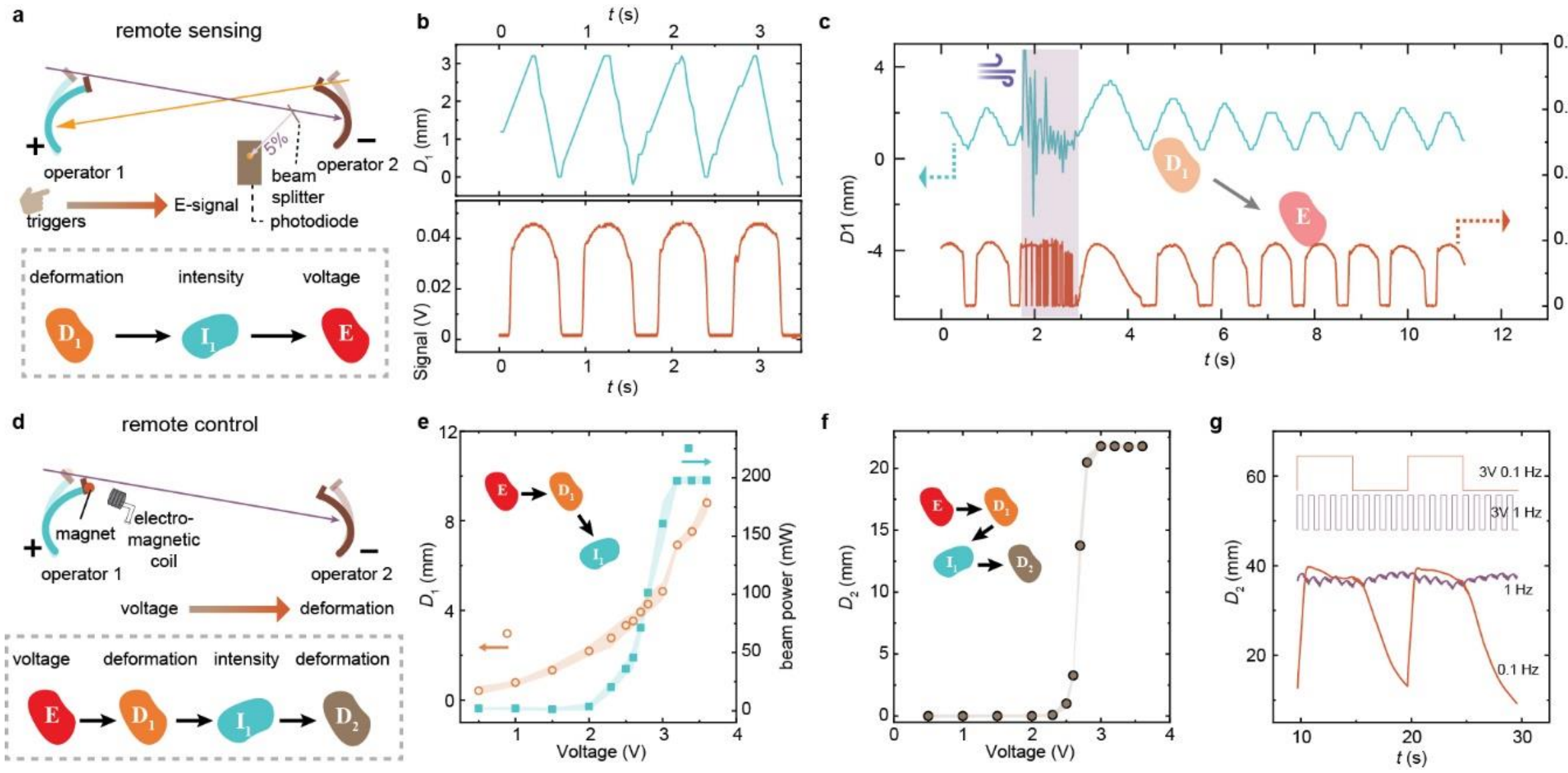

**Figure 2.8. Remote sensing and robotic control.** (a) Top: Schematic illustration of the two-unit network connected via two light beams. Bottom: Mechanism of signal transmission, showing how material deformation is converted into an electrical signal at a remote distance. (b) Displacement data from operator 1 and the corresponding electric signal received near operator 2. (c) Detection of wind disturbance at operator 1 and subsequent signal transition observed at the detector near operator 2. (d) Top: Schematic of the setup for remote deformation control. Bottom: Explanation of the remote-control mechanism. (e) Displacement of operator 1 and transmitted light power at various applied voltages. (f) Displacement of operator 2 in response to voltage changes applied to the electromagnetic coil near operator 1. (g) Modulation of operator 2 via the application of a square wave signal voltage to the electromagnetic coil. *E*, electric voltage generated by a photodiode detector or voltage applied to an electromagnetic coil. $I_1$, intensity of beam 1. $D_1$, displacement of operator 1. $D_2$, displacement of operator 2. (b, c) Laser 1: 532 nm, 256 mW, 3 mm. Laser 2: 532 nm, 200 mW, 2 mm. LCE sample dimensions: $24 \times 2 \times 0.1$ mm$^3$, baffle size: $5 \times 20 \times 0.01$ mm$^3$. (e-g) Laser: 532 nm, 200 mW, 2 mm.

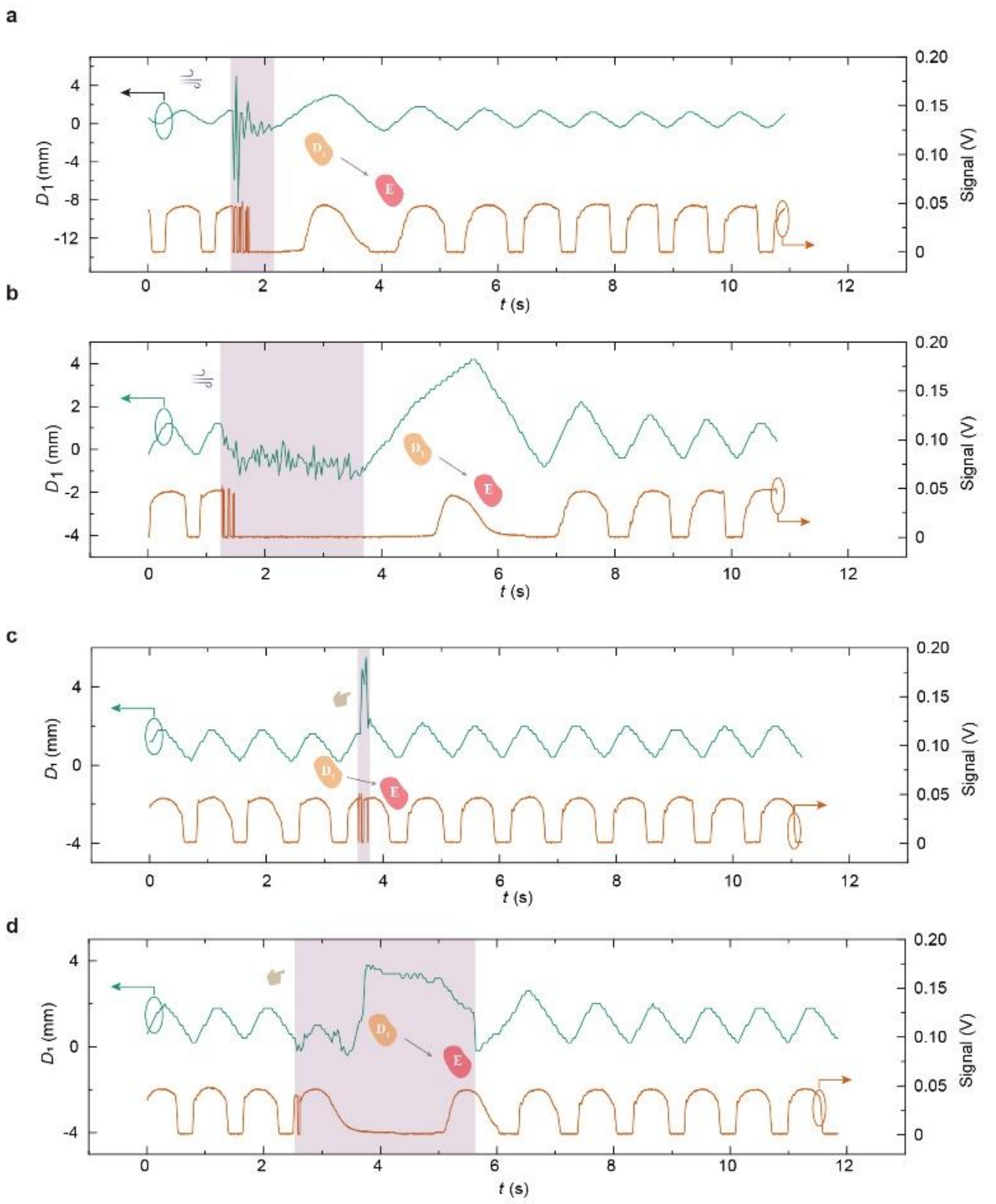

**Supplementary Figure 2.9. Remote sensing.** Oscillation data of the coupled oscillators in response to external wind disturbance applied to operator 1 for (a) short and (b) long durations, and the received electrical signal at a position near operator 2. Oscillation data of the coupled oscillators in response to external mechanical disturbance applied to operator 1 for (c) short and (d) long durations, and the received electrical signal at a position near operator 2. Laser 1: 532 nm, 256 mW, 3 mm. Laser 2: 532 nm, 200 mW, 2 mm. LCE sample dimensions: $24 \times 2 \times 0.1$ mm$^3$, baffle size: $5 \times 20 \times 0.01$ mm$^3$. Scale bar: 2 cm.

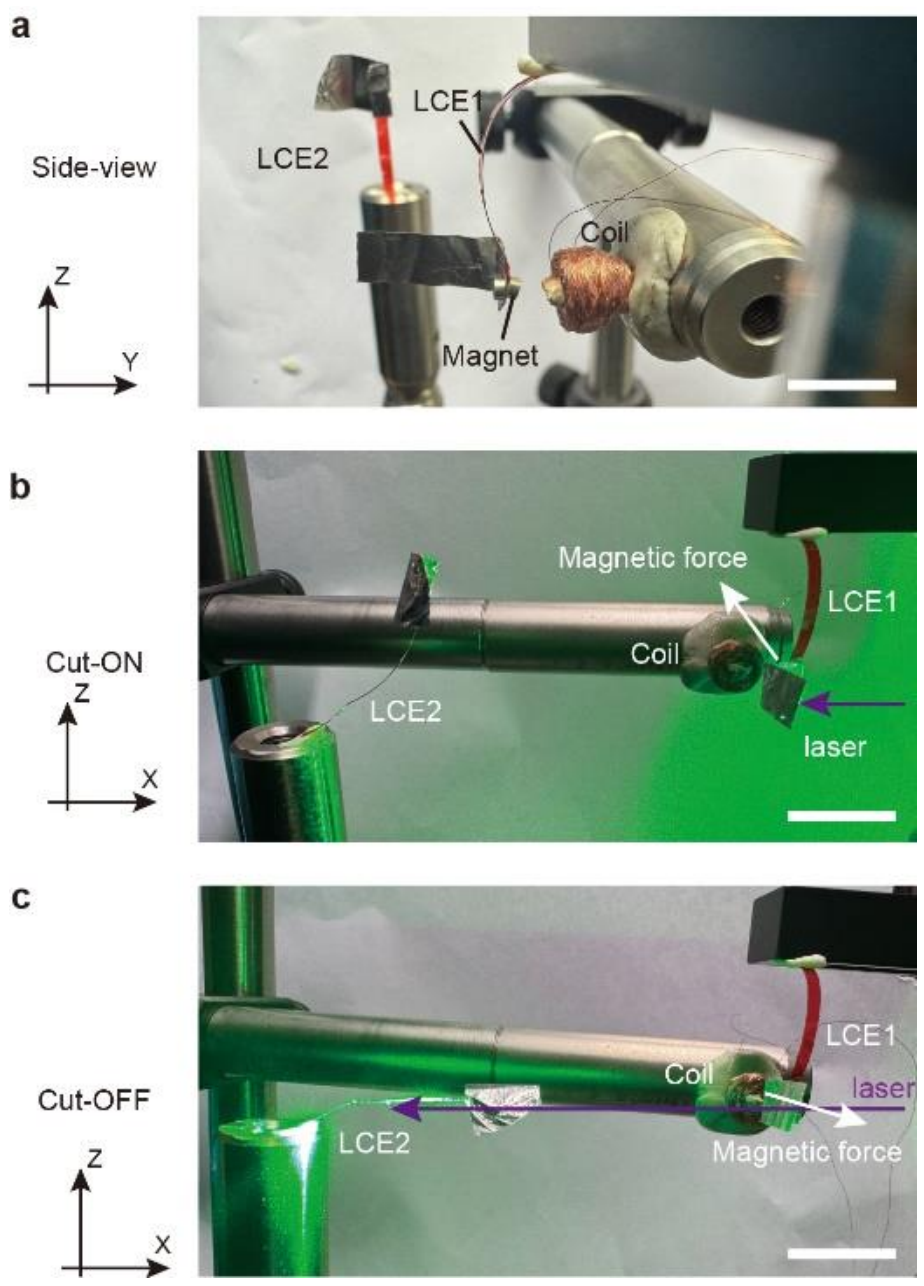

**Supplementary Figure 2.10. Photograph of the electric control system.** (a) Side view showing the light-coupled components and electromagnetic coil. Front view images of the system with (b) the 'cut-ON' state activated by a positive voltage and the 'cut-OFF' state activated by a negative voltage signal. Two magnets are mounted on the baffle of one component near the coil. Scale bar is 1 cm. Magnet: Neodymium 50 magnet, 2mm diameter × 1mm thick, 23mg weight. The coil is made of copper wire with 1,000 turns.

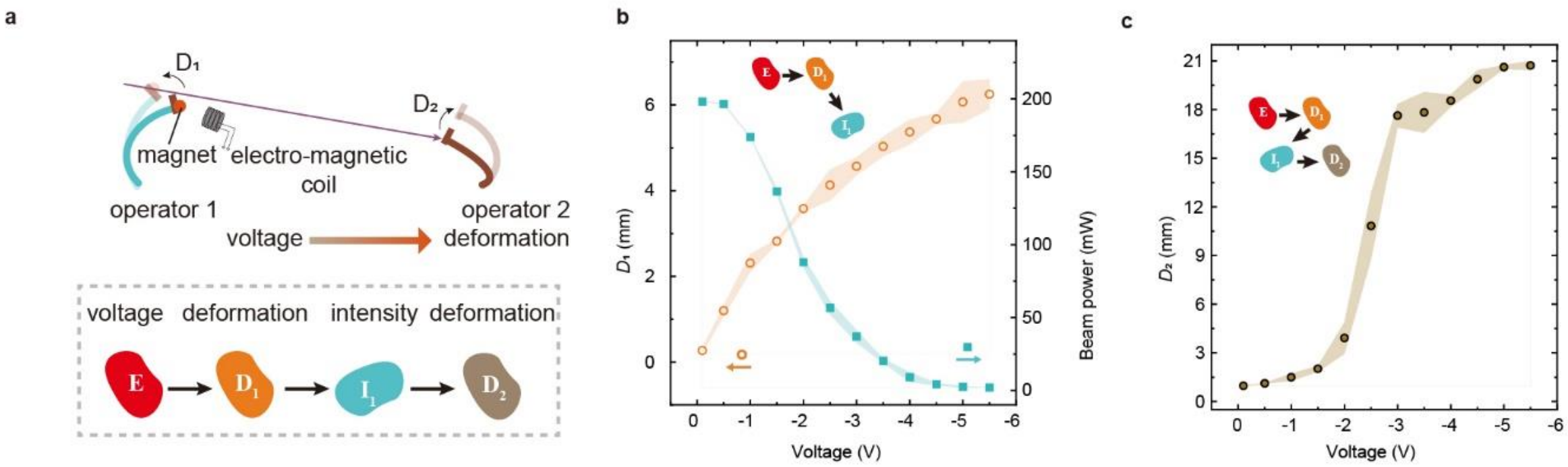

**Supplementary Figure 2.11. Remote control system.** (a) Top: Schematic of the setup for remote deformation control. Bottom: Explanation of the remote-control mechanism. (b) Displacement of operator 1 and transmitted light power at various applied voltages. (f) Displacement of operator 2 in response to voltage changes applied to the electromagnetic coil near operator 1. $E$, electric voltage applied to the electromagnetic coil. $I_1$, intensity of beam 1. $D_1$, displacement of operator 1. $D_2$, displacement of operator 2. Laser 1: 532 nm, 200 mW, 2 mm. LCE sample dimensions: $24 \times 2 \times 0.1$ mm$^3$, baffle size: $5 \times 20 \times 0.01$ mm$^3$.

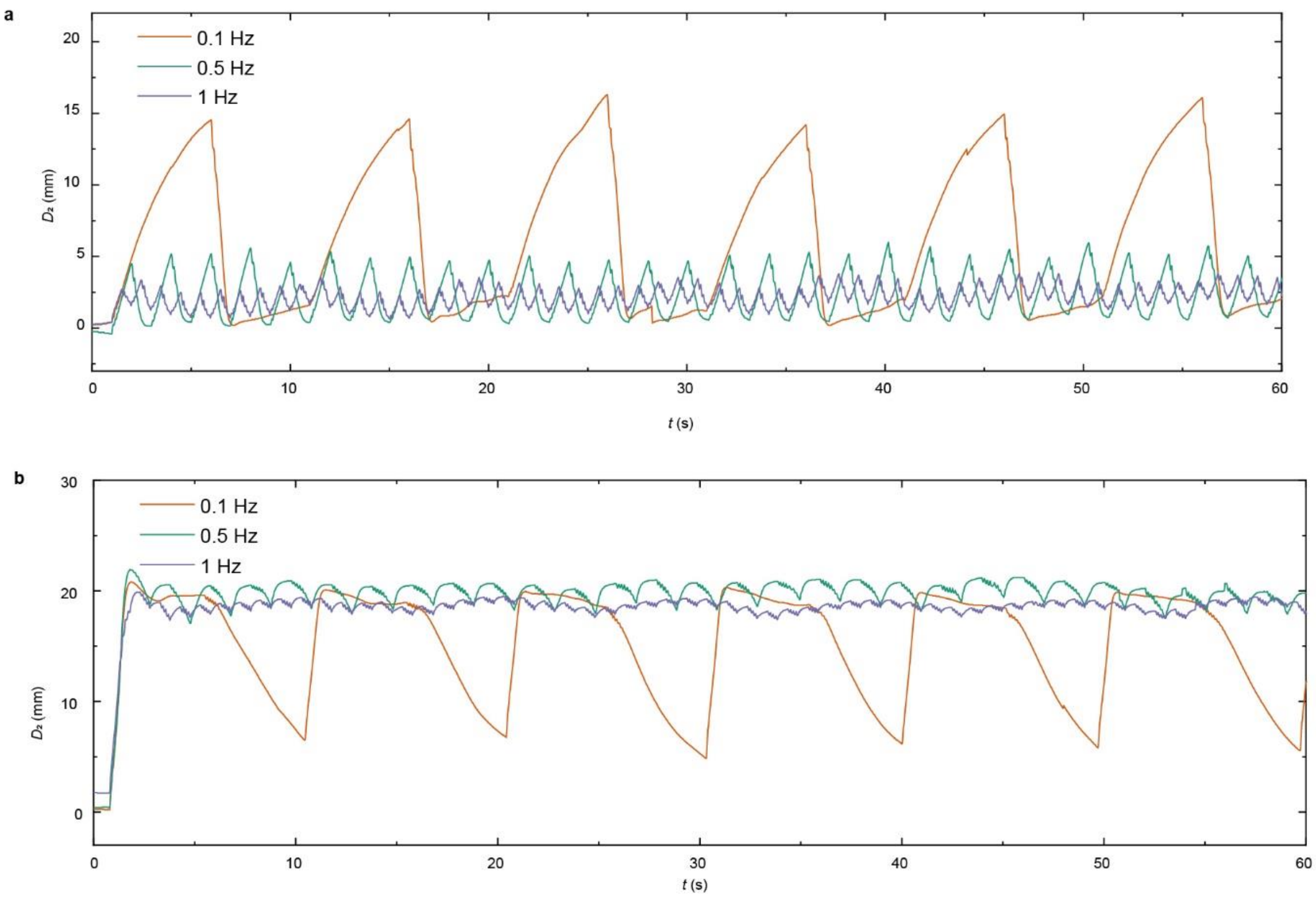

**Supplementary Figure 2.12. Remote control data.** Modulation of operator 2 via the application of a square wave voltage signal. Voltage: (a) 0 to -5 V, (b) 0 to 3 V. 50% duty cycle. $D_2$, displacement of operator 2.

## 4. Supplementary Video captions

**Supplementary Movie 1. Coupled self-oscillators.**
This real-time video shows continuous oscillations of two coupled operators. The black markings facilitate position tracking during the oscillation. Laser 1 (right to left): 532 nm, 150 mW, 2 mm spot size. Laser 2 (left to right): 635 nm, 200 mW, 3 mm spot size. LCE actuator dimensions: $24 \times 2 \times 0.1$ mm$^3$. Baffle dimension: $5 \times 20 \times 0.01$ mm$^2$. The optical pathway follows the same design in Fig. 1f.

**Supplementary Movie 2. Coupling between two isolated self-oscillators.**
This real-time video shows continuous oscillations of two coupled operators with a screen in the middle. The laser beams are guided around the screen and reflected onto the sample through mirrors. Laser 1 (exciting on left sample): 532 nm, 150 mW, 2 mm spot size. Laser 2 (exciting on right sample): 635 nm, 300 mW, 3 mm spot size. LCE actuator dimensions: $24 \times 2 \times 0.1$ mm$^3$. Baffle dimension: $5 \times 20 \times 0.01$ mm$^2$. The optical pathway follows the same design in Fig. 2a.

**Supplementary Movie 3. Coupling between two self-oscillators through optical fibers.**
This real-time video shows continuous oscillations of two coupled operators through optical fibers. Laser 1 (exciting on upper sample): 532 nm, 256 mW, 3 mm spot size. Laser 2 (exciting on bottom sample): 532 nm, 200 mW, 2 mm spot size. LCE actuator dimensions: $24 \times 2 \times 0.1$ mm$^3$. Baffle dimension: $5 \times 20 \times 0.01$ mm$^2$. The samples and connections to fibers are shown in Fig. 2h.

**Supplementary Movie 4. Coupled network composed of three units.**
This video shows continuous oscillations of three coupled operators. The laser beams are guided through optical fibers. All laser beams: 532 nm, 320 mW, 2 mm spot size at the sample positions. LCE actuator dimensions: $24 \times 2 \times 0.1$ mm$^3$. Baffle dimension: $5 \times 20 \times 0.01$ mm$^2$. The movie is played with 4× accelerated speed. The optical pathway is schematically shown in Supplementary Figure 18c.

**Supplementary Movie 5. Coupled network composed of four units.**
This video shows continuous oscillations of four coupled operators. The laser beams are guided through optical fibers. All laser beams: 532 nm, 320 mW, 2 mm spot size at the sample positions. LCE actuator dimensions: $24 \times 2 \times 0.1$ mm$^3$. Baffle dimension: $5 \times 20 \times 0.01$ mm$^2$. A red coloured filter is used to block laser wavelengths for video recording. The movie is played with 4× accelerated speed. The optical pathway is schematically shown in Supplementary Fig. 18b.

**Supplementary Movie 6. The coupling between a thermometer and LCE actuator.**
This video shows continuous oscillations of the coupled system. Laser 1 (exciting on LCE): 532 nm, 44 mW, 2 mm spot size. Laser 2 (exciting on thermometer): 532 nm, 930 mW, 3 mm spot size. LCE actuator dimensions: $24 \times 2 \times 0.1$ mm$^3$. Baffle dimension: $5 \times 20 \times 0.01$ mm$^2$. The movie is played with 2× accelerated speed. The optical pathway is schematically shown in Supplementary Fig. 24c.

**Supplementary Movie 7. The coupling between the paraffin and LCE actuator.**

This video shows continuous oscillations of the coupled system. Laser 1 (exciting on LCE): 532 nm, 80 mW, 1.2 mm spot size. Laser 2 (exciting on paraffin): 532 nm, 1440 mW, 2 mm spot size. LCE actuator dimensions: 24 × 2 × 0.1 $mm^3$. Baffle dimension: 5 × 20 × 0.01 $mm^2$. The movie is played with 16× accelerated speed. The optical pathway follows the same design in Supplementary Fig. 26.

**Supplementary Movie 8. Cascading transition.**

This real-time video shows the cascading transition from ON to OFF state and then from OFF to ON state. All Laser: 532 nm, 100 mW, 2 mm spot size. LCE actuator dimensions: 16 × 2 × 0.1 $mm^3$. Baffle dimension: 4 × 12 × 0.01 $mm^2$. The optical pathway follows the same design in Fig. 4e.

**Supplementary Movie 9. Dual rhythm in self-oscillation.**

This real-time video shows the mechanical trigger induced transition between two oscillation frequencies. Laser 1 excited on operator 1: 532 nm, 180 mW, 2 mm; Laser 2 spot on operator 2: 532 nm, 180 mW, 2 mm; Laser 3 spot on operator 1: 532 nm, 160 mW, 2 mm. LCE fiber dimensions: 1.6 cm in length and 1 mm in diameter. Baffle dimension: 4 × 12 × 0.01 $mm^2$. The optical pathway follows the same design in Fig. 4h.